\documentclass[11pt]{article}
\usepackage[utf8]{inputenc}
\usepackage{caption}
\usepackage{braket,slashed,bm}
\usepackage{jheppub}
\usepackage{simplewick}
\usepackage{array,multirow}
\usepackage{amsmath} \DeclareMathAlphabet{\mathpzc}{OT1}{pzc}{m}{it}
\usepackage[normalem]{ulem}
\usepackage{calligra}
\usepackage{floatrow}
\DeclareMathAlphabet{\mathpzc}{OT1}{pzc}{m}{it}
\usepackage{mathrsfs}
\usepackage{tikz}

\usepackage{subcaption}
\usepackage{lscape}
\usepackage{hyperref}
\usepackage{graphicx}
\usepackage{booktabs}
\newfloatcommand{capbtabbox}{table}[][\FBwidth]

\graphicspath{{./Figures/}}
\newcommand{\titlemath}{\texorpdfstring}
\def\beq{\begin{equation}}
\def\eeq{\end{equation}}
\def\bea{\begin{eqnarray}}
\def\eea{\end{eqnarray}}
\def\nn{\nonumber \\}
\def\hyp{\mathsf{y}}
\renewcommand{\a}{\alpha}

\newcommand{\gsm}[2]{g^{#1,{\rm SM}}_{#2}}

\newcommand{\Lagr}{\mathcal{L}}
\renewcommand{\to}{\rightarrow}

\begin{document}
\title{Exact SMEFT formulation and expansion to ${\cal{O}}(v^4/\Lambda^4)$}

\author[a]{Chris Hays,}

\author[b]{Andreas Helset,}

\author[c]{Adam Martin,}

\author[b]{Michael Trott}

\affiliation[a]{Department of Physics, University of Oxford, Oxford OX1 3RH, UK}

\affiliation[b]{Niels Bohr Institute, University of Copenhagen,
Blegdamsvej 17, DK-2100, Copenhagen, Denmark}

\affiliation[c]{Department of Physics, University of Notre Dame, Notre Dame, IN, 46556, USA}

\abstract{
	The Standard Model Effective Field Theory (SMEFT) theoretical framework is increasingly used to interpret particle physics measurements and constrain physics
beyond the Standard Model.  We investigate the truncation of the effective-operator expansion using the geometric formulation of the SMEFT, which
allows exact solutions, up to mass-dimension eight.  Using this construction, we compare the
exact solution to the expansion at ${\cal{O}}(v^2/\Lambda^2)$, partial ${\cal{O}}(v^4/\Lambda^4)$ using a subset of terms
with dimension-6 operators, and full ${\cal{O}}(v^4/\Lambda^4)$, where $v$ is the vacuum expectation value and $\Lambda$ is the
scale of new physics.  This comparison is performed for general values of the coefficients, and for the specific model of
a heavy U(1) gauge field kinetically mixed with the Standard Model.  We additionally determine the input-parameter scheme dependence
at all orders in $v/\Lambda$, and show that this dependence increases at higher orders in $v/\Lambda$.
}

\maketitle
\section{Introduction}

With the proliferation of precise experimental results at the Large Hadron Collider (LHC) and other facilities, and the
lack of observed particles beyond the Standard Model (SM), data analysis and theoretical developments in the framework
of the Standard Model Effective Field Theory (SMEFT) are of increasing interest.  The SMEFT parameterizes the effects
of high-scale phenomena as effective operators with dimension $d>4$ suppressed by a factor $1/\Lambda^{d-4}$,
where $\Lambda$ is the new-physics energy scale.  The truncation of the effective field theory (EFT) typically leads to relative
errors of ${\cal{O}}(Q^2/\Lambda^2)$ on the operator coefficients, where $Q^2$ is the square of the momentum transfer
in a process.  Given the wide range of $Q^2$ probed by LHC measurements, a systematic accounting of these errors is
central to the result \cite{Passarino:2012cb,Passarino:2016pzb,Passarino:2019yjx,David:2015waa,Ghezzi:2015vva,Dawson:2019clf}.  They are relevant even when purely resonance observables are considered ($Q^2 \lesssim v^2$),
as the measurements typically constrain scales only within an order of magnitude of the process.
Here $v \equiv \bar{v}_T = \sqrt{2 \langle H^\dagger H \rangle}$ is the vacuum expectation value of $H^\dagger H$,
with $H$ the $\rm SU(2)_L$ scalar doublet Higgs field.

For LHC resonance processes, the leading non-SM contribution occurs at ${\cal{O}}(v^2/\Lambda^2)$ and is described by the
interference between SM operators and dimension-6 operators in the effective Lagrangian.  Given the historical lack of
a complete formulation relating processes to dimension-8 operator coefficients,\footnote{Recently, a complete dimension-8 operator basis for the SMEFT was reported \cite{Murphy:2020rsh,Li:2020gnx}.} several approaches have been proposed to address
unknown contributions at ${\cal{O}}(v^4/\Lambda^4)$: (1) a generic relative $v^2/\Lambda^2$ uncertainty, which assumes
the higher-dimensional coefficients are of similar magnitude to the leading coefficients; (2) an uncertainty estimated
from the squared amplitude of diagrams containing dimension-6 coefficients, which accounts for the next order in the
coefficients present at ${\cal{O}}(v^2/\Lambda^2)$ as well as new coefficients from dimension-6 operators that do not
interfere with the SM operators, or are otherwise suppressed; and (3) an uncertainty based on a scan of dimension-8 operator
coefficients present in a specific process, which allows a more complete accounting of the missing ${\cal{O}}(v^4/\Lambda^4)$
effects.  Each approach has its limitations, which should be considered when translating any coefficient constraints to a
particular model.

The recent development of the geometric formulation of the SMEFT (geoSMEFT~\cite{Helset:2020yio}), which builds on an
extensive theoretical
foundation~\cite{Vilkovisky:1984st,Burgess:2010zq,Alonso:2015fsp,Alonso:2016btr,Alonso:2016oah,Helset:2018fgq,Corbett:2019cwl},
allows a comprehensive analysis of the truncation of the SMEFT expansion.  The geoSMEFT approach provides results
not only for operators up to dimension 8, but at all orders in the $v/\Lambda$ expansion for several observables.
For these observables we can now quantitatively compare different orders in the SMEFT truncation, including partial
higher-order contributions.  In this paper, we study the observables $\Gamma(h \rightarrow \gamma \gamma)$,
$\Gamma(h \rightarrow \mathcal{Z} \, \gamma)$, and $\Gamma(\mathcal{Z} \rightarrow \bar{\psi} \, \psi)$, and determine
the input-parameter dependence in two schemes to all orders in $v/\Lambda$.  We further match the SMEFT coefficients
to an underlying U(1) kinetic mixing model up to dimension eight, quantifying the differences in inferred model
parameters using different EFT truncation prescriptions.

\section{SMEFT and geoSMEFT}\label{setup}

The SMEFT Lagrangian is defined as
\begin{align}
	\Lagr_{\textrm{SMEFT}} &= \Lagr_{\textrm{SM}} + \Lagr^{(d)}, &   \Lagr^{(d)} &= \sum_i \frac{C_i^{(d)}}{\Lambda^{d-4}}\mathcal{Q}_i^{(d)}
	\quad \textrm{ for } d>4.
\end{align}
The particle spectrum includes an $\rm SU(2)_L$ scalar doublet ($H$) with hypercharge
$\hyp_h = 1/2$. The higher-dimensional operators $\mathcal{Q}_i^{(d)}$ in the SMEFT are constructed out of the SM fields.
Our SM Lagrangian and conventions are consistent with Ref.~\cite{Brivio:2017vri}.
The operators $\mathcal{Q}_i^{(d)}$ are labelled with a mass dimension $d$ superscript
and multiply unknown Wilson coefficients $C_i^{(d)}$. We use the Warsaw basis~\cite{Grzadkowski:2010es} for
$\mathcal{L}^{(6)}$ and Refs.~\cite{Helset:2020yio,Hays:2018zze} for $\mathcal{L}^{(8)}$ results, with the geoSMEFT
conventions taking definitional precedence in the case of modified conventions or notation.
We define $\tilde{C}^{(d)}_i \equiv C^{(d)}_i \bar{v}_T^{d-4}/\Lambda^{d-4}$.
Our remaining notation is defined in Refs.~\cite{Alonso:2013hga,Brivio:2017vri}.

The parameter $\bar{v}_T$ in the SMEFT is defined as the minimum of the potential, including
corrections due to higher-dimensional operators. The value of $\bar{v}_T^2$ represents a tower of higher-order
corrections in the SMEFT, but we do not expand out $\bar{v}_T^2$ explicitly in terms of its SM value plus
$1/\Lambda$ corrections, as this same tower of higher order effects is present in all instances of
$\bar{v}_T^2$ in numerical predictions.\footnote{We nevertheless present in Appendix \ref{allordersvev}
an iterative solution for the vev in terms of $1/\Lambda$ corrections for completeness.}

The geometric formulation of the SMEFT organizes the theory in terms of field-space connections $G_i$
multiplying composite operator forms $f_i$, represented schematically by
\bea\label{basicdecomposition}
\Lagr_{\textrm{SMEFT}} = \sum_i G_i(I,A,\phi \dots) \, f_i ,
\eea
where $G_i$ depend on the group indices $A,I$ of the (non-spacetime) symmetry groups,
and the scalar field coordinates of the composite operators, except powers of $D^\mu H$, which are grouped into $f_i$.
The connections can be thought of as background-field form factors.
The field-space connections depend on the coordinates of the Higgs scalar doublet expressed in terms of
real scalar field coordinates, $\phi_I = \{\phi_1,\phi_2,\phi_3,\phi_4\}$, with normalization
\begin{align}
H(\phi_I) = \frac{1}{\sqrt{2}}\begin{bmatrix} \phi_2+i\phi_1 \\ \phi_4 - i\phi_3\end{bmatrix}.
\end{align}
When considering the vacuum expectation value (vev), we note that $\phi_4 \rightarrow \phi_4 + \bar{v}_T$.
The gauge boson field coordinates are defined as $\mathcal{W}^A = \{W^1,W^2,W^3,B\}$ with $A =\{1,2,3,4\}$.
The corresponding general coupling in the SM is $\alpha_A = \{g_2, g_2, g_2, g_1\}$.  The mass eigenstate
field coordinates are $\mathcal{A}^A = \{\mathcal{W}^+,\mathcal{W}^-,\mathcal{Z},\mathcal{A}\}$, and final-state
photons are represented by $\gamma$.

In the observables we examine, the field-space connections $(h_{IJ},g_{AB},k_{IJA},L^{\cdots}_{I,A})$
are used:
\bea
&h_{IJ}& \hspace{-0.3cm}(\phi)(D_\mu\phi)^I (D^\mu \phi)^J, \\
&g_{AB}&\hspace{-0.3cm}(\phi)\mathcal{W}_{\mu\nu}^A \mathcal{W}^{B,\mu\nu}, \\
&k_{IJA}&\hspace{-0.2cm}(\phi)(D^\mu \phi)^I(D^\nu\phi)^J \mathcal{W}_{\mu\nu}^A, \\
&L^{\psi_{pr}^{R/L}}_{IA}&\hspace{-0.2cm} (\phi) (D^\mu \phi)^J  (\bar{\psi}_p^{R/L} \gamma_\mu \tau_A \psi_r^{R/L}),
\eea
each of which is defined to all orders in the $\sqrt{2 \langle H^\dagger H \rangle}/\Lambda$ expansion
in Ref.~\cite{Helset:2020yio}. The geometric Lagrangian parameters are functions of the field-space connections
$h_{IJ},g_{AB}$, in particular the matrix square roots of these field space connections $\sqrt{g}_{AB} = \langle g_{AB} \rangle^{1/2}$,
and $\sqrt{h}_{IJ} = \langle h_{IJ} \rangle^{1/2}$.\footnote{Note that $\sqrt{g}^{AB} \sqrt{g}_{BC} \equiv \delta^A_C$ and
$\sqrt{h}^{IJ} \sqrt{h}_{JK} \equiv \delta^I_K$.}
As the SMEFT perturbations are small corrections to the SM, the field-space connections are positive semi-definite matrices,
with unique square roots.

\section{Partial vs full ${\cal{O}}(v^4/\Lambda^4)$}\label{analyticexpressions}
The standard procedure for evaluating partial ${\cal{O}}(v^4/\Lambda^4)$ corrections is to include the squared
amplitude of diagrams with a linear dependence on dimension-6 Wilson coefficients for an observable $\mathcal{O}_i$:
\bea\label{partialsquare}
\langle \mathcal{O}_i \rangle^{\it p.s.} &\simeq& \int [{\rm dps}] \left| A_{\rm SM} + A_{\rm SMEFT}(\tilde{C}^{(6)}_i)\right|^2 \\
						&\simeq&\int [{\rm dps}] \left( \left| A_{\rm SM}\right|^2 + A_{\rm SM} \, A^\star_{\rm SMEFT}(\tilde{C}^{(6)}_i)
+ A_{\rm SM}^\star \, A_{\rm SMEFT}(\tilde{C}^{(6)}_i) + \left|A_{\rm SMEFT}(\tilde{C}^{(6)}_i)\right|^2\right), \nonumber
\eea
where $\int [\rm dps]$ indicates an integral over phase space.
We define this calculation to be the {\it partial-square} procedure, which we compare to the full
${\cal{O}}(v^4/\Lambda^4)$ result.  In the above calculation the SMEFT amplitude correction includes corrections to
the SM amplitude from dimension-six operators, and novel contributions to the Wick expansion without an SM equivalent.
The dependence on the full set of dimension-six Wilson coefficients is indicated by $\tilde{C}^{(6)}_i$,
with the sum over $i$ suppressed.  When an observable is predicted using a reference set of observables to numerically
fix Lagrangian parameters, i.e. an input-parameter set, the corrections to an SM amplitude in the SMEFT also include
redefinitions of this mapping.  We defer a discussion on input-parameter effects to Appendix~\ref{inputs} and first
compare results analytically.  We restrict our analysis to $\rm CP$-even operators, approximating
$A^\star_{\rm SMEFT}= A_{\rm SMEFT}$.

The full result in the SMEFT to ${\cal{O}}(v^4/\Lambda^4)$ is
\bea\label{correctsquare}
\langle \mathcal{O}_i \rangle^{\rm SMEFT} &=&
\int [{\rm dps}]  \left( \left| A_{\rm SM}\right|^2 + 2 \,{\rm Re}(A_{\rm SM}) \, A_{\rm SMEFT}(\tilde{C}^{(6)}_i)
+ \left|A_{\rm SMEFT}(\tilde{C}^{(6)}_i)\right|^2\right)  \nn
&+& \int [{\rm dps}] \left(2 \, {\rm Re}(A_{\rm SM}) \, A_{\rm SMEFT}(\tilde{C}^{(8)}_i)\right).
\eea
This expression incorporates not only the $\mathcal{L}^{(8)}$ coefficients, but importantly also terms
quadratic in the $\mathcal{L}^{(6)}$ coefficients that are missing in the partial-square procedure, as
we will discuss below.

Due to the large number of operators at $\mathcal{L}^{(8)}$, it has not been possible
to perform practical calculations until recently.  However, the geoSMEFT formalism now defines corrections to all
orders in the $\bar{v}_T/\Lambda$ expansion for several observables.  Here we compare results for the
partial-square procedure and the full ${\cal{O}}(v^4/\Lambda^4)$ calculation for
$\Gamma(h \rightarrow \gamma \gamma)$, $\Gamma(h \rightarrow \mathcal{Z} \, \gamma)$, and
$\Gamma(\mathcal{Z} \rightarrow \bar{\psi} \, \psi)$, and comment on other observables.

\subsection{$\Gamma(h \rightarrow \gamma \gamma)$}
In the SM, $\Gamma(h \rightarrow \gamma \gamma)$ is loop-suppressed, and the leading-order result was developed in
Refs.~\cite{Ellis:1975ap,Shifman:1979eb,Bergstrom:1985hp}. Defining
\begin{align}
&i \,\mathcal{A}_{\rm SM}^{h \gamma \gamma} = \frac{i\, g_2 \, e^2}{16 \, \pi^2 \, m_W} \int_0^1 dx \int_0^{1-x} dy
\Bigg( \frac{-4\, m_W^2 + 6 \, x\, y\, m_W^2 + x\ y\, m_h^2}{m_W^2 - x \, y \, m_h^2} +  \sum_f \, N_c\, Q_f^2 \, \frac{m_f^2 \, (1-4\, x\, y)}{m_f^2 - x\, y\, m_h^2} \Bigg), \nn
&\langle h \mathcal{A}^{\mu\nu} \mathcal{A}_{\mu\nu}\rangle = \langle h | h \, \mathcal{A}^{\mu \, \nu} \, \mathcal{A}_{\mu \, \nu}| \gamma(p_a),\gamma(p_b) \rangle
= - 4 \, \left(p_a \cdot p_b \, g^{\alpha \, \beta} - p_a^\beta \, p_b^\alpha \right) \epsilon_\alpha \epsilon_\beta,
\end{align}
the three-point function $h-\gamma-\gamma$ in the SMEFT is~\cite{Helset:2020yio}
\begin{align}
	\langle h|\gamma \, \gamma \rangle =&
	-\langle h \mathcal{A}^{\mu\nu} \mathcal{A}_{\mu\nu}\rangle
	\frac{\sqrt{h}^{44}}{4} \left[
	\left\langle \frac{\delta g_{33}(\phi)}{\delta \phi_4}\right\rangle \frac{\overline{e}^2}{g_2^2}
	+2 \left\langle \frac{\delta g_{34}(\phi)}{\delta\phi_4}\right\rangle \frac{\overline{e}^2}{g_1 g_2}
	+ \left\langle \frac{\delta g_{44}(\phi)}{\delta\phi_4}\right\rangle \frac{\overline{e}^2}{g_1^2}
	\right]
	\nonumber \\ &+ \langle h \mathcal{A}^{\mu\nu} \mathcal{A}_{\mu\nu}\rangle \mathcal{A}_{\rm SM}^{h\gamma\gamma}.
\end{align}
Here we have used the geometric electric charge gauge coupling $\bar{e}$ and Weinberg angle $s_{\bar{\theta}}$~\cite{Helset:2020yio}
defined in Appendix~\ref{appendixdefinitions}.

We write the ${\cal{O}}(v^2/\Lambda^2)$ correction to the $h-\gamma-\gamma$ function as
$\langle h \mathcal{A}^{\mu\nu} \mathcal{A}_{\mu\nu} \, \rangle \langle h|\gamma \gamma\rangle_{\mathcal{L}^{(6)}}/\bar{v}_T$,
with
\begin{align}
	\label{eq:hggDim6}
\langle h|\gamma \gamma\rangle_{\mathcal{L}^{(6)}} &= \left[\frac{g_2^2 \, \tilde C_{HB}^{(6)}
  + g_1^2 \, \tilde C_{HW}^{(6)} - g_1 \, g_2 \, \tilde C_{HWB}^{(6)}}{({g}^{\rm SM}_Z)^2} \right],
\end{align}
where $({g}^{\rm SM}_Z)^2 \equiv g_1^2+g_2^2$.  To ${\cal{O}}(v^4/\Lambda^4)$ the full three-point function is
\begin{align}
	\label{eq:hggDim8}
	\langle h|\gamma \gamma \rangle_{{\rm{to}} \, {\cal{O}}(v^4/\Lambda^4)} =
	\frac{\langle h \mathcal{A}^{\mu\nu} \mathcal{A}_{\mu\nu}\rangle}{\bar{v}_T}&\left[\bar{v}_T \mathcal{A}_{\rm SM}^{h\gamma\gamma}+
  \left(1 +	\langle \sqrt{h}^{44}\rangle_{{\cal{O}}(v^2/\Lambda^2)} \right) \, \langle h|\gamma \gamma\rangle_{\mathcal{L}^{(6)}} \right. \nn
  &+
   \left.\left. 2  \, \langle h|\gamma \gamma\rangle_{\mathcal{L}^{(6)}}^2 + 2 \, (\langle h|\gamma \gamma\rangle_{\mathcal{L}^{(6)}})\right|_{C_i^{(6)} \rightarrow C_i^{(8)}} \right].
\end{align}
Here
\begin{align}
	\langle \sqrt{h}^{44}\rangle_{{\cal{O}}(v^2/\Lambda^2)}
	= \tilde C_{H\Box}^{(6)} - \frac{1}{4}\tilde C_{HD}^{(6)},
\end{align}
and we have used the short-hand notation $C_i^{(6)} \rightarrow C_i^{(8)}$ for the replacements
\begin{align}
	C_{HB}^{(6)} &\rightarrow \frac{1}{2} C_{HB}^{(8)},  & \quad
	C_{HW}^{(6)} &\rightarrow \frac{1}{2} \left(  C_{HW}^{(8)}+C_{HW,2}^{(8)}\right), &  \quad
	C_{HWB}^{(6)} &\rightarrow \frac{1}{2} C_{HWB}^{(8)}.
\end{align}
Squaring the amplitude at ${\cal{O}}(v^2/\Lambda^2)$ gives the partial-square result \\
$|\langle h \mathcal{A}^{\mu\nu} \mathcal{A}_{\mu\nu} \, \rangle \langle h| \gamma \gamma\rangle_{{\rm{to}} \, {\cal{O}}(v^2/\Lambda^2)} /\bar{v}_T|^2$,
where
\bea\label{pshgamgam}
|\langle h| \gamma \gamma\rangle_{{\rm{to}} \, {\cal{O}}(v^2/\Lambda^2)}|^2 =
\bar{v}_T^2 \bigg|\mathcal{A}_{\rm SM}^{h\gamma\gamma}  \bigg|^2+ 2\bar{v}_T \,{\rm Re}(\mathcal{A}_{\rm SM}^{h\gamma\gamma}) \, \langle h|\gamma \gamma\rangle_{\mathcal{L}^{(6)}}
 + \langle h|\gamma \gamma\rangle_{\mathcal{L}^{(6)}}^2,
\eea
while the square of the amplitude with $\mathcal{L}^{(8)}$ operators can be expanded to give the full
${\cal{O}}(v^4/\Lambda^4)$ result
$|\langle h \mathcal{A}^{\mu\nu} \mathcal{A}_{\mu\nu} \, \rangle|^2 |\langle h| \gamma \gamma\rangle|^2_{{\rm{to}} \, {\cal{O}}(v^4/\Lambda^4)} /\bar{v}^2_T$,
with:
\bea\label{correctsquarehgamgam}
|\langle h| \gamma \gamma\rangle|^2_{{\rm{to}} \, {\cal{O}}(v^4/\Lambda^4)} &=&  \bar{v}_T^2\bigg|\mathcal{A}_{\rm SM}^{h\gamma\gamma}  \bigg|^2 + 2\bar{v}_T \,{\rm Re}(\mathcal{A}_{\rm SM}^{h\gamma\gamma})
(1+\langle \sqrt{h}^{44}\rangle_{{\cal{O}}(v^2/\Lambda^2)}) \, \langle h|\gamma \gamma\rangle_{\mathcal{L}^{(6)}} \\
&+&(1+ 4 \,\bar{v}_T \, {\rm Re}(\mathcal{A}_{\rm SM}^{h\gamma\gamma})) \, \langle h|\gamma \gamma\rangle_{\mathcal{L}^{(6)}}^2
+ 4\bar{v}_T \,{\rm Re}(\mathcal{A}_{\rm SM}^{h\gamma\gamma})
\left.(\langle h|\gamma \gamma\rangle_{\mathcal{L}^{(6)}})\right|_{C_i^{(6)} \rightarrow C_i^{(8)}}. \nonumber
\eea
The dependence on $\langle h|\gamma \gamma\rangle_{\mathcal{L}^{(6)}}^2$, which one might
expect to correctly determine in the partial-square procedure, is not correctly predicted by
Eqn~\ref{pshgamgam}.
This arises from a modification to the couplings in the transformation to the mass-eigenstate field basis.
The relationship between $\bar{e}^2$ in the SMEFT and $(e^2)^{\rm SM}$ is
\bea
\bar{e}^2 = (e^2)^{\rm SM} \left[1+ 2 \,  \langle h|\gamma \gamma\rangle_{\mathcal{L}^{(6)}}
	+4 \, \left( \langle h|\gamma \gamma\rangle_{\mathcal{L}^{(6)}}\right)^2
\left.	+2 \, \left( \langle h|\gamma \gamma\rangle_{\mathcal{L}^{(6)}}\right)\right|_{C_i^{(6)} \rightarrow C_i^{(8)}}
\right].
\eea
As a result, when expanding to ${\cal{O}}(v^4/\Lambda^4)$, the dependence on $\langle h|\gamma \gamma\rangle_{\mathcal{L}^{(6)}}^2$
is not correctly predicted by the partial-square procedure.  The procedure has further inconsistencies if the
$\mathcal{L}^{(6)}$ operators are rescaled by powers of the gauge couplings of the theory, as in Ref.~\cite{Contino:2013kra}.

In addition to the incorrect coefficient of $\langle h|\gamma \gamma\rangle_{\mathcal{L}^{(6)}}^2$ in the
partial-square procedure, there are missing quadratic $\mathcal{L}^{(6)}$ coefficients due to the normalization
of the Higgs field, which modifies the coefficient of $\langle h|\gamma \gamma\rangle_{\mathcal{L}^{(6)}}$ in
Eqn.~\ref{correctsquarehgamgam}.

\subsection{$\Gamma(h \rightarrow \mathcal{Z} \gamma)$}
For $\Gamma(h \rightarrow \mathcal{Z} \gamma)$ the differences between the partial-square procedure and the full
${\cal{O}}(v^4/\Lambda^4)$ result are similar to those for $\Gamma(h \rightarrow \gamma \gamma)$.
The SM result for this decay was developed in Refs.~\cite{Cahn:1978nz,Bergstrom:1985hp},\footnote{Note the sign correction
to the results in Ref.~\cite{Bergstrom:1985hp} pointed out in Ref.~\cite{Manohar:2006gz}.} and is
\begin{align}
&i \, \mathcal{A}_{\rm SM}^{h \mathcal{Z} \gamma} = \frac{i\, g_2 \, e^2}{16 \, \pi^2 \, m_W}
\Bigg( I_W^\mathcal{Z}(\frac{m_h^2}{4 m_W^2},\frac{m_Z^2}{4 m_W^2}) +  \int_0^1 dx \int_0^{1-x} \hspace{-0.2cm}dy \sum_f \, \frac{4 m_f^2 \, (1-4\, x\, y) \, N_c\, Q_f  \, g_V^\psi/s_{2 \bar{\theta}}}{m_f^2 - (m_h^2-m_Z^2) x\, y - m_Z^2 y(1-y)} \Bigg), \nn
&I_W^{\mathcal{Z}}(a,b)= \frac{2}{t_{\bar{\theta}}} \, \int_0^1 dx \int_0^{1-x} dy \frac{(5 - t_{\bar{\theta}}^2 + 2 a (1- t_{\bar{\theta}}^2)) x y - (3 - t_{\bar{\theta}}^2)}{1 - 4(a-b) x y - 4 b y (1-y) - i 0^+}, \nn
&\langle h \mathcal{A}^{\mu\nu} \mathcal{Z}_{\mu\nu}\rangle = \langle h | h \, \mathcal{A}^{\mu \, \nu} \, \mathcal{Z}_{\mu \, \nu}| \gamma(p_a),\mathcal{Z}(p_b) \rangle
=  -2 \left(p_a \cdot p_b \, g^{\alpha \, \beta} - p_a^\beta \, p_b^\alpha \right) \epsilon^A_\alpha \epsilon^Z_\beta.
\end{align}
Here we have used $\gsm{\psi}{V} = T_3/2 - Q_\psi  (s^{\rm SM}_{\theta_Z})^2$. For $\psi = \{u,\nu, d, e\}$
we have $2 \, T_3(\psi) = \{1,1,-1,-1\}$ and $Q_\psi= \{2/3, 0, -1/3, -1 \}$.

The three-point function $h-\mathcal{Z}-\gamma$ in the SMEFT is~\cite{Helset:2020yio}
\begin{align}
	\langle h|\gamma \mathcal{Z}\rangle =&
	-\langle h \mathcal{A}^{\mu\nu} \mathcal{Z}_{\mu\nu}\rangle
	\frac{\sqrt{h}^{44}}{2}\overline{e}\,\overline{g}_Z \left[
		\left\langle \frac{\delta g_{33}(\phi)}{\delta \phi_4}\right\rangle \frac{c_{\theta_Z}^2}{g_2^2}
		+ \left\langle \frac{\delta g_{34}(\phi)}{\delta\phi_4}\right\rangle \frac{c_{\theta_Z}^2-s_{\theta_Z}^2}{g_1 g_2}
		- \left\langle \frac{\delta g_{44}(\phi)}{\delta\phi_4}\right\rangle \frac{s_{\theta_Z}^2}{g_1^2}
	\right]
	\nonumber\\ & +\langle h \mathcal{A}^{\mu\nu} \mathcal{Z}_{\mu\nu}\rangle \mathcal{A}_{\rm SM}^{h\gamma Z},
\end{align}
which depends on the geometric rotation angle $s_{\theta_Z}^2$ and $\mathcal{Z}$ effective gauge coupling $\bar{g}_Z$
defined in Appendix~\ref{appendixdefinitions}.
Expanding out the $h-\mathcal{Z}-\gamma$ three-point function to order $v^2/\Lambda^2$, we have
\begin{align}
	\label{eq:hgZDim6}
	\langle h|\gamma \mathcal{Z}\rangle_{{\rm{to}} \, {\cal{O}}(v^2/\Lambda^2)} &=
	\langle h \mathcal{A}^{\mu\nu} \mathcal{Z}_{\mu\nu} \rangle \times \frac{1}{\bar{v}_T}\left[\bar{v}_T \mathcal{A}_{\rm SM}^{h\gamma \mathcal{Z}}+ \langle h|\gamma \mathcal{Z}\rangle_{\mathcal{L}^{(6)}}\right],
\end{align}
where
\begin{align}\label{hzgamdim6}
	\langle h|\gamma \mathcal{Z}\rangle_{\mathcal{L}^{(6)}} =
\left[	 \frac{ 2 g_1 g_2 \left(\tilde{C}^{(6)}_{HW}-\tilde{C}^{(6)}_{HB}\right)+\left(g_1^2-g_2^2\right) \tilde{C}^{(6)}_{HWB}}{({g}^{\rm SM}_Z)^2}\right].
\end{align}
Expanding out to order $v^4/\Lambda^4$ yields
\begin{align}\label{eq:hgzDim8}
	\langle h|\gamma \mathcal{Z} \rangle_{{\rm{to}} \, {\cal{O}}(v^4/\Lambda^4)} &=
	\langle h \mathcal{A}^{\mu\nu} \mathcal{Z}_{\mu\nu}\rangle\frac{1}{\bar{v}_T} \left[\bar{v}_T \mathcal{A}_{\rm SM}^{h\gamma \mathcal{Z}}+
  \left(1 +	\langle \sqrt{h}^{44}\rangle_{{\cal{O}}(v^2/\Lambda^2)} \right) \, \langle h|\gamma \mathcal{Z}\rangle_{\mathcal{L}^{(6)}}
+ \frac{2 \,g_1 \, g_2 }{g_2^2-g_1^2} \langle h|\gamma \mathcal{Z}\rangle_{\mathcal{L}^{(6)}}^2 \right] \nn
&+\langle h \mathcal{A}^{\mu\nu} \mathcal{Z}_{\mu\nu}\rangle\frac{1}{\bar{v}_T} \langle h|\gamma \mathcal{Z}\rangle_{\mathcal{L}^{(6)}} \left[ \frac{\tilde{C}_{HB}^{(6)} \, g_1^2-\tilde{C}_{HW}^{(6)} \, g_2^2
+ 3 (\tilde{C}_{HW}^{(6)} \, g_1^2-\tilde{C}_{HB}^{(6)} \, g_2^2) }{g_1^2-g_2^2} \right] \nn
&+ \langle h \mathcal{A}^{\mu\nu} \mathcal{Z}_{\mu\nu} \rangle \frac{1}{\bar{v}_T}
\left.2	\left(	\langle h|\gamma \mathcal{Z}\rangle_{\mathcal{L}^{(6)}}\right)\right|_{C_i^{(6)}\rightarrow C_i^{(8)}}.
\end{align}
The difference between the partial-square procedure and the full ${\cal{O}}(v^4/\Lambda^4)$ result then follows from
\bea
|\langle h| \mathcal{Z} \gamma\rangle_{{\rm{to}} \, {\cal{O}}(v^2/\Lambda^2)}|^2 =
\bar{v}_T^2 \bigg|\mathcal{A}_{\rm SM}^{h\gamma Z}  \bigg|^2+ 2\bar{v}_T  \,{\rm Re}(\mathcal{A}_{\rm SM}^{h\gamma Z})  \, \langle h|\gamma \mathcal{Z}\rangle_{\mathcal{L}^{(6)}}
 + \langle h|\gamma \mathcal{Z}\rangle_{\mathcal{L}^{(6)}}^2,
\eea
and
\begin{align}
|\langle h| \mathcal{Z} \gamma\rangle|^2_{{\rm{to}} \, {\cal{O}}(v^4/\Lambda^4)} =
	&\bar{v}_T^2 \bigg|\mathcal{A}_{\rm SM}^{h\gamma \mathcal{Z}}  \bigg|^2+ 2\bar{v}_T \,{\rm Re}(\mathcal{A}_{\rm SM}^{h\gamma \mathcal{Z}}) \,  \left(1 + \langle \sqrt{h}^{44}\rangle_{{\cal{O}}(v^2/\Lambda^2)} \right) \, \langle h|\gamma \mathcal{Z}\rangle_{\mathcal{L}^{(6)}} \nn
&+ 2\bar{v}_T \,{\rm Re}(\mathcal{A}_{\rm SM}^{h\gamma \mathcal{Z}}) \langle h|\gamma \mathcal{Z}\rangle_{\mathcal{L}^{(6)}} \left[ \frac{\tilde{C}_{HB}^{(6)} \, g_1^2-\tilde{C}_{HW}^{(6)} \, g_2^2
+ 3 (\tilde{C}_{HW}^{(6)} \, g_1^2-\tilde{C}_{HB}^{(6)} \, g_2^2) }{g_1^2-g_2^2} \right] \nn
&+\left(1+ \frac{4 \,g_1 \, g_2 \, \bar{v}_T}{g_2^2-g_1^2} \,{\rm Re}(\mathcal{A}_{\rm SM}^{h\gamma Z})\right) \, \langle h|\gamma Z\rangle_{\mathcal{L}^{(6)}}^2 \nn
&+ 4\bar{v}_T \,{\rm Re}(\mathcal{A}_{\rm SM}^{h\gamma \mathcal{Z}})
\left.	\left(	\langle h|\gamma \mathcal{Z}\rangle_{\mathcal{L}^{(6)}}\right)\right|_{C_i^{(6)}\rightarrow C_i^{(8)}}.
\end{align}
The quadratic dependence on the ${\mathcal{L}^{(6)}}$ coefficients again differs due to contributions from the Higgs field
normalization and the coupling expansion to ${\cal{O}}(v^4/\Lambda^4)$, which includes an additional term due to electroweak
mixing.  We have normalized these expressions to cancel the dimensions in $\mathcal{A}_{\rm SM}^{h\gamma \mathcal{Z}}$.

\subsection{$\Gamma(\mathcal{Z} \rightarrow \bar{\psi} \psi)$}
We now consider a process that is present at tree level in the SM: the decay of a $\mathcal{Z}$ boson into a pair of fermions.
This decay can be defined at all orders of the $v/\Lambda$ expansion via
\bea\label{Zdecay}
\bar{\Gamma}_{\mathcal{Z} \rightarrow \bar{\psi}_p \psi_r} =\frac{N_c^\psi}{24 \pi} \sqrt{\bar{m}_Z^2} |g_{\rm eff,pr}^{\mathcal{Z},\psi}|^2 \left(1- \frac{4 \bar{M}_{\psi,p}^2}{\bar{m}_Z^2} \right)^{3/2},
\eea
where
\bea
g_{\rm eff,pr}^{\mathcal{Z},\psi} = \frac{\bar{g}_Z}{2} \left[(2 s_{\theta_Z}^2 \, Q_\psi -\sigma_3)\delta_{pr}+ \bar{v}_T \langle L_{3,4}^{\psi,pr} \rangle + \sigma_3
\bar{v}_T \langle L_{3,3}^{\psi,pr} \rangle  \right].
\eea
Here $\psi = \{q_L,u_R,d_R,\ell_L,e_R\}$, with $\sigma_3 = 1$ for $u_L, \nu_L$ and $\sigma_3 = -1$ for $d_L, e_L$.
The decay width depends on the Lagrangian parameters defined in the previous sections, supplemented with the masses
\begin{align}
\bar{m}_Z^2 &= \frac{\bar{g}_Z^2}{4}  \sqrt{h_{33}}^2 \bar{v}_T^2, \nn
\bar{m}^\psi_{pr} &= \langle (Y^\psi_{pr})^\dagger \rangle,
\end{align}
with $\bar{M}_{\psi,i}$ the vector of eigenvalues of the $\bar{m}^\psi_{pr}$ matrix.  The generalized Yukawa couplings
($Y^\psi_{pr}$) are defined in Ref.~\cite{Helset:2020yio}.  Focusing on the effective coupling contributing to the
decay width, the expressions at each order are:
\begin{align}
\langle g_{\rm eff,pr}^{\mathcal{Z},\psi}\rangle_{\rm SM} &= \bar{g}^{\rm SM}_Z \left[(s^{\rm SM}_{\theta})^2 \, Q_\psi -\frac{\sigma_3}{2}\right] \, \delta_{pr},\\
\label{gzexpansion}
\langle g_{\rm eff,pr}^{\mathcal{Z},\psi}\rangle_{{\cal{O}}(v^2/\Lambda^2)} &= \frac{\langle {\bar{g}}_Z \rangle_{{\cal{O}}(v^2/\Lambda^2)}}{\bar{g}^{\rm SM}_Z} \langle g_{\rm eff,pr}^{\mathcal{Z},\psi}\rangle_{\rm SM} \, \delta_{pr}
+ \bar{g}^{\rm SM}_Z \, Q_\psi \, \langle s_{\theta_Z}^2 \rangle_{{\cal{O}}(v^2/\Lambda^2)} \, \delta_{pr}
+ \frac{\bar{g}^{\rm SM}_Z }{2}\left[\tilde{C}^{1,(6)}_{\substack{H \psi\\ pr}}- \sigma_3 \, \tilde{C}^{3,(6)}_{\substack{H \psi\\ pr}}\right],
\end{align}
\begin{align}\label{gzexpansion2}
\hspace{-1cm}
\langle g_{\rm eff,pr}^{\mathcal{Z},\psi}\rangle_{{\cal{O}}(v^4/\Lambda^4)} &=
\frac{\langle {\bar{g}}_Z \rangle_{{\cal{O}}(v^4/\Lambda^4)}}{\bar{g}^{\rm SM}_Z} \langle g_{\rm eff,pr}^{\mathcal{Z},\psi}\rangle_{\rm SM} \, \delta_{pr}
+ \bar{g}^{\rm SM}_Z \, Q_\psi \, \langle s_{\theta_Z}^2 \rangle_{{\cal{O}}(v^4/\Lambda^4)} \, \delta_{pr}
+ \langle {\bar{g}}_Z \rangle_{{\cal{O}}(v^2/\Lambda^2)}  \, \langle s_{\theta_Z}^2 \rangle_{{\cal{O}}(v^2/\Lambda^2)} \, Q_\psi \delta_{pr} \nn
&+ \frac{\langle {\bar{g}}_Z \rangle_{{\cal{O}}(v^2/\Lambda^2)}}{2}\left[\tilde{C}^{1,(6)}_{\substack{H \psi\\ pr}}- \sigma_3 \, \tilde{C}^{3,(6)}_{\substack{H \psi\\ pr}}\right]
+\frac{{g}^{\rm SM}_Z }{4}\left[\tilde{C}^{1,(8)}_{\substack{H \psi\\ pr}}- \sigma_3 \, \tilde{C}^{2,(8)}_{\substack{H \psi\\ pr}}
- \sigma_3 \, \tilde{C}^{3,(8)}_{\substack{H \psi\\ pr}}\right].
\end{align}
Expressions for $\langle s_{\theta_Z}^2 \rangle_{{\cal{O}}(v^2/\Lambda^2)}$,$\langle s_{\theta_Z}^2 \rangle_{{\cal{O}}(v^4/\Lambda^4)}$,
$\langle {\bar{g}}_Z \rangle_{{\cal{O}}(v^2/\Lambda^2)},$ and $\langle {\bar{g}}_Z \rangle_{{\cal{O}}(v^4/\Lambda^4)}$, are given in Ref.~\cite{Helset:2020yio}
and summarized in Appendix \ref{appendixdefinitions}.  In addition, there are scheme-dependent
corrections due to the mapping of redefined Lagrangian parameters to measured input parameters.

As can be seen from Eqs.~\ref{gzexpansion} and~\ref{gzexpansion2}, the partial-square and full
${\cal{O}}(v^4/\Lambda^4)$ results differ extensively.  As an illustration we consider the dependence on the
Wilson coefficient $(\tilde{C}_{HWB}^{(6)})^2$, which corresponds to the (squared) $S$ parameter in the Warsaw basis.
The partial-square procedure yields a dependence of
\bea
|g_{\rm eff,pr}^{\mathcal{Z},\psi}|^2_{\rm partial \, square} \supset \frac{g_1^2 \, g_2^2 \, (\tilde{C}_{HWB}^{(6)})^2}{({g}^{\rm SM}_Z)^6} \delta_{pr} \left[{g}^{\rm SM}_Z \, \langle g_{\rm eff,pr}^{\mathcal{Z},\psi}\rangle_{\rm SM} +(g_2^2-g_1^2) \, Q_\psi) \right]^2,
\eea
while the full ${\cal{O}}(v^4/\Lambda^4)$ result yields a dependence of
\bea
|g_{\rm eff,pr}^{\mathcal{Z},\psi}|^2_{{\cal{O}}(v^4/\Lambda^4)} \supset
\frac{g_1^2 \, g_2^2 \, (\tilde{C}_{HWB}^{(6)})^2 \, (g_2^2-g_1^2)^2 \, Q_\psi^2}{({g}^{\rm SM}_Z)^6} \delta_{pr}
+ (\tilde{C}_{HWB}^{(6)})^2 \langle g_{\rm eff,pr}^{\mathcal{Z},\psi}\rangle_{\rm SM}^2 \delta_{pr}.
\eea
The partial-square result contains a term proportional to $Q_\psi$ that is not present in the full result
due to cancellations.  We examine the numerical difference between these results in Sec.~\ref{numerics}.

\subsection{$h \rightarrow \mathcal{Z} \mathcal{Z}^*$}
Although the $h \rightarrow \mathcal{Z} \mathcal{Z}^*$ final state includes an off-shell particle and is
not directly observable, it is of interest to examine the structure of corrections to the $h-\mathcal{Z}-\mathcal{Z}$
three-point function.  The geoSMEFT result is~\cite{Helset:2020yio}
\begin{align}
	\langle h|\mathcal{Z}\mathcal{Z}\rangle &=
	-\langle h \mathcal{Z}^{\mu\nu} \mathcal{Z}_{\mu\nu}\rangle
	\frac{\sqrt{h}^{44}}{4}\overline{g}^2_Z \left[
		\left\langle \frac{\delta g_{33}(\phi)}{\delta \phi_4}\right\rangle \frac{c_{\theta_Z}^4}{g_2^2}
		-2 \left\langle \frac{\delta g_{34}(\phi)}{\delta\phi_4}\right\rangle \frac{c_{\theta_Z}^2s_{\theta_Z}^2}{g_1 g_2}
		+ \left\langle \frac{\delta g_{44}(\phi)}{\delta\phi_4}\right\rangle \frac{s_{\theta_Z}^4}{g_1^2}
	\right] \nonumber \\
	&+
	\langle h \mathcal{Z}_\mu \mathcal{Z}^\mu \rangle\sqrt{h}^{44} \frac{\overline{g}_Z^2}{2} \left[
	\left\langle \frac{\delta h_{33}(\phi)}{\delta \phi_4} \right\rangle \left( \frac{\overline{v}_T}{2}\right)^2
	+\langle h_{33}(\phi)\rangle \frac{\overline{v}_T}{2}
	\right] \nonumber \\
	&+ \langle \partial_\nu h \mathcal{Z}_\mu \mathcal{Z}^{\mu\nu}\rangle \sqrt{h}^{44} \overline{g}_Z^2 \overline{v}_T
	\left[
		\langle k^3_{34}\rangle \frac{c_{\theta_Z}^2}{g_2} - \langle k^{4}_{34}\rangle
		\frac{s_{\theta_Z}^2}{g_1}
	\right].
\end{align}
In the SM, we have:
\begin{align}
\langle h|\mathcal{Z}^{\mu} \mathcal{Z}_{\mu} \rangle_{\rm SM} &= \frac{(\bar{g}^{\rm SM}_Z)^2}{4} \, \bar{v}_T, \\
\langle h|\mathcal{Z}^{\mu \nu} \mathcal{Z}_{\mu \nu} \rangle_{\rm SM} &= 0,\\
\langle \partial_\nu h | \mathcal{Z}_\mu \mathcal{Z}^{\mu\nu}\rangle_{\rm SM} &=0,
\end{align}
where the notation is such that $\langle h|\mathcal{Z}^{\mu} \mathcal{Z}_{\mu} \rangle$ represents the term multiplying
$\langle h \mathcal{Z}^{\mu\nu} \mathcal{Z}_{\mu\nu}\rangle$ and so forth.  Expanding the geoSMEFT result to
${\cal{O}}(v^2/\Lambda^2)$ gives:
\begin{align}
\langle h|\mathcal{Z}^{\mu} \mathcal{Z}_{\mu} \rangle_{{\cal{O}}(v^2/\Lambda^2)} &=
 \frac{(\bar{g}^{\rm SM}_Z)^2 \,\bar{v}_T}{4} \left[ 2 \frac{\langle \bar{g}_Z \rangle_{{\cal{O}}(v^2/\Lambda^2)}}{\bar{g}^{\rm SM}_Z}
  + \tilde{C}_{HD} +  \langle \sqrt{h}^{44}\rangle_{{\cal{O}}(v^2/\Lambda^2)} \right],\\
\langle h|\mathcal{Z}^{\mu \nu} \mathcal{Z}_{\mu \nu} \rangle_{{\cal{O}}(v^2/\Lambda^2)} &= \frac{\langle \bar{g}_Z \rangle_{{\cal{O}}(v^2/\Lambda^2)}}{\bar{g}^{\rm SM}_Z \, \bar{v}_T},\\
\langle \partial_\nu h | \mathcal{Z}_\mu \mathcal{Z}^{\mu\nu}\rangle_{{\cal{O}}(v^2/\Lambda^2)} &=0,
\end{align}
and expanding to ${\cal{O}}(v^4/\Lambda^4)$ one has
\begin{align}
	\langle h|\mathcal{Z}^{\mu} \mathcal{Z}_{\mu} \rangle_{{\cal{O}}(v^4/\Lambda^4)} = &\nn
\left[\langle \sqrt{h}^{44}\rangle_{{\cal{O}}(v^4/\Lambda^4)}
\right. & \left.+ 2 \, \frac{\langle {\bar{g}}_Z \rangle_{{\cal{O}}(v^4/\Lambda^4)}}{\bar{g}^{\rm SM}_Z} - \frac{3 (\langle {\bar{g}}_Z \rangle_{{\cal{O}}(v^2/\Lambda^2)})^2}{(\bar{g}^{\rm SM}_Z)^2} - (\tilde{C}_{HD}^{(6)})^2
- 2 \frac{\tilde{C}_{HD}^{(6)} \, \langle {\bar{g}}_Z \rangle_{{\cal{O}}(v^2/\Lambda^2)}}{\bar{g}^{\rm SM}_Z} \right] \, \langle h|\mathcal{Z}^{\mu} \mathcal{Z}_{\mu} \rangle_{\rm SM} \nn
&+\left[\tilde{C}_{HD}^{(6)} + 2 \frac{\langle {\bar{g}}_Z\rangle_{{\cal{O}}(v^2/\Lambda^2)}}{\bar{g}^{\rm SM}_Z} \right]\,\langle h|\mathcal{Z}^{\mu} \mathcal{Z}_{\mu} \rangle_{{\cal{O}}(v^2/\Lambda^2)}
+ \frac{3 \,(\bar{g}^{\rm SM}_Z)^2\,(\tilde{C}^{(8)}_{\text{HD}}+ \tilde{C}^{(8)}_{\text{H,D2}})\, \bar{v}_T}{16},\\
	\langle h|\mathcal{Z}^{\mu \nu} \mathcal{Z}_{\mu \nu} \rangle_{{\cal{O}}(v^4/\Lambda^4)} &=
	\left[\langle\sqrt{h}^{44}\rangle_{{\cal{O}}(v^2/\Lambda^2)} +
2 \frac{\langle \bar{g}_Z \rangle_{{\cal{O}}(v^2/\Lambda^2)}}{\bar{g}^{\rm SM}_Z} \right] \, \langle h|\mathcal{Z}^{\mu \nu} \mathcal{Z}_{\mu \nu} \rangle_{{\cal{O}}(v^2/\Lambda^2)}
	\nn &+ 2 \left.\left(\langle h|\mathcal{Z}^{\mu \nu} \mathcal{Z}_{\mu \nu} \rangle_{{\cal{O}}(v^2/\Lambda^2)}\right)\right|_{C_i^{(6)}\rightarrow C_i^{(8)}}
+ \frac{4 \, (\langle h|\gamma \mathcal{Z}\rangle_{{\cal{O}}(v^2/\Lambda^2)})^2}{\bar{v}_T}, \\
\langle \partial_\nu h | \mathcal{Z}_{\mu} \mathcal{Z}^{\mu\nu}\rangle_{{\cal{O}}(v^4/\Lambda^4)} &= \frac{g_1 \, \tilde{C}^{(8)}_{HDHB}
+ g_2 \, \tilde{C}^{(8)}_{HDHW}}{4 \, \bar{v}_T},
\end{align}
where
\begin{align}
\langle \sqrt{h}^{44}\rangle_{{\cal{O}}(v^4/\Lambda^4)} &= -\frac{1}{8}  (\tilde{C}_{H,D2}^{(8)} +  \tilde{C}_{HD}^{(8)}) +
 \frac{3}{4} \tilde C_{H\Box}^{(6)} (2 \tilde C_{H\Box}^{(6)} -  \tilde C_{HD}^{(6)}) + \frac{3}{32}  (\tilde C_{HD}^{(6)})^2 .
\end{align}
By direct inspection, it is clear that the differences between the partial-square and full results for the $h-\mathcal{Z}-\mathcal{Z}$ three-point
function are extensive.

\section{Numerical results}\label{numerics}
It is now possible to quantitatively compare the predictions of processes in the SMEFT using a partial-square result
and a full (CP-even) SMEFT result up to order $v^4/\Lambda^4$. In this section we perform this comparison for the first time
using an exact SMEFT formulation to ${\cal{O}}(v^4/\Lambda^4)$.

Accounting for SMEFT corrections to Lagrangian parameters is a precursor to an exact SMEFT calculation of observables to
sub-leading order.
In the SM a key set of Lagrangian parameters are the gauge couplings and the vacuum expectation value.
In the SMEFT the inference of these parameters from well-measured observables is modified by the presence of
higher-dimensional operators.  When a Lagrangian parameter in a prediction is not accompanied by the same set
of SMEFT corrections as in the observable used to fix the parameter, it is necessary to correct for this
difference.  This is the case for the numerical extractions of the gauge couplings.  The electroweak (EW) vacuum in the SMEFT
$\sqrt{2 H^\dagger H} \equiv \bar{v}_T$ is a common parameter for all instances when the Higgs vev appears in predictions.
Formally, this parameter includes an infinite tower of $1/\Lambda^n$ corrections.  There is no need to re-expand out
$\bar{v}_T$ in terms of an SM vev and these $1/\Lambda^n$ corrections when the same combination of higher-order terms
is present in all instances of $\bar{v}_T$ in a prediction.  This is the case for the set of higher-dimensional operators
that define the minimum of the potential, but this is
{\it not the case} for a) four-fermion operators and b) modifications of the $\mathcal{W}^{\pm}$ couplings to fermions
when this parameter is extracted from muon decay. These effects must be corrected for when making predictions using a
value of the vev.

Two popular input-parameter schemes are the $\{\hat{\alpha}_{ew}, \hat{m}_Z, \hat{G}_F\}$ and $\{\hat{m}_{W}, \hat{m}_Z, \hat{G}_F \}$
schemes.  Results to $\mathcal{L}^{(6)}$ for these schemes were developed in
Refs.~\cite{Brivio:2017bnu,Brivio:2017vri,Grinstein:1991cd,Alonso:2013hga,Berthier:2015oma,Berthier:2015gja,Bjorn:2016zlr,Berthier:2016tkq}\footnote{A
self-contained and up-to-date summary of these
effects, in both schemes, is included in Ref.~\cite{Brivio:2019myy}.} \footnote{It is important to note, when considering
scheme dependence in the SMEFT, that such scheme dependence is due to the effects of physics beyond the SM being absorbed
into a set of low-energy parameters. This is distinct physics from the scheme dependence associated with
a perturbative expansion in a renormalizable model, such as the SM, though the same label of ``scheme dependence"
is used in both cases.}.
Extending these schemes to $\mathcal{L}^{(8)}$ was explored in Ref.~\cite{Hays:2018zze}.
Here we build on these results, with some differences due to the consistent formulation of the SMEFT at $\mathcal{L}^{(8)}$
using the geoSMEFT \cite{Helset:2020yio}.  We provide numerical relations between partial widths and Wilson coefficients in the
$\{\hat{m}_{W}, \hat{m}_Z, \hat{G}_F \}$ scheme in this section, and the corresponding relations in the
$\{\hat{\alpha}_{ew}, \hat{m}_Z, \hat{G}_F\}$ scheme in Appendix~\ref{zwidthalphaem}.
Formulas for the inference of Lagrangian parameters at all orders in $\bar{v}_T/\Lambda$ are provided in Appendix \ref{inputs}.

\subsection{Order-of-magnitude estimates}

The numerical impact of $\mathcal{L}^{(6)}$ and $\mathcal{L}^{(8)}$ corrections on decay widths can be categorized using
several factors, such as whether the correction is part of an input parameter shift, whether it comes from interference
with the SM or between different $\mathcal{L}^{(6)}$ terms, and whether the corresponding SM amplitudes are tree- or
loop-level.  It is worthwhile to estimate the order-of-magnitude effect in each of these categories before diving into
numerics, in order to develop some intuition for the hierarchy of effects.

Considering first loop-suppressed SM amplitudes, a SMEFT correction will have an equivalent loop suppression for terms that
arise from input-parameter $\mathcal{L}^{(6)}$ and $\mathcal{L}^{(8)}$ corrections.  In general, an $\mathcal{L}^{(4+ 2 n)}$
correction to the input parameters will give a correction to the squared loop-suppressed amplitude of order
\bea\label{gam0}
\sim \frac{(\bar{g}^{\rm SM})^4}{(16 \pi^2)^2} \, \left(\frac{\bar{v}_T}{\Lambda}\right)^{2 \, n} C_i^{(4+ 2 n)}.
\eea
Similarly, a one-loop correction in the SMEFT with a higher-dimensional operator inserted in a loop
will have an effect of this order.  Such loop corrections are generally not included in
partial-square estimates, as they are available for a limited set of processes.  These results
are neglected here, although they are available in the literature for
$\Gamma(h \rightarrow \gamma \gamma)$~\cite{Hartmann:2015oia,Ghezzi:2015vva,Hartmann:2015aia,Dedes:2018seb,Dawson:2018liq},
$\Gamma(h \rightarrow \mathcal{Z} \gamma)$~\cite{Dawson:2018pyl,Dedes:2019bew}, and
$\Gamma(\mathcal{Z} \rightarrow \bar{\psi} \psi)$~\cite{Hartmann:2016pil,Dawson:2019clf}.  These
calculations could be used to study the important issue of perturbative uncertainties in the SMEFT.

In the case where a loop-suppressed SM amplitude interferes with an $\mathcal{L}^{(4 + 2 n)}$ amplitude, the
correction will be of order
\bea\label{gam1}
\sim \frac{(\bar{g}^{\rm SM})^2}{(16 \pi^2)} \, \left(\frac{\bar{v}_T}{\Lambda}\right)^{2 \, n} C_i^{(4+ 2 n)}.
\eea
Higher-order terms arise when a loop-suppressed SM amplitude with an input-parameter correction interferes with an
$\mathcal{L}^{(4+ 2 n)}$ amplitude,
\bea\label{gam2}
\sim \frac{(\bar{g}^{\rm SM})^2}{(16 \pi^2)} \, \left(\frac{\bar{v}_T}{\Lambda}\right)^{2m} \left(\frac{\bar{v}_T}{\Lambda}\right)^{2n} \, C_i^{(4+2n)}\, C_j^{(4+2m)},
\eea
or when two $\mathcal{L}^{(4+ 2 n)}$ amplitudes interfere,
\begin{align}\label{gam3}
&\sim \left(\frac{\bar{v}_T}{\Lambda}\right)^{4 \, n} (C_i^{(4+ 2 n)})^2,& \quad
{\rm or} & \quad
	 &\sim \left(\frac{\bar{v}_T}{\Lambda}\right)^{2 \, n} \left(\frac{\bar{v}_T}{\Lambda}\right)^{2 \, m} (C_i^{(4+ 2 n)}) (C_k^{(4+ 2 m)}).
\end{align}

The effect of input-parameter corrections (and corresponding scheme dependence) is more
significant for decays that occur at tree level in the SM.
In a tree-level SM decay, an input-parameter correction due to $\mathcal{L}^{(4+ 2 n)}$ operators gives a width correction of order
\begin{align}
&\sim (\bar{g}^{\rm SM}_{Z})^2 \, \left(\frac{\bar{v}_T}{\Lambda}\right)^{2 \, n} C_i^{(4+ 2 n)},
\end{align}
while the direct interference between $\mathcal{L}^{(4+ 2 n)}$ and SM amplitudes gives a width correction of
\begin{align}
&\sim \bar{g}^{\rm SM}_{Z} \, \left(\frac{\bar{v}_T}{\Lambda}\right)^{2 \, n} C_i^{(4+ 2 n)}.
\end{align}

These scalings dictate the numerical size of the corrections we report below.

\subsection{$\Gamma(h \rightarrow \gamma \gamma)$}
Using the input parameters in Table~\ref{tab:inputs}, we find the following SM leading-order
$h \rightarrow \gamma \gamma$ partial width in the $\hat{m}_W$ scheme:
\bea
\Gamma^{ \hat{m}_W}_{\rm SM}(h \rightarrow \gamma \gamma) &=& \frac{\hat{m}_h^3}{4 \pi} \, \bigg|\mathcal{A}_{\rm SM}^{h\gamma\gamma}  \bigg|^2 \nn
&=& 1.00 \times 10^{-5}\, {\rm GeV}.
\eea
The corresponding value in the $\hat{\alpha}_{ew}$ scheme is
$\Gamma^{\rm \hat{\alpha}_{ew}}_{\rm SM}(h \rightarrow \gamma \gamma) = 1.08 \times 10^{-5} \, {\rm GeV}$.
The differences in the SM results for different schemes are reduced at higher order in pertubation theory.
Since we use leading-order results when evaluating $\mathcal{L}^{(6)}$ and $\mathcal{L}^{(8)}$
corrections, we list here the SM leading-order results for completeness.
Results in the SMEFT are quoted as ratios with respect to the SM, which can be
applied to the highest-order result known in perturbation theory (in the SM).

\begin{table}[t]\centering
\renewcommand{\arraystretch}{1.2}
 \begin{tabular}{crlc}\hline
  $\hat{m}_{W}$ & 80.387 &GeV &  \cite{Aaltonen:2013iut} \\
  $\hat\alpha_{ew}(M_Z)$ & 1/127.950& & \cite{Olive:2016xmw}\\\hline
  $\hat{m}_Z$ &  91.1876 &GeV & \cite{Z-Pole,Olive:2016xmw,Mohr:2012tt} \\
  $\hat{G}_F$ & 1.1663787 $\cdot 10^{-5}$ &GeV$^{-2}$ &  \cite{Olive:2016xmw,Mohr:2012tt} \\

  $\hat{m}_h$ & 125.09 &GeV & \cite{Aad:2015zhl} \\
  $\hat{\alpha}_s(\hat m_Z)$& 0.1181 & &\cite{Olive:2016xmw}\\
  \hline
  $\hat{m}_t$&     173.21   &GeV & \cite{Olive:2016xmw}\\
  $\hat{m}_b$&     4.18 &GeV & \cite{Olive:2016xmw}\\
  $\hat{m}_c$&     1.28  &GeV & \cite{Olive:2016xmw}\\
  $\hat{m}_\tau$&  1.77686  &GeV & \cite{Olive:2016xmw}\\
  \hline
  \end{tabular}
\caption{Numerical central values of the relevant SM parameters used as inputs.
Only one of $\hat m_W$ or $\hat\alpha_{ew}$ is used as input depending on the scheme adopted. 
The remaining SM inputs are taken from the central values in the PDG \cite{Olive:2016xmw}.}\label{tab:inputs}
\end{table}

The squared amplitudes $|\langle h| \gamma \gamma\rangle_{{\rm{to}} \, {\cal{O}}(v^2/\Lambda^2)}|^2$ and
$|\langle h| \gamma \gamma \rangle|^2_{{\rm to} \, \mathcal{O}(v^4/\Lambda^4)}$ give the partial-square and full ${\cal{O}}(v^4/\Lambda^4)$
SMEFT corrections, respectively, to the partial decay width.  Explicitly,
\bea
\Gamma_{p.s.}(h \rightarrow \gamma \gamma) &\simeq& \frac{\hat{m}_h^3}{4 \pi \, \bar{v}_T^2}
|\langle h| \gamma \gamma\rangle_{{\rm{to}} \, {\cal{O}}(v^2/\Lambda^2)}|^2 \nonumber \\
&\simeq& \frac{\hat{m}_h^3}{4 \pi \, \bar{v}_T^2} \left[\bar{v}_T^2 \bigg|\mathcal{A}_{\rm SM}^{h\gamma\gamma}
\bigg|^2+ 2\bar{v}_T \,{\rm Re}(\mathcal{A}_{\rm SM}^{h\gamma\gamma}) \, \langle h|\gamma \gamma\rangle_{\mathcal{L}^{(6)}}
 + \langle h|\gamma \gamma\rangle_{\mathcal{L}^{(6)}}^2\right],
\eea
while in the case of the full CP-even ${\cal{O}}(v^4/\Lambda^4)$ SMEFT result one has
\begin{align}
\Gamma_{\rm SMEFT}(h \rightarrow \gamma \gamma) &= \frac{\hat{m}_h^3}{4 \pi \, \bar{v}_T^2}
\left|\langle h| \gamma \gamma \rangle \right|^2_{{\rm{to}} \, {\cal{O}}(v^4/\Lambda^4)} \nn
&= \frac{\hat{m}_h^3}{4 \pi \, \bar{v}_T^2}
\left[\bar{v}_T^2\bigg|\mathcal{A}_{\rm SM}^{h\gamma\gamma}  \bigg|^2 + 2\bar{v}_T \,{\rm Re}(\mathcal{A}_{\rm SM}^{h\gamma\gamma})
(1+\langle \sqrt{h}^{44}\rangle_{{\cal{O}}(v^2/\Lambda^2)}) \, \langle h|\gamma \gamma\rangle_{\mathcal{L}^{(6)}} \right.\\
&+
\left.
(1+ 4 \,\bar{v}_T \, {\rm Re}(\mathcal{A}_{\rm SM}^{h\gamma\gamma})) \, (\langle h|\gamma \gamma\rangle_{\mathcal{L}^{(6)}})^2
+ 4\bar{v}_T \,{\rm Re}(\mathcal{A}_{\rm SM}^{h\gamma\gamma})
\left.(\langle h|\gamma \gamma\rangle_{\mathcal{L}^{(6)}})\right|_{C_i^{(6)} \rightarrow C_i^{(8)}}\right]. \nonumber
\end{align}
Restricting the analysis to corrections scaling as Eqns.~\eqref{gam1}-\eqref{gam3} and neglecting corrections
$\propto~\frac{(\bar{g}^{\rm SM})^4}{(16 \pi^2)^2}$ as in Eqn.~\eqref{gam0}, the partial-square correction is
\bea
\frac{\Gamma_{p.s.}^{\hat{m}_{W}}(h \rightarrow \gamma \gamma)}{ \Gamma^{\hat{m}_{W}}_{\rm SM}(h \rightarrow \gamma \gamma)} &\simeq&
1 -  788 f^{\hat{m}_W}_1
+394^2 \, (f^{\hat{m}_W}_1)^2
- 351 \, (\tilde{C}_{HW}^{(6)} - \tilde{C}_{HB}^{(6)})\, f^{\hat{m}_W}_3 \\
&+&  979 \, \tilde{C}_{HD}^{(6)}(\tilde{C}_{HB}^{(6)} +0.80\, \, \tilde{C}_{HW}^{(6)} -1.02  \, \tilde{C}_{HWB}^{(6)})
+ 2228 \, \delta G_F^{(6)} \, f^{\hat{m}_W}_1 \nn
&+& 2283 \, \tilde{C}_{HWB}^{(6)}(\tilde{C}_{HB}^{(6)} +0.66 \, \, \tilde{C}_{HW}^{(6)} -0.88  \, \tilde{C}_{HWB}^{(6)}),\nonumber
\label{eq:hgamgamPSmW}
\eea
where
\bea
f^{\hat{m}_W}_1 &\simeq& f^{\hat{\alpha}_{ew}}_1= \left[\tilde{C}_{HB}^{(6)} +0.29 \, \, \tilde{C}_{HW}^{(6)} -0.54  \, \tilde{C}_{HWB}^{(6)}\right],\\
f^{\hat{m}_W}_2 &\simeq& f^{\hat{\alpha}_{ew}}_2 = \left[\tilde{C}_{HB}^{(8)} +0.29 \, \, (\tilde{C}_{HW}^{(8)}+ \tilde{C}_{HW,2}^{(8)}) -0.54  \, \tilde{C}_{HWB}^{(8)}\right],\\
f^{\hat{m}_W}_3 &\simeq& f^{\hat{\alpha}_{ew}}_3 = \left[\tilde{C}_{HW}^{(6)} - \tilde{C}_{HB}^{(6)} -0.66  \, \tilde{C}_{HWB}^{(6)}\right],
\eea
in both input-parameter schemes. The corresponding (CP-even) ${\cal{O}}(v^4/\Lambda^4)$ SMEFT result in the $\hat{m}_{W}$ scheme
is
\bea \label{eq:hgamgamSMEFT}
\frac{\Gamma_{\rm SMEFT}^{\hat{m}_{W}}(h \rightarrow \gamma \gamma)}{ \Gamma^{\hat{m}_{W}}_{\rm SM}(h \rightarrow \gamma \gamma)}
=\frac{\Gamma_{p.s.}^{\hat{m}_{W}}(h \rightarrow \gamma \gamma)}{ \Gamma^{\hat{m}_{W}}_{\rm SM}(h \rightarrow \gamma \gamma)} -788 \left[ \left(\tilde C_{H\Box}^{(6)} - \frac{\tilde C_{HD}^{(6)}}{4}\right) \, f^{\hat{m}_W}_1+ f^{\hat{m}_W}_2\right]
- 1224 \, (f^{\hat{m}_W}_1)^2. \nn
\eea
We numerically analyse the difference between the SMEFT result and the partial-square result in Secs.~\ref{sec:MCscan} and \ref{sec:UVmodel}.

\subsection{$\Gamma(h \rightarrow \mathcal{Z} \gamma)$}
A similar analysis for $\Gamma(h \rightarrow \mathcal{Z} \gamma)$ begins with the SM result
\begin{align}
\Gamma^{ \hat{m}_W}_{\rm SM}(h \rightarrow \mathcal{Z} \gamma) = \frac{\hat{m}_h^3}{8 \pi} \, \left(1- \frac{\hat{m}_Z^2}{\hat{m}_h^2} \right)^3 \bigg|\mathcal{A}_{\rm SM}^{hZ\gamma}  \bigg|^2 =
6.5 \times 10^{-6}\, {\rm GeV}.
\end{align}
Again neglecting corrections $\propto~\frac{(\bar{g}^{\rm SM})^4}{(16 \pi^2)^2}$, the partial-square correction is
\begin{align}
&\frac{\Gamma_{p.s.}^{\hat{m}_{W}}(h \rightarrow \mathcal{Z} \gamma)}{ \Gamma^{\hat{m}_{W}}_{\rm SM}(h \rightarrow \mathcal{Z} \gamma)} \simeq
1 -237 f^{\hat{m}_W}_3
+118^2 \, (f^{\hat{m}_W}_3)^2
-131 \, (\tilde{C}_{HB}^{(6)} - \,\tilde{C}_{HW}^{(6)})^2 -670 \, \delta G_F^{(6)} \, f^{\hat{m}_W}_3 \nn
 &\qquad\qquad -616 \,\tilde{C}_{HWB}^{(6)} \, (\tilde{C}_{HB}^{(6)} - \,\tilde{C}_{HW}^{(6)}+ 0.02 \tilde{C}_{HWB}^{(6)})
-265\,  \tilde{C}_{HD}^{(6)}\, (\tilde{C}_{HW}^{(6)} - \,\tilde{C}_{HB}^{(6)} -0.54 \tilde{C}_{HWB}^{(6)}).\nonumber
\end{align}
Finally, the full (CP-even) SMEFT result in the $\hat{m}_{W}$ scheme to ${\cal{O}}(v^4/\Lambda^4)$ is
\bea
\frac{\Gamma_{\rm SMEFT}^{\hat{m}_{W}}(h \rightarrow \mathcal{Z} \gamma)}{ \Gamma^{\hat{m}_{W}}_{\rm SM}(h \rightarrow \mathcal{Z} \gamma)}
&=&\frac{\Gamma_{p.s.}^{\hat{m}_{W}}(h \rightarrow  \mathcal{Z} \gamma)}{ \Gamma^{\hat{m}_{W}}_{\rm SM}(h \rightarrow \mathcal{Z} \gamma)}
-237 \left[\left(\tilde C_{H\Box}^{(6)} - \frac{\tilde C_{HD}^{(6)}}{4}\right) \, f^{\hat{m}_W}_3+ f^{\hat{m}_W}_4\right] \\
 &-& 296 \, (f^{\hat{m}_W}_3)^2 -237 f^{\hat{m}_W}_3 \left(3.8 \tilde{C}_{HB}^{(6)}+ 0.20 \tilde{C}_{HW}^{(6)} \right), \nonumber
\eea
where
$f^{\hat{m}_W}_4 \simeq f^{\hat{\alpha}_{ew}}_4 = \left[\tilde{C}_{HW}^{(8)}+ \tilde{C}_{HW,2}^{(8)} - \tilde{C}_{HB}^{(8)} -0.66  \, \tilde{C}_{HWB}^{(8)}\right]$.

\subsection{$\Gamma(\mathcal{Z} \rightarrow\bar{\psi} \psi)$}
For $\Gamma(\mathcal{Z} \rightarrow\bar{\psi} \psi)$ the difference between the partial-square and ${\cal{O}}(v^4/\Lambda^4)$ results
is dictated by the difference in $\left|g_{\rm eff,pr}^{\mathcal{Z},\psi} \right|^2$ for each input-parameter case.
We add the two chiral final states for each fermion pair to obtain a partial-square correction of
\begin{align}
\frac{\bar{\Gamma}^{p.s.}_{\mathcal{Z} \rightarrow \bar{\psi}_p \psi_p}}{\bar{\Gamma}^{\rm SM}_{\mathcal{Z} \rightarrow \bar{\psi}_p \psi_p}}
&\simeq 1 + 2  \frac{\rm{Re} \, \left[\langle g_{\rm SM,pp}^{\mathcal{Z},\psi_L}\rangle \,  \langle g_{\rm eff,pp}^{\mathcal{Z},\psi_L}\rangle_{{\cal{O}}(v^2/\Lambda^2)}
+ \langle g_{\rm SM,pp}^{\mathcal{Z},\psi_R}\rangle \,  \langle g_{\rm eff,pp}^{\mathcal{Z},\psi_R}\rangle_{{\cal{O}}(v^2/\Lambda^2)} \right]}{|\langle g_{\rm SM,pp}^{\mathcal{Z},\psi_L}\rangle|^2+ |\langle g_{\rm SM,pp}^{\mathcal{Z},\psi_R}\rangle|^2} \nn
& \hspace{1cm} + \frac{|\langle g_{\rm eff,pp}^{\mathcal{Z},\psi_R}\rangle_{{\cal{O}}(v^2/\Lambda^2)}|^2+ |\langle g_{\rm eff,pp}^{\mathcal{Z},\psi_L}\rangle_{{\cal{O}}(v^2/\Lambda^2)}|^2}{|\langle g_{\rm SM,pp}^{\mathcal{Z},\psi_L}\rangle|^2+ |\langle g_{\rm SM,pp}^{\mathcal{Z},\psi_R}\rangle|^2},
\end{align}
while the full ${\cal{O}}(v^4/\Lambda^4)$ result is
\begin{align}
\frac{\bar{\Gamma}^{\rm SMEFT}_{\mathcal{Z} \rightarrow \bar{\psi}_p \psi_p}}{\bar{\Gamma}^{\rm SM}_{\mathcal{Z} \rightarrow \bar{\psi}_p \psi_p}}
 &= \frac{\bar{\Gamma}^{p.s.}_{\mathcal{Z} \rightarrow \bar{\psi}_p \psi_p}}{\bar{\Gamma}^{\rm SM}_{\mathcal{Z} \rightarrow \bar{\psi}_p \psi_p}}
 + 2 \frac{\rm{Re} \, \left[\langle g_{\rm SM,pp}^{\mathcal{Z},\psi_L}\rangle \, \langle g_{\rm eff,pp}^{\mathcal{Z},\psi_L}\rangle_{{\cal{O}}(v^4/\Lambda^4)}
 + \langle g_{\rm SM,pp}^{\mathcal{Z},\psi_R}\rangle \,  \langle g_{\rm eff,pp}^{\mathcal{Z},\psi_R}\rangle_{{\cal{O}}(v^4/\Lambda^4)} \right]}{|\langle g_{\rm SM,pp}^{\mathcal{Z},\psi_L}\rangle|^2+ |\langle g_{\rm SM,pp}^{\mathcal{Z},\psi_R}\rangle|^2}.
\end{align}
In the $\{\hat{m}_{W}, \hat{m}_Z, \hat{G}_F \}$ input-parameter scheme, the leading-order SM results are
\begin{align}
\bar{\Gamma}^{\rm SM}_{\mathcal{Z} \rightarrow \bar{u} u} &= 0.29 \, {\rm GeV}, & \quad
\bar{\Gamma}^{\rm SM}_{\mathcal{Z} \rightarrow \bar{d} d} &= 0.37 \, {\rm GeV}, \\
\bar{\Gamma}^{\rm SM}_{\mathcal{Z} \rightarrow \bar{\ell} \ell} &= 0.08 \, {\rm GeV}, & \quad
\bar{\Gamma}^{\rm SM}_{\mathcal{Z} \rightarrow \bar{\nu} \nu} &= 0.17 \, {\rm GeV}.
\end{align}
For up-type quarks, we find the following expressions for $\langle g_{\rm eff,pp}^{\mathcal{Z},\psi_{L/R}}\rangle$:
\begin{align}
\langle g_{\rm SM,pp}^{\mathcal{Z},u_{L}}\rangle &= -0.26, \\
\langle g_{\rm eff,pp}^{\mathcal{Z},u_{L}}\rangle_{{\cal{O}}(v^2/\Lambda^2)}
&= -0.13 \tilde{C}_{HD}^{(6)} -0.21 \,  \tilde{C}_{HWB}^{(6)} + 0.18 \, \delta G_{F}^{(6)}
+ 0.37 \, (\tilde{C}_{\substack{Hq\\pp}}^{(6)}- \tilde{C}_{\substack{Hq\\pp}}^{3,(6)}), \\
\langle g_{\rm eff,pp}^{\mathcal{Z},u_{L}}\rangle_{{\cal{O}}(v^4/\Lambda^4)}
&= -\left(\frac{\tilde{C}_{HD}^{(6)}}{4}+ \frac{\delta G_{F}^{(6)}}{\sqrt{2}}\right) \langle g_{\rm eff,pp}^{\mathcal{Z},u_{L}}\rangle_{{\cal{O}}(v^2/\Lambda^2)}
\\ &+  \tilde{C}_{HWB}^{(6)} \left(0.13 \tilde{C}_{HD}^{(6)}
-0.21 (\tilde{C}_{HB}^{(6)} +\tilde{C}_{HW}^{(6)})  \right)\nn
&-0.01 (\tilde{C}_{HD}^{(6)})^2 + 0.05 \tilde{C}_{HD}^{(6)} \, \delta G_{F}^{(6)} +0.03 \tilde{C}_{HD}^{(8)}
-0.16 \tilde{C}_{H,D2}^{(8)} -0.10 \tilde{C}_{HWB}^{(8)} \nn
&- 0.38 \tilde{C}_{HW,2}^{(8)} - \frac{0.37}{2} (\tilde{C}_{\substack{Hq\\pp}}^{2,(8)} + \tilde{C}_{\substack{Hq\\pp}}^{3, (8)}- \tilde{C}_{\substack{Hq\\pp}}^{(8)})  -0.07 (\delta G_{F}^{(6)})^2 +0.18 \delta G_{F}^{(8)}, \nn
\langle g_{\rm SM,pp}^{\mathcal{Z},u_{R}}\rangle &=  0.11,\\
\langle g_{\rm eff,pp}^{\mathcal{Z},u_{R}}\rangle_{{\cal{O}}(v^2/\Lambda^2)}&=
-0.22 \, \tilde{C}_{HD}^{(6)} - 0.21 \,  \tilde{C}_{HWB}^{(6)} -0.08 \, \delta G_{F}^{(6)}
+ 0.37 \, \tilde{C}_{\substack{Hu\\pp}}^{(6)}, \\
\langle g_{\rm eff,pp}^{\mathcal{Z},u_{R}}\rangle_{{\cal{O}}(v^4/\Lambda^4)}
&= -\left(\frac{\tilde{C}_{HD}^{(6)}}{4}+ \frac{\delta G_{F}^{(6)}}{\sqrt{2}}\right) \langle g_{\rm eff,pp}^{\mathcal{Z},u_{R}}\rangle_{{\cal{O}}(v^2/\Lambda^2)}
\\ &+  \tilde{C}_{HWB}^{(6)} \left(0.13 \tilde{C}_{HD}^{(6)} -0.21 (\tilde{C}_{HB}^{(6)} +\tilde{C}_{HW}^{(6)})  \right) \nn
&+0.003 (\tilde{C}_{HD}^{(6)})^2 - 0.02 \tilde{C}_{HD}^{(6)} \, \delta G_{F}^{(6)} -0.01 \tilde{C}_{HD}^{(8)}
-0.21 \tilde{C}_{H,D2}^{(8)} -0.10 \tilde{C}_{HWB}^{(8)} \nn
&- 0.38 \tilde{C}_{HW,2}^{(8)} + \frac{0.37}{2} \, \tilde{C}_{\substack{Hu\\pp}}^{(8)} +0.03 (\delta G_{F}^{(6)})^2  -0.08  \delta G_{F}^{(8)}.\nonumber
\end{align}
The remaining expressions for final-state fermions are listed in Appendix~\ref{zwidthmw}. The total width is
the linear sum of the partial widths.


\section{Coefficient sampling analysis}\label{sec:MCscan}

The final step required to numerically compare the full ${\cal{O}}(v^4/\Lambda^4)$ result to the partial-square calculation for each
process is to choose the coefficients $C^{(d)}$.  The processes examined in Sec.~\ref{numerics} depend on $\mathcal O(30)$
coefficients\footnote{This number assumes flavor universality and treats $\delta G^{(d)}_F$ as a single coefficient rather than
separating out the different contributions as shown in Appendix~\ref{inputs}. Relaxing either of these assumptions will change the
count by a small amount.}. While 30 is far less than the $\mathcal O(1000)$ coefficients in the full dimension-6 plus dimension-8
SMEFT ($N_f = 1$), it is still too many to analyze coherently without making further assumptions (outside of a global fit). We explore
two options for choosing coefficients: a sampling approach (this section), and an ultraviolet (UV) model-based approach (Sec.~\ref{sec:UVmodel}).

In a sampling study, coefficient values are drawn from assumed distributions. This approach treats the SMEFT as a bottom-up
effective field theory, irrespective of a particular UV completion of the SM. The decoupling theorem~\cite{Appelquist:1974tg,Symanzik:1973vg}
establishes the SMEFT as a distinct theory, so this approach is favored in EFT studies of experimental data.  Absent any UV model
constraint on the parameters, the simplest assumed distribution is a uniform flat distribution with all coupling values
equally likely, consistent with perturbation theory.  One can add a mild assumption by noting that UV models typically introduce
particles whose parameters can be mapped to a few coefficients, and in these cases the majority of the coefficients will have small
values.  This scenario can be approximated by a gaussian distribution for the coefficient values.  We find that the differences
between the uniform and gaussian distributions are imperceptible, and we sample from a gaussian distribution for the results in
this section.

We start with a qualitative assessment of the variations in the partial widths as terms at ${\cal{O}}(v^4/\Lambda^4)$ are included
in the calculation (Sec.~\ref{sec:MCscanwidths}).  We then turn to a study of procedures for coefficient uncertainty estimates in
Sec.~\ref{sec:MCscancoefficients}.

\subsection{Partial-width variations}\label{sec:MCscanwidths}

In order to compare partial-square and full ${\cal{O}}(v^4/\Lambda^4)$ results for the partial widths, we proceed as follows:
\begin{enumerate}
\item[1.)] We sample coefficients affecting the calculation to ${\cal{O}}(v^2/\Lambda^2)$.  Coefficients of tree-level operators are
	drawn from a gaussian (or uniform) distribution with a mean of zero and a root mean square (r.m.s.) equal to one.  For loop-level
coefficients the r.m.s. and range are
reduced by a factor of 100, since larger values give large relative corrections to the SM predictions that are inconsistent with
experimental results.  The categorization of operator coefficients as tree-level or loop-level is determined using the results of
Ref.~\cite{Arzt:1994gp, Jenkins:2013fya, Craig:2019wmo}.  For example, the $h \to \gamma\gamma$ tree-level matching coefficients for
SMEFT operators are $C^{(6)}_{HD}, C^{(6)}_{H\Box}, C^{1,(6)}_{H\psi}$, $C^{3,(6)}_{H\psi}$, and $\delta G^{(6)}_{F}$, while the loop-level
coefficients are $C^{(6)}_{HB}, C^{(6)}_{HW},$ and $C^{(6)}_{HWB}$.  We draw random values for the $C^{(6)}$ coefficients, and the
$\tilde C^{(6)}$ coefficients that appear in Secs.~\ref{setup}~and~\ref{analyticexpressions} can be obtained by multiplying
by $\bar v^2_T/\Lambda^2$.

With these coefficients and using the procedure outlined in Appendix~\ref{inputs} to connect to experimental EW inputs, the
${\cal{O}}(v^2/\Lambda^2)$ and partial-square calculations are determined up to the value of $\Lambda$.  The result after this step
is schematically
\begin{align}
\Gamma = \Gamma_{\rm SM} + \frac{\#}{\Lambda^2} + \frac{\#'}{\Lambda^4}.
\end{align}

\item[2.)] For each set of coefficients obtained above, we perform 10,000 separate samplings of the remaining coefficients affecting the
the full ${\cal{O}}(v^4/\Lambda^4)$ result.  The coefficients are again separated into tree- and loop-level, with dimension-8 operators with
Higgs and gauge field strengths (e.g. $C^{(8)}_{HB}, C^{(8)}_{HW}, C^{(8)}_{HWB}$) classified as tree-level following~\cite{Craig:2019wmo}
(their dimension-6 counterparts are classified as loop-level for the Higgs boson partial widths we consider).

\item[3.)] We calculate the deviation from the SM for each $\Gamma_{\rm SMEFT}(\Lambda)$,
\begin{align}
\delta_{\rm SMEFT}(\Lambda) = \frac{\Gamma_{\rm SMEFT}(\Lambda) - \Gamma_{\rm SM}}{\Gamma_{\rm SM}}
\label{eq:deltaXS}
\end{align}
and determine the standard deviation $\sigma_\delta$ of the $\delta_{\rm SMEFT}(\Lambda)$ distribution.

\item[4.)] We compare the $\delta_{{\cal{O}}(v^2/\Lambda^2)}(\Lambda)$ and $\delta_{p.s.}(\Lambda)$ curves, defined in the same way as
Eqn.~\eqref{eq:deltaXS}, to the $\pm 1,2, 3\,\sigma_\delta$ $\delta_{\rm SMEFT}$ curves.
\end{enumerate}

\begin{figure*}[!tp]
\centering
\includegraphics[width=0.49\textwidth]{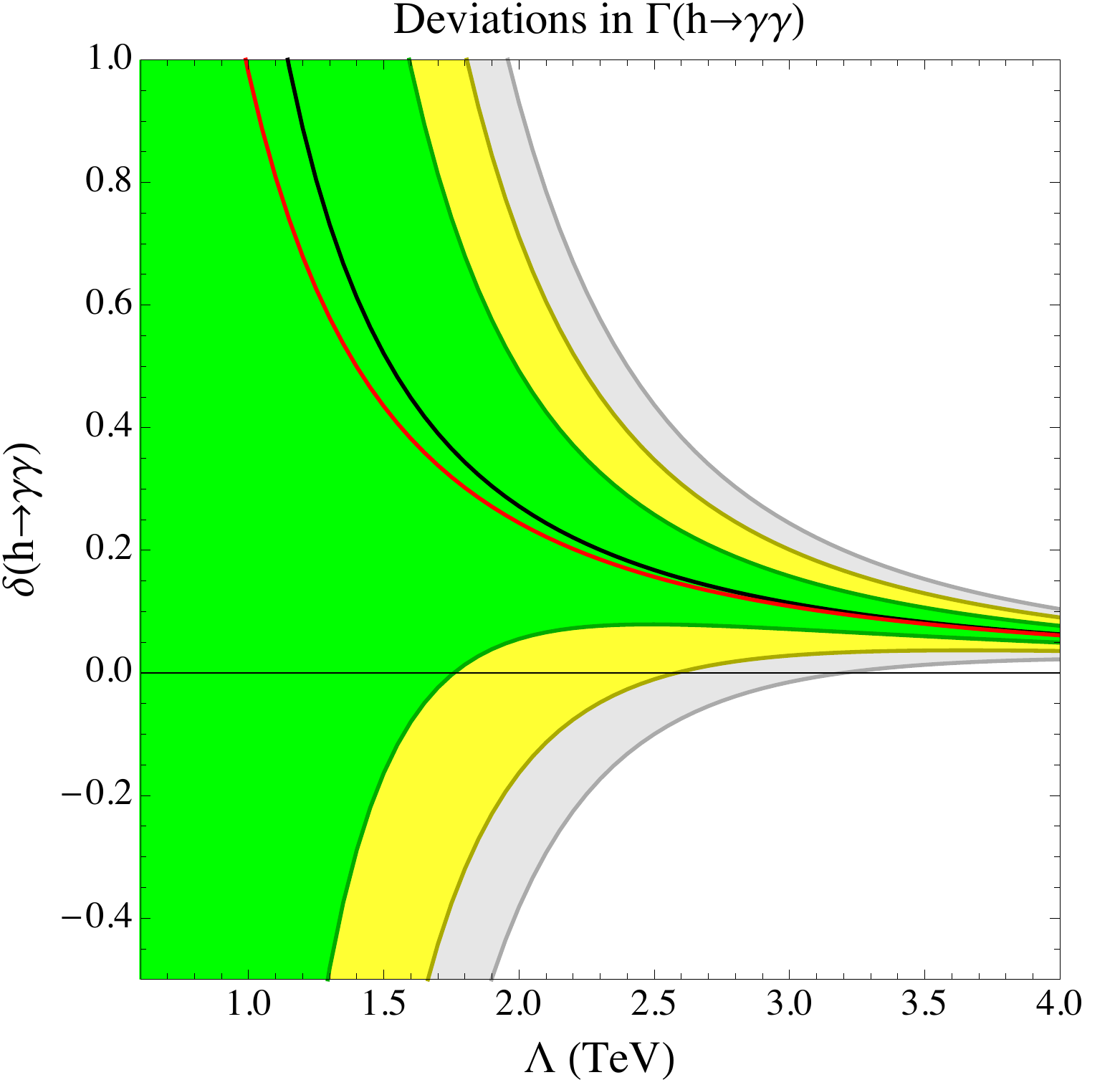}
\includegraphics[width=0.49\textwidth]{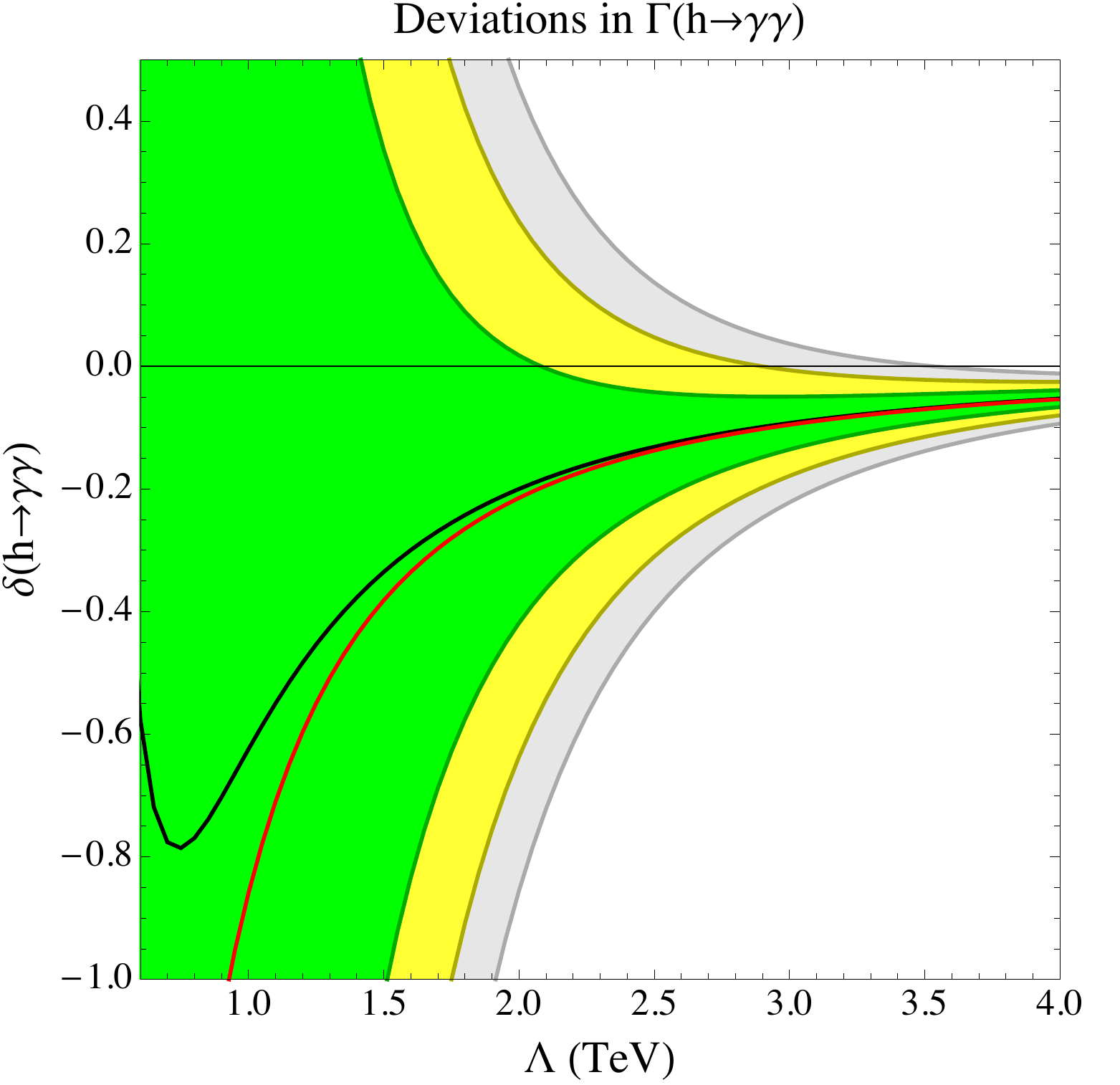}
\caption{
The deviations in $h \to \gamma\gamma$ from the ${\cal{O}}(v^2/\Lambda^2)$ (red line) and partial-square (black line)
results, and the full ${\cal{O}}(v^4/\Lambda^4)$ results (green $\pm 1\,\sigma_\delta$, yellow $\pm 2\,\sigma_\delta$, and grey
$\pm 3\,\sigma_\delta$ regions).
In the left panel the coefficients determining the ${\cal{O}}(v^2/\Lambda^2)$ and partial-square
results are $C^{(6)}_{HB} = -0.01,\, C^{(6)}_{HW} = 0.004, C^{(6)}_{HWB} = 0.007, C^{(6)}_{HD} = -0.74,$ and $\delta G^{(6)}_{F} = -1.6$.
In the right panel they are $C^{(6)}_{HB} = 0.007,\, C^{(6)}_{HW} = 0.007, C^{(6)}_{HWB} = -0.015, C^{(6)}_{HD} = 0.50,$ and
$\delta G^{(6)}_{F} = 1.26$.
}
\label{fig:MCgammagamma}
\end{figure*}

Following this procedure for $h\to \gamma \gamma$ in the $\hat m_W$ scheme gives the results in Figure~\ref{fig:MCgammagamma}
for two sets of ${\cal{O}}(v^2/\Lambda^2)$ coefficients.  The green shaded region shows the $\pm 1\, \sigma_\delta$ deviations
of the partial width from the SM prediction with the full ${\cal{O}}(v^4/\Lambda^4)$ SMEFT calculation, for given
${\cal{O}}(v^2/\Lambda^2)$ (red line) and partial-square (black line) results.  The $\pm 2, 3\, \sigma_\delta$ regions are
shown in yellow and gray, respectively.  The figure shows the expected dependence of the partial width on ${\cal{O}}(v^4/\Lambda^4)$: as
$\Lambda$ increases the impact of these terms decreases.  For $\Lambda \lesssim 1$~TeV neither the ${\cal{O}}(v^2/\Lambda^2)$ nor the
partial-square calculation provides a good approximation.  Equivalently, for a given measured partial width the inferred coupling or scale
is affected by higher order terms if the scale is low.  A 10\% deviation in the partial width corresponds to a scale of $\approx 3$~TeV
for the chosen coefficients at ${\cal{O}}(v^2/\Lambda^2)$, but the scale can be much lower if there are cancellations from higher-order
terms.

\begin{figure*}[htp!]
\centering
\includegraphics[width=0.49\textwidth]{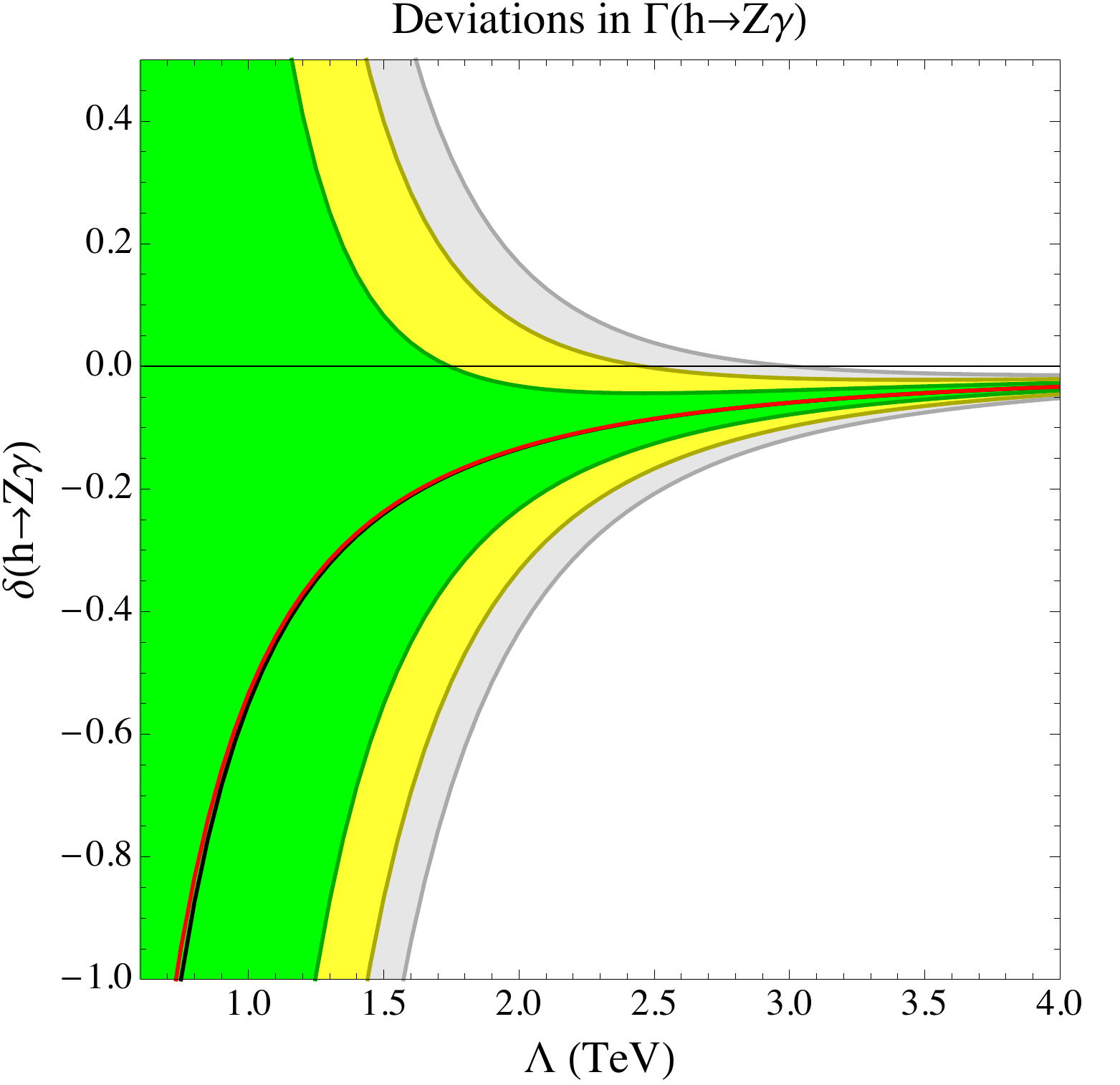}
\includegraphics[width=0.49\textwidth]{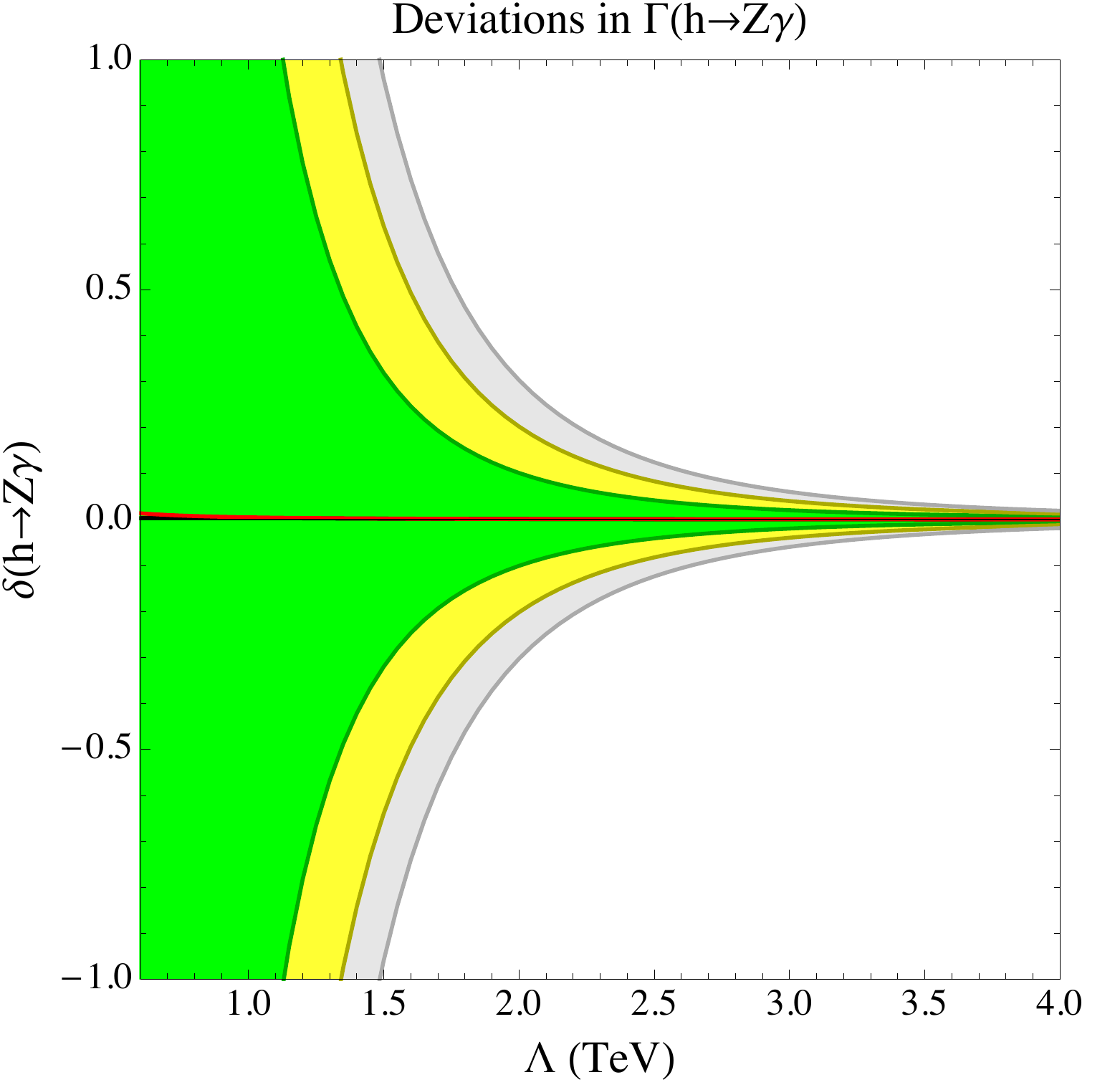}
\caption{
The deviations in $h \to \mathcal{Z}\gamma$ from the ${\cal{O}}(v^2/\Lambda^2)$ (red line) and partial-square (black line)
results, and the full ${\cal{O}}(v^4/\Lambda^4)$ results (green $\pm 1\,\sigma_\delta$, yellow $\pm 2\,\sigma_\delta$, and grey
$\pm 3\,\sigma_\delta$ regions).
In the left panel the coefficients determining the ${\cal{O}}(v^2/\Lambda^2)$ and partial-square results are
$C^{(6)}_{HB} = -0.01,\, C^{(6)}_{HW} = 0.02, C^{(6)}_{HWB} = -0.011, C^{(6)}_{HD} = 0.53,$ and $\delta G^{(6)}_{F} = 0.13$.
In the right panel they are $C^{(6)}_{HB} = 0.002,\, C^{(6)}_{HW} = 0.001, C^{(6)}_{HWB} = -0.001, C^{(6)}_{HD} = 0.28,$
and $\delta G^{(6)}_{F} = -1.15$.
}
\label{fig:MCZgamma}
\end{figure*}

\begin{figure*}[htp!]
\centering
\includegraphics[width=0.49\textwidth]{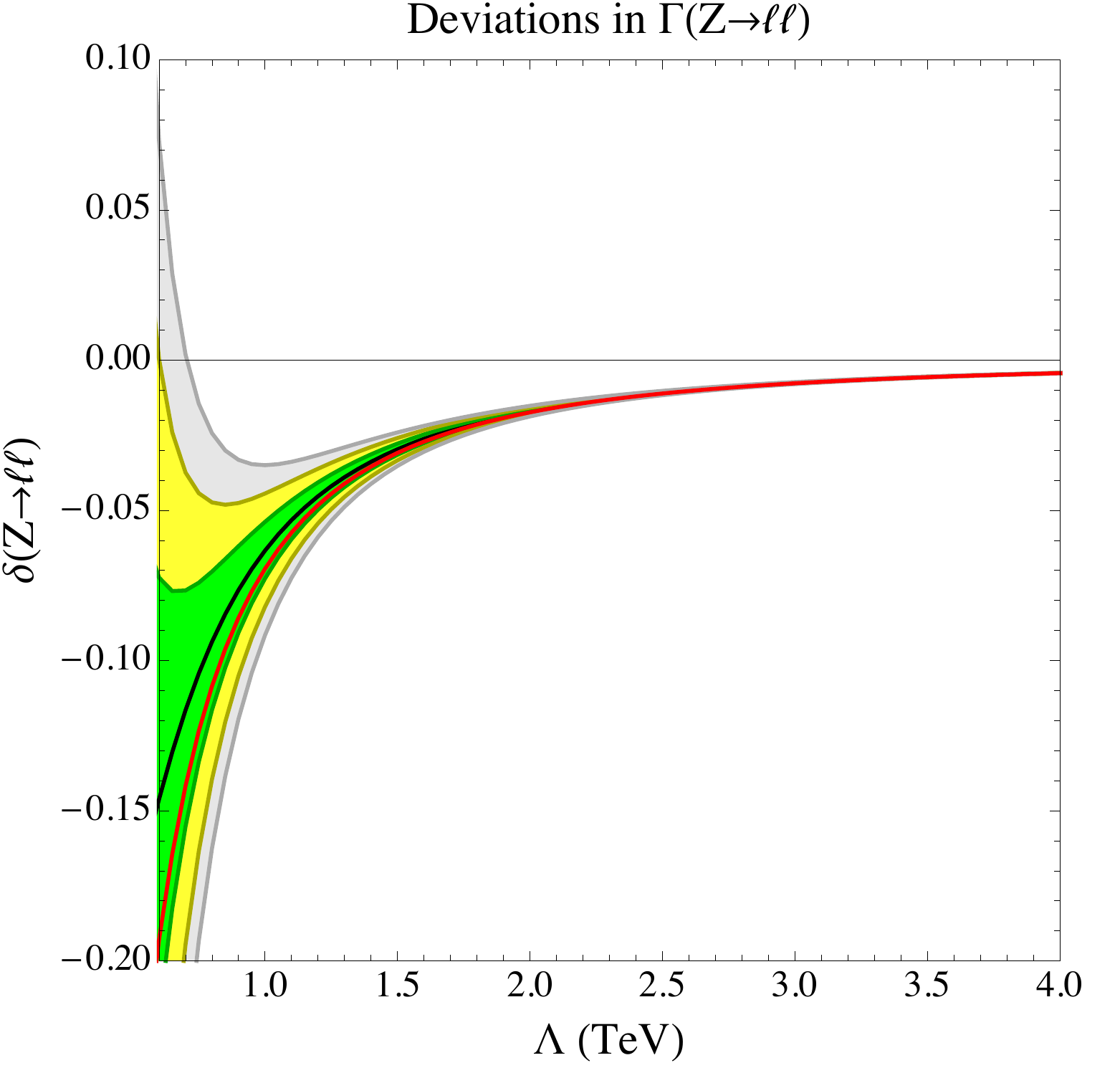}
\includegraphics[width=0.49\textwidth]{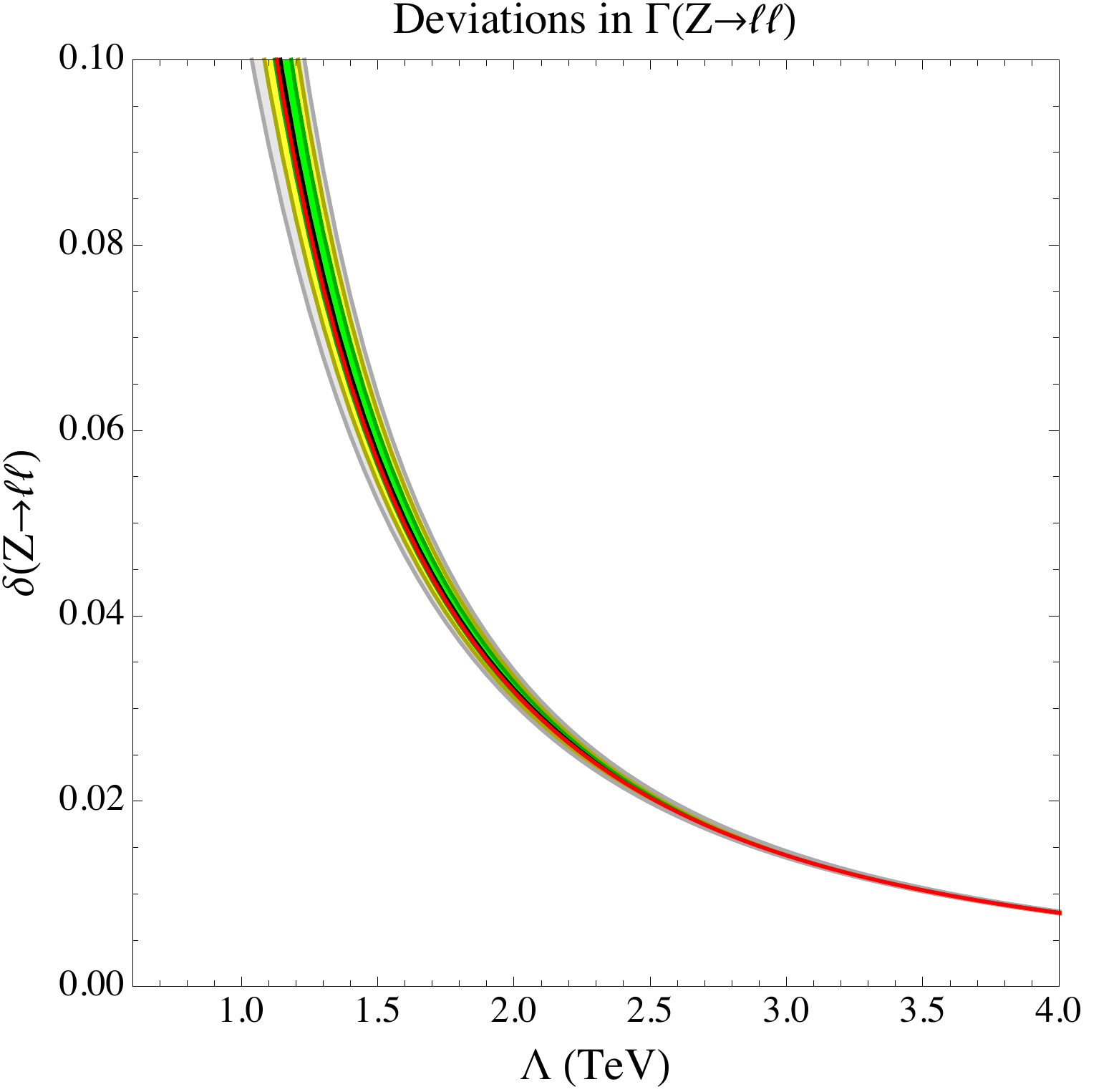}
\caption{
The deviations in $\mathcal{Z} \to \ell\ell$ from the ${\cal{O}}(v^2/\Lambda^2)$ (red line) and partial-square (black line)
results, and the full ${\cal{O}}(v^4/\Lambda^4)$ results (green $\pm 1\,\sigma_\delta$, yellow $\pm 2\,\sigma_\delta$, and grey
$\pm 3\,\sigma_\delta$ regions).
In the left panel the coefficients determining the ${\cal{O}}(v^2/\Lambda^2)$ and partial-square results are
$C^{1,(6)}_{H\ell} = -0.46, C^{3,(6)}_{H\ell} = 1.24, C^{(6)}_{He} = 1.53, C^{(6)}_{HD} = -0.79, C^{(6)}_{HWB} = 0.007,$ and
$\delta G^{(6)}_F = 0.16$.  In the right panel they are
$C^{1,(6)}_{H\ell} = 1.55, C^{3,(6)}_{H\ell} = -0.71, C^{(6)}_{He} = 0.23, C^{(6)}_{HD} = -0.51, C^{(6)}_{HWB} = -0.008,$ and
$\delta G^{(6)}_F = -0.44$. }
\label{fig:MCZll}
\end{figure*}

Figures~\ref{fig:MCZgamma}~and~\ref{fig:MCZll} show the results of similar coefficient sampling studies for $h \to \mathcal{Z}\gamma$
and $\mathcal{Z} \to \ell\ell$ in the $\hat m_W$ scheme.
The loop-level $h \to \mathcal{Z}\gamma$ and $h \to \gamma\gamma$ processes are similar, with a broad band of deviations from
${\cal{O}}(v^4/\Lambda^4)$ contributions at low scales.  In the right panel of Fig.~\ref{fig:MCZgamma} an accidental cancellation
in the partial-square result leads to essentially zero deviation, while the ${\cal{O}}(v^4/\Lambda^4)$ band is just as broad as
in the left panel.
The band of deviations is narrower for the $\mathcal{Z} \to \ell\ell$ partial width, which is tree-level in the SM.

\subsection{Coefficient variations}\label{sec:MCscancoefficients}

In global fits for SMEFT coefficients at ${\cal{O}}(v^2/\Lambda^2)$, it is appropriate to consider the
effect of the EFT truncation on the extracted values.  We investigate two possible procedures for
estimating this effect: (1) using the difference between the partial-square result and the ${\cal{O}}(v^2/\Lambda^2)$
SMEFT result as an estimate of a `truncation uncertainty'; and (2) taking the fractional uncertainty on each
coefficient to be $v^2/\Lambda^2$.  The former procedure uses the partial ${\cal{O}}(v^4/\Lambda^4)$ information
in the $\mathcal{L}^{(6)}$ operators to take all the calculable terms when complete higher orders are not available.
The latter procedure instead only scales the measured coefficient by the ratio of dimensionful parameters.

We test the uncertainty procedures by taking the full ${\cal{O}}(v^4/\Lambda^4)$ SMEFT calculation to provide
the `true' value of a given coefficient.  The shift in the partial width relative to the SM is calculated for a set
of coefficients drawn from a gaussian distribution.  Fixing the value of this shift and taking a given value of $\Lambda$,
we determine the change in one of the coefficients when calculating the partial width at ${\cal{O}}(v^2/\Lambda^2)$,
or with the partial-square procedure.  The deviation in the coefficient value relative to its initial value is taken
as the `truncation error'.

\begin{figure*}[tp!]
\centering
\includegraphics[width=0.49\textwidth]{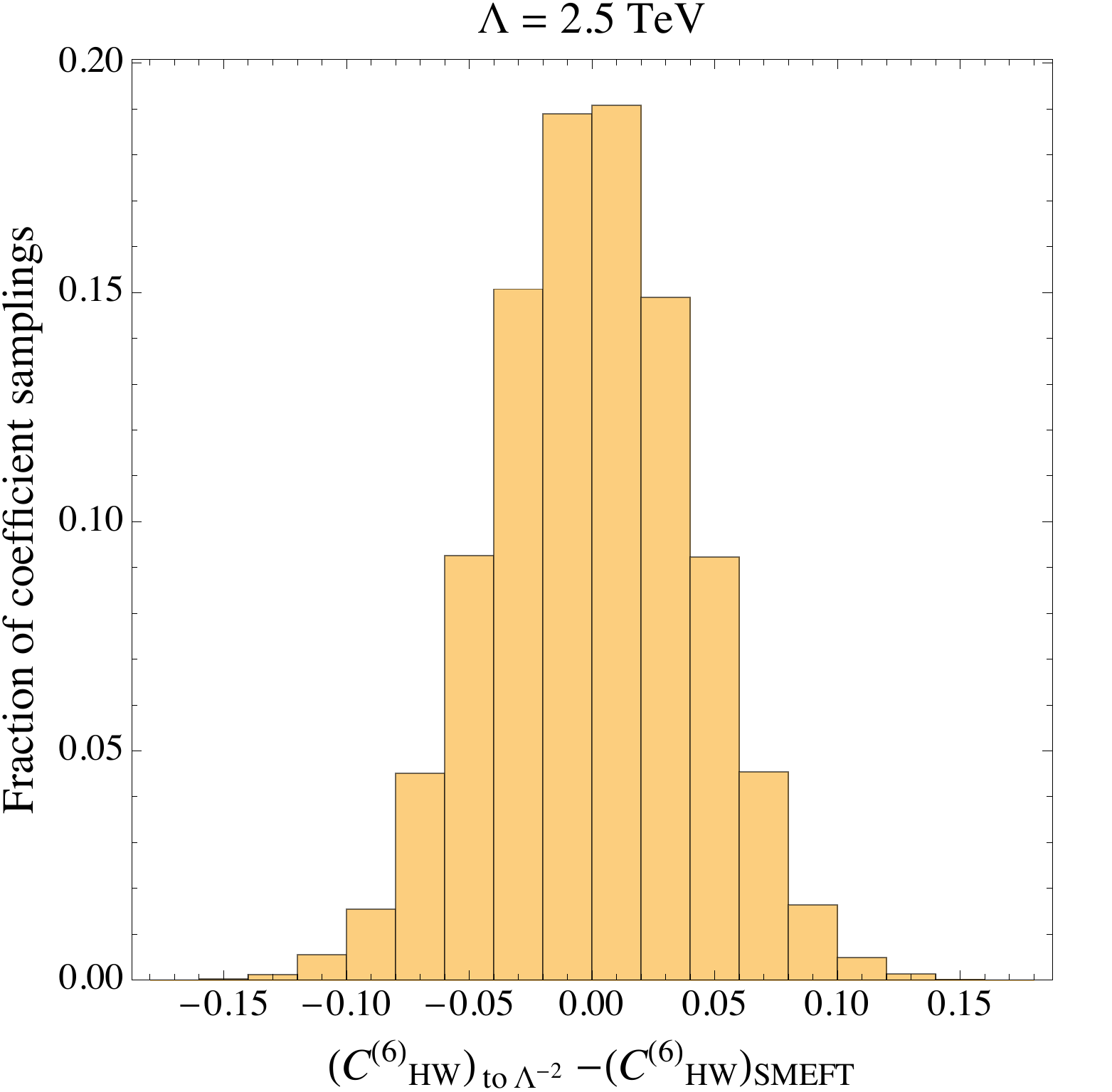}
\includegraphics[width=0.49\textwidth]{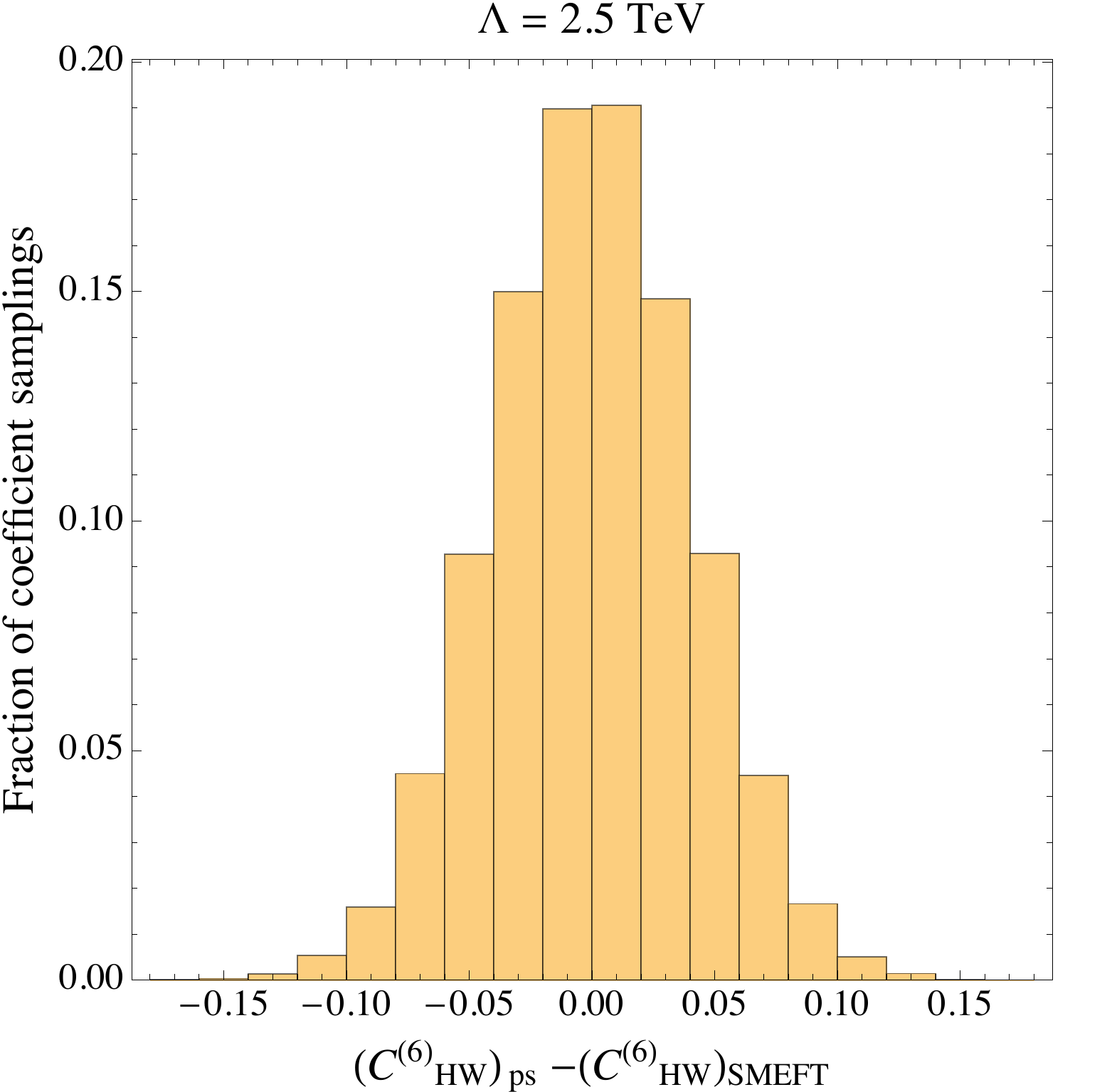}
\caption{The distribution of $C^{(6)}_{HW}$ deviations between ${\cal{O}}(v^4/\Lambda^4)$ and either ${\cal{O}}(v^2/\Lambda^2)$ (left)
or partial-square (right) calculations, where the impact on $\Gamma(h \rightarrow \gamma \gamma)$ relative to the SM is fixed to the
value obtained from the ${\cal{O}}(v^4/\Lambda^4)$ calculation.  These deviations represent the truncation errors on the $C^{(6)}_{HW}$
coefficients extracted from the $\mathcal{L}^{(6)}$ calculations. }
\label{fig:truncerr}
\end{figure*}

\begin{figure*}[tp!]
\centering
\includegraphics[width=0.49\textwidth]{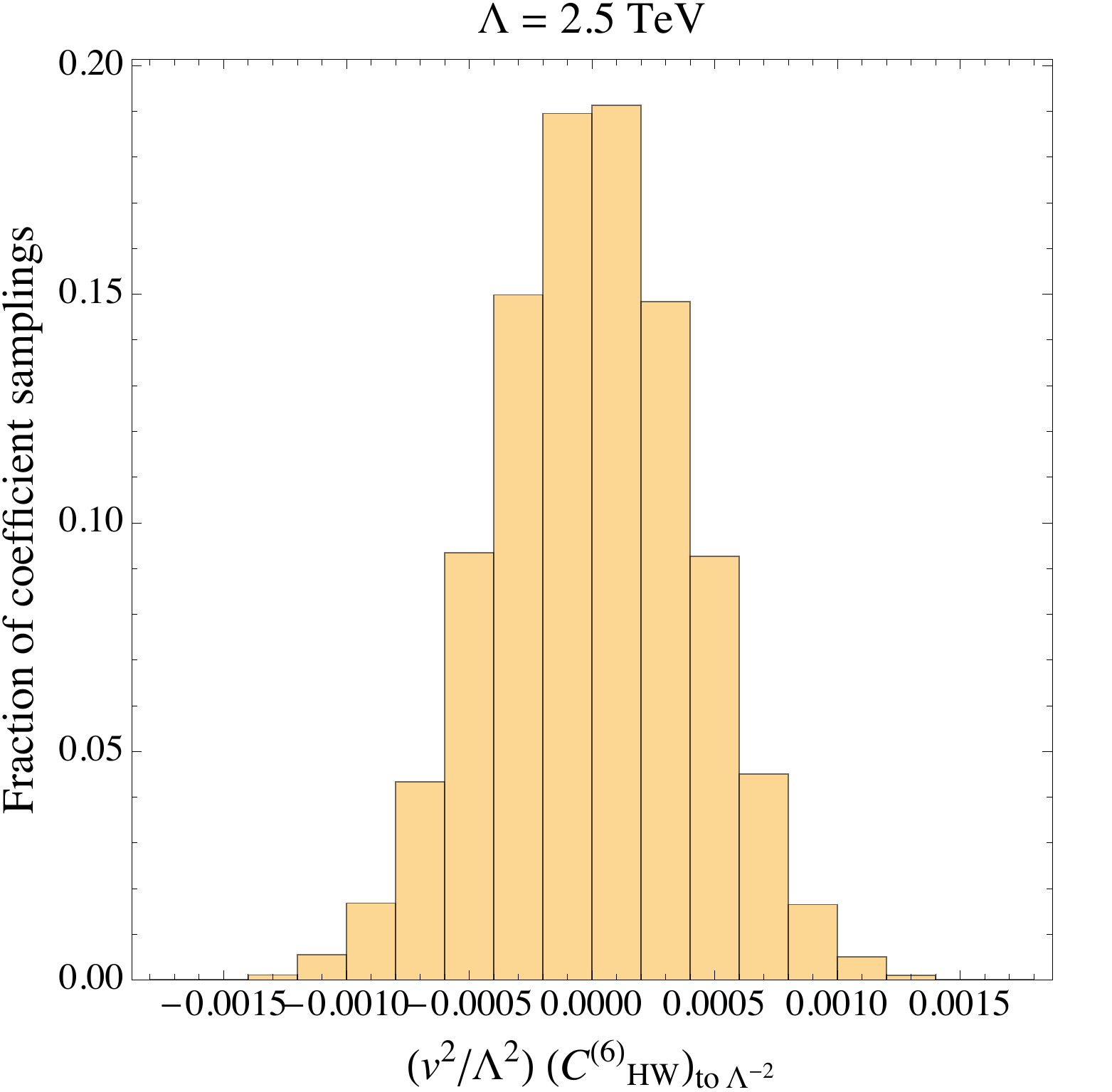}
\includegraphics[width=0.49\textwidth]{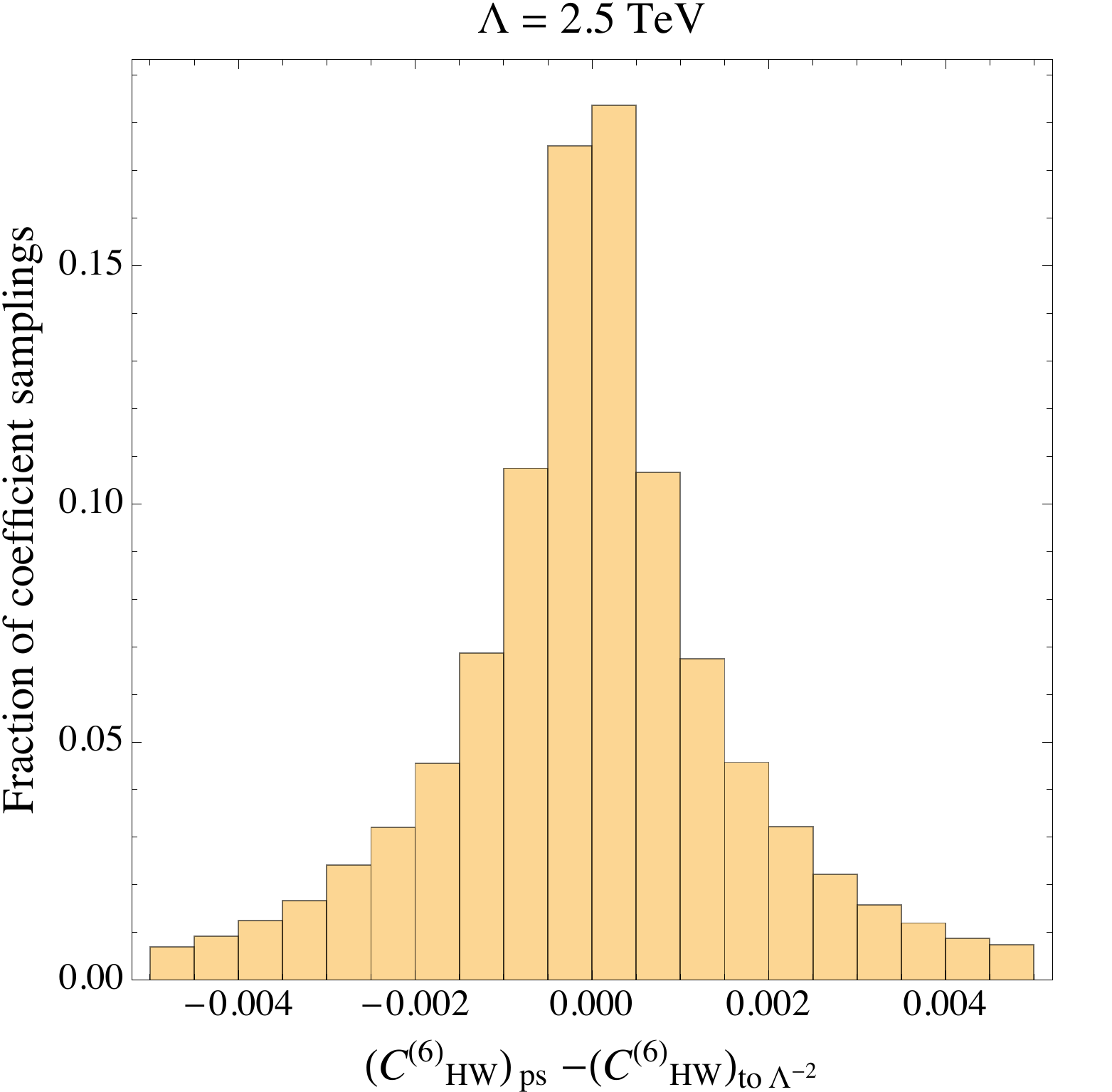}
\caption{Left: The distribution of the $C^{(6)}_{HW}$ deviation between ${\cal{O}}(v^2/\Lambda^2)$ and partial-square calculations,
where the impact on $\Gamma(h \rightarrow \gamma \gamma)$ relative to the SM is fixed.  Right: The distribution of $(v^2/\Lambda^2) C^{(6)}_{HW}$
for the same parameter sets.  The two calculations estimate the difference between the calculations at
${\cal{O}}(v^4/\Lambda^4)$ and ${\cal{O}}(v^2/\Lambda^2)$, and we consider their applicability as truncation
uncertainties. }
\label{fig:truncunc}
\end{figure*}

Figure~\ref{fig:truncerr} shows the distribution of this error for $C^{(6)}_{HW}$ in the ${\cal{O}}(v^2/\Lambda^2)$ (left)
and partial-square (right) calculations of $\Gamma(h \rightarrow \gamma \gamma)$ using 50,000 samplings of the coefficients
and taking $\Lambda=2.5$~TeV.  This error distribution can be compared to the distribution of uncertainty estimates shown in
Fig.~\ref{fig:truncunc}, where the distribution in the left panel is the difference between the ${\cal{O}}(v^2/\Lambda^2)$
and partial-square calculations, and in the right panel it is $v^2/\Lambda^2$ times the coefficient.  The uncertainty
estimate is 1-2 orders of magnitude smaller than the error, with the $v^2/\Lambda^2$ distribution narrower by a factor of
a few.

\begin{figure*}[tp!]
\centering
\includegraphics[width=0.49\textwidth]{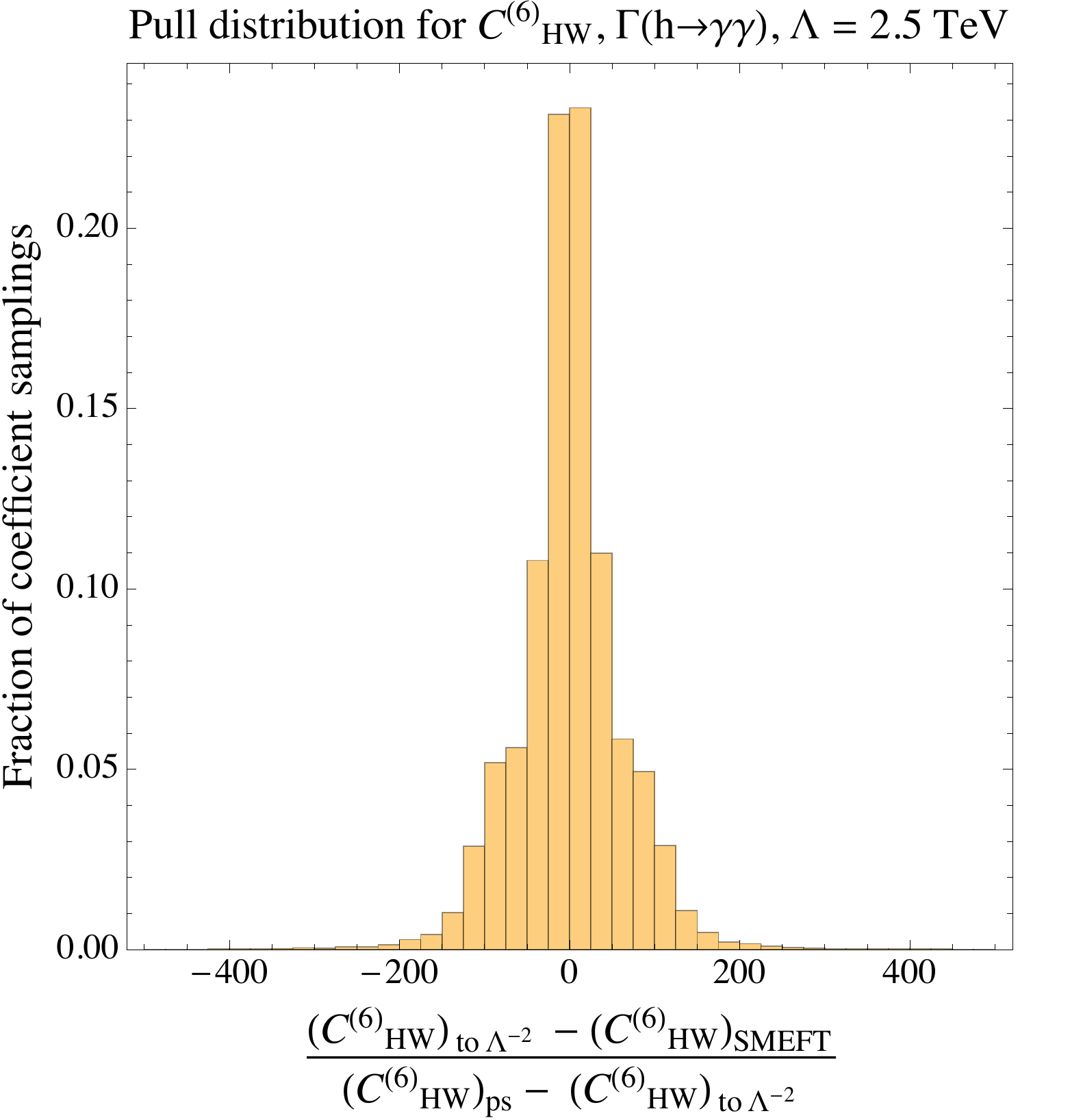}
\includegraphics[width=0.49\textwidth]{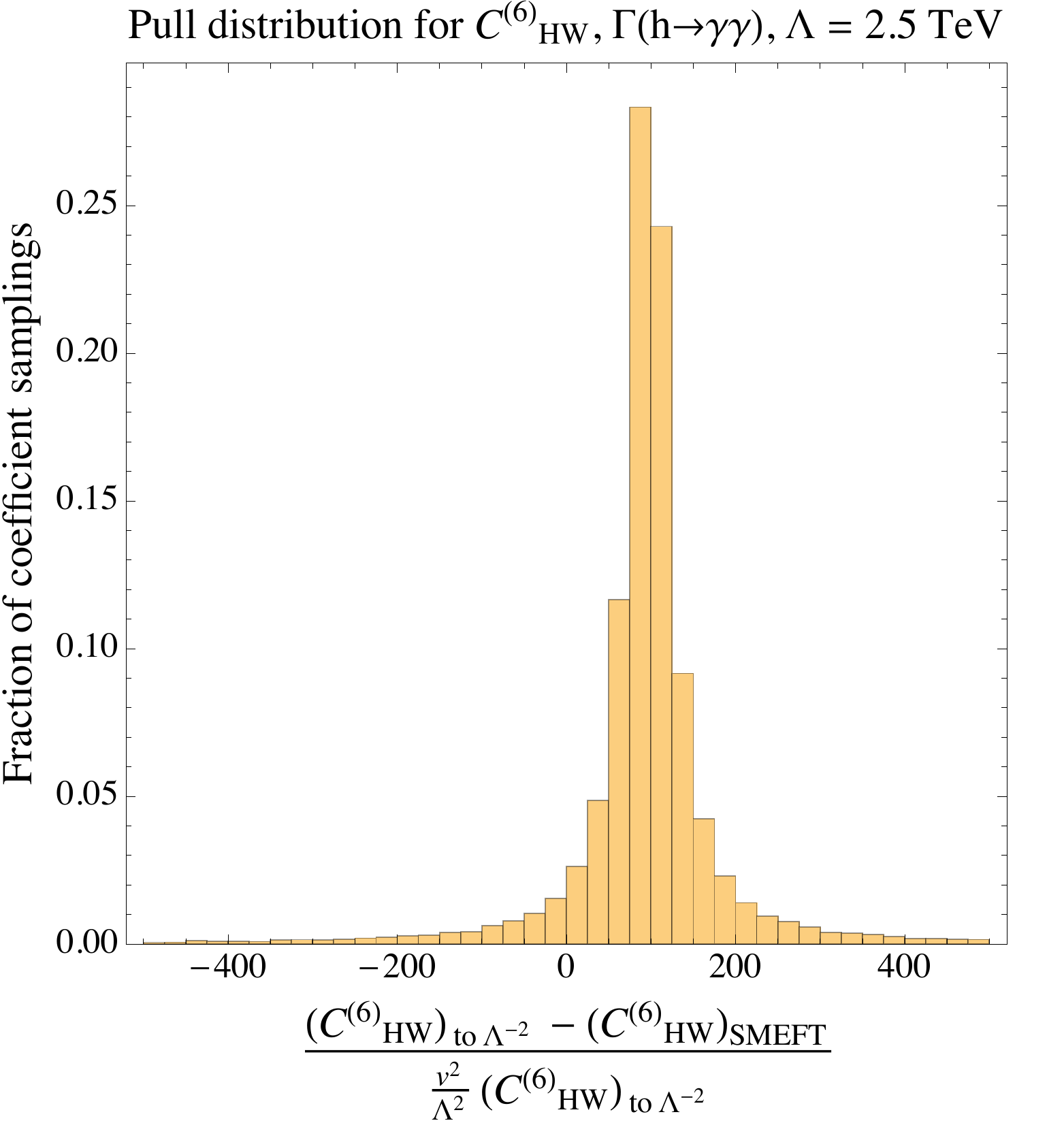}
\includegraphics[width=0.49\textwidth]{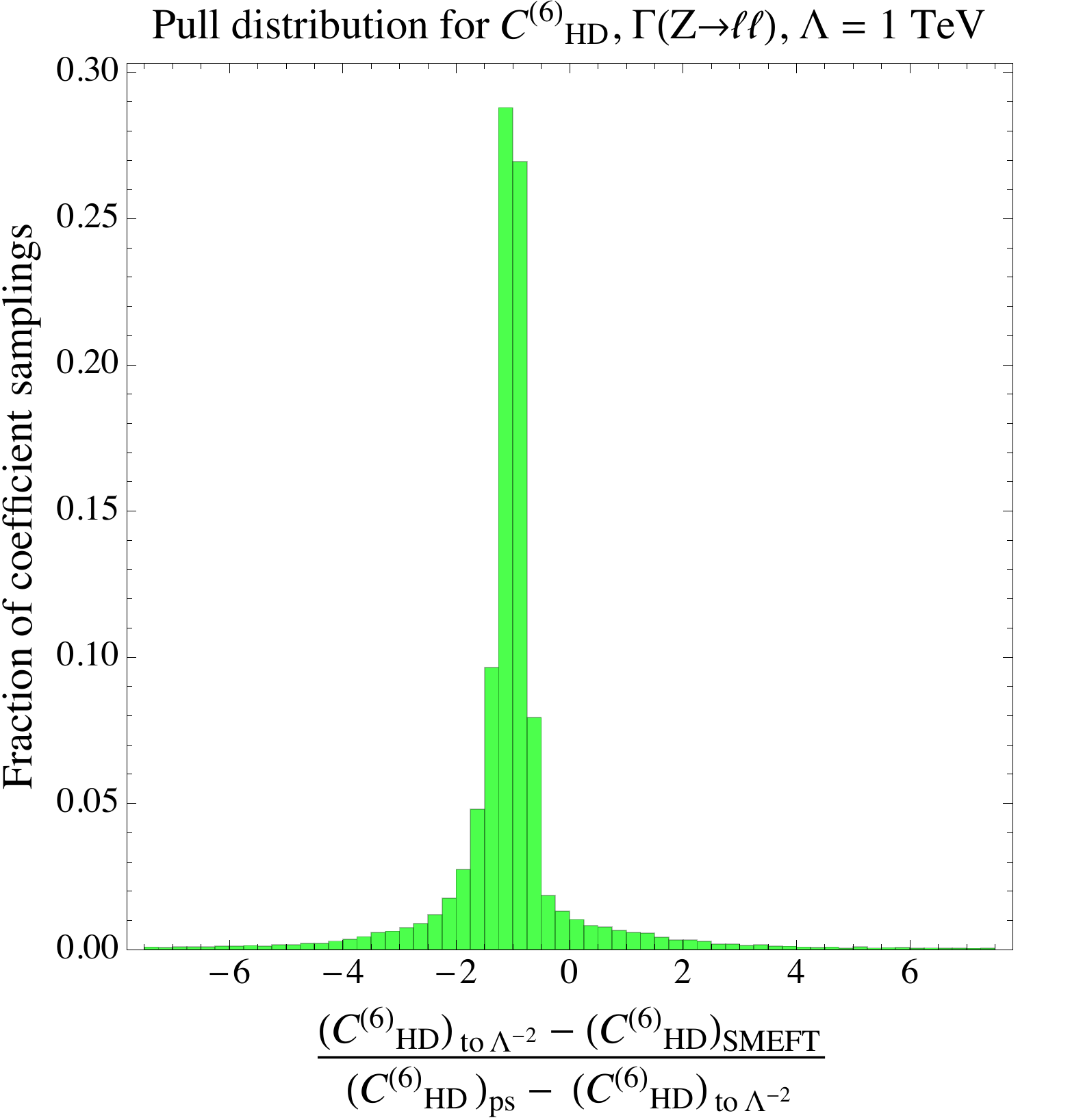}
\includegraphics[width=0.49\textwidth]{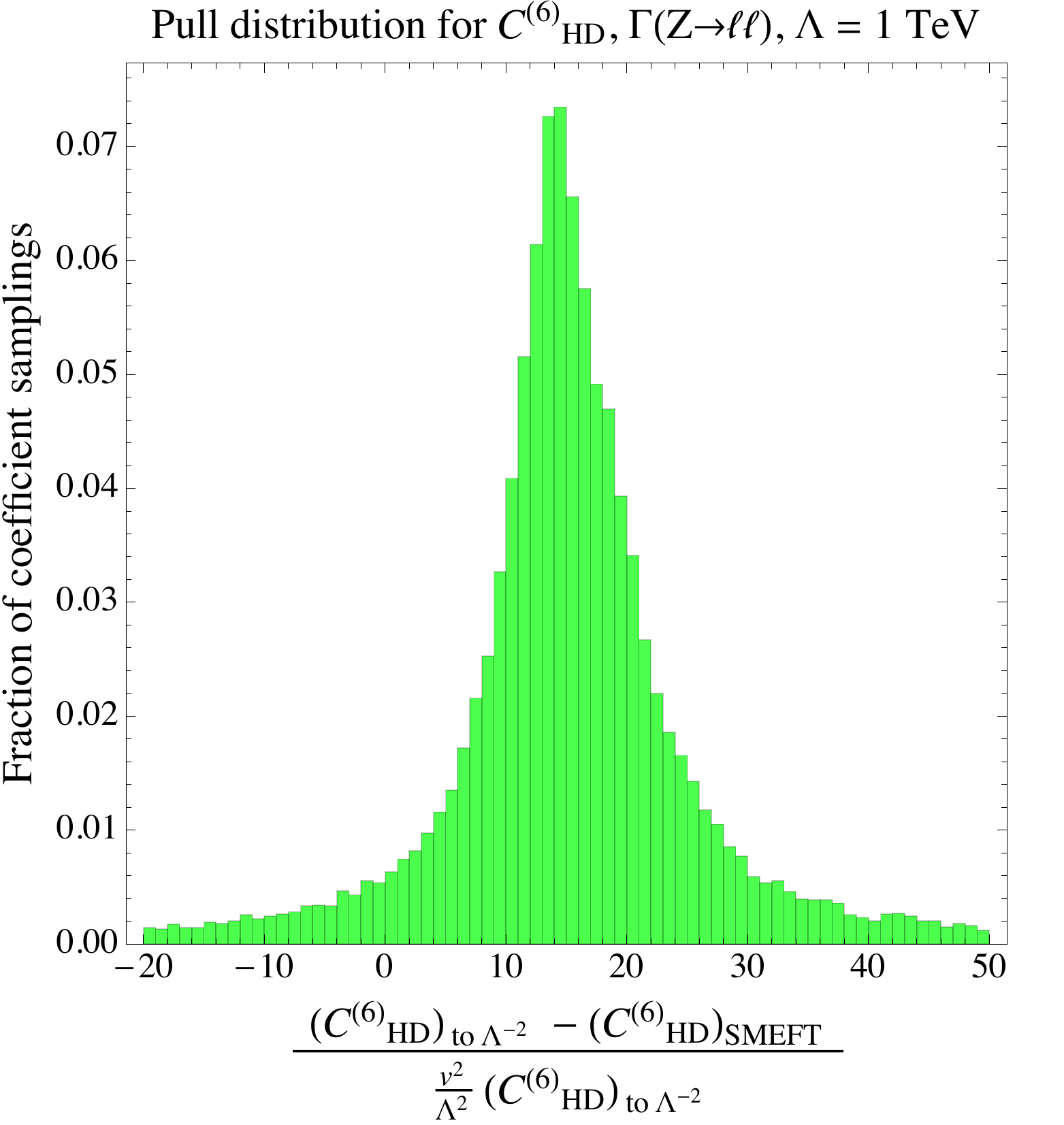}
\caption{The pull distributions for coefficients affecting $\Gamma(h \rightarrow \gamma \gamma)$ (top) and
$\Gamma(Z \to \ell\ell)$ (bottom), using as the uncertainty the difference between the partial-square and
${\cal{O}}(v^2/\Lambda^2)$ calculations (left) or $(v^2/\Lambda^2) C^{(6)}_i$ (right).  }
\label{fig:pulls}
\end{figure*}

The validity of an uncertainty estimate is typically demonstrated by the pull distribution, defined as the error divided
by the uncertainty.  An unbiased estimate of the central value and uncertainty would have a pull distribution with a mean
of zero and a standard deviation of one.  Figure~\ref{fig:pulls} shows this distribution for $\Gamma(h \rightarrow \gamma \gamma)$
(top) and $\Gamma(Z \to \ell\ell)$ (bottom) for the two estimates of the uncertainty using the ${\cal{O}}(v^2/\Lambda^2)$
calculation for the central value of the coefficient.  The least biased estimate of the uncertainty comes from the
partial-square calculation, and has an ${\cal{O}}(1)$ width when applied to the tree-level $\Gamma(Z \to \ell\ell)$ process.
An uncertainty of $10 (v^2/\Lambda^2) C^{(6)}_i$ could give a reasonable estimate, as it would scale down the entire
$x$-axis by a factor of 10.  Such an uncertainty would imply that a scale of $\gtrsim 1$~TeV would be required to reduce
the truncation uncertainty to $\lesssim 100\%$.

We do not address here the case where measurements do not have sensitivity to the true values of the coefficients, which
are thus consistent with zero within experimental uncertainties.  In this situation the procedures discussed here would be
dominated by the noise in the measurement and would not provide an accurate estimate of the uncertainty.  When using these
results to constrain specific models, a truncation uncertainty based on the measurement uncertainty may be sufficient, e.g.
$a (v^2/\Lambda^2) \sigma_{C^{(6)}_i}$ with $a$ of order 1.

\section{Model example: Kinetic mixing of gauge bosons}\label{sec:UVmodel}

In the previous section we investigated the numerical differences between the various calculations
using coefficient sampling.  Here we examine the differences that arise when using experimental results
to infer UV model parameters.  To this end, we explore a simple two-parameter model where
a heavy 
U(1) gauge boson $K_\mu$ with mass $m_K$ kinetically mixes with $B_\mu$,
the 
U(1)$_{\rm Y}$ gauge boson in the SM.

\subsection{Matching to $\mathcal{L}^{(8)}$}
We follow and extend the treatment of this model in Ref.~\cite{Henning:2014wua}, where the SM Lagrangian
is supplemented with the UV Lagrangian
\begin{align}
	\Delta \Lagr = -\frac{1}{4} K_{\mu\nu} K^{\mu\nu} + \frac{1}{2}m_K^2 K_\mu K^\mu
	- \frac{k}{2} B^{\mu\nu}K_{\mu\nu},
\end{align}
where the field strength is $K_{\mu\nu} = \partial_\mu K_\nu - \partial_\nu K_\mu$.
Integrating out the heavy $K$ state, a particular matching pattern results in the SMEFT.
The equation of motion for $K_\mu$ is
\begin{align}
	\partial_\nu K^{\mu\nu} - m_K^2 K^\mu + k \left(\partial_\nu B^{\mu\nu}\right) = 0,
\end{align}
which can be split into two equations:
\begin{align}
	\partial_\mu K^\mu &= 0,  \\
	\left(\partial^2 + m_K^2\right) K^\mu &= k\left(\partial_\nu B^{\mu\nu}\right).
\end{align}
To find the tree-level matching, it is sufficient to insert the solution for the equation of motion for
$K^\mu$ back into the Lagrangian. The classical solution is
\begin{align}
	K_{\it cl}^\mu = \frac{k}{\partial^2+m_K^2} \left(\partial_\nu B^{\mu\nu}\right)
	= \frac{k}{m_K^2}\left(\partial_\nu B^{\mu\nu}\right) - \frac{k}{m_K^4}\left(\partial^2 \partial_\nu B^{\mu\nu}\right) + \dots
\end{align}
Plugging this solution back into the Lagrangian, we find \cite{Henning:2014wua}
\begin{align}
	\Delta \Lagr =& \frac{1}{2} K_{cl,\mu}\left[  \partial^2 + m_K^2 \right]K_{cl}^\mu
	- \frac{1}{2}K_{cl}^\mu\partial_{\mu}\partial_\nu K_{cl}^\nu
	-\frac{k}{2} B_{\mu\nu}  K_{cl}^{\mu\nu} \nonumber \\
	=& - \frac{k}{2} \left( \partial_\nu B^{\mu\nu}\right) K_{cl,\mu}\nonumber \\
	=& - \frac{k^2}{2m_K^2} \left( \partial_\nu B^{\mu\nu}\right)\left(\partial^\alpha B_{\mu\alpha}\right)
	 + \frac{k^2}{2m_K^4} \left( \partial_\nu B^{\mu\nu}\right)\left(\partial^2\partial^\alpha B_{\mu\alpha}\right).
\end{align}
The induced operators are reducible by the equations of motion.
The relevant terms in the Lagrangian are
\begin{align}
	\Lagr_B = -\frac{1}{4} B_{\mu\nu} B^{\mu\nu} + \sum_\psi \bar{\psi}\gamma^\mu iD_\mu \psi
	+ \left(D_\mu H\right)^\dagger \left(D^\mu H\right) + \Delta \Lagr,
\end{align}
with $\psi = \{q,\ell,u,d,e \}$ possessing hypercharges $\hyp_\psi = \{1/6,-1/2,2/3,-1/3,-1\}$. By redefining the field,
\begin{align}
	B_\mu \rightarrow &B_\mu + \frac{k^2}{2m_K^2} \left[ \left(\partial^\nu B_{\nu\mu}\right)
		- j_\mu \right] \nonumber \\
		&+ \frac{1}{m_K^4}\left[
		\left(-\frac{k^2}{2} + \frac{3k^4}{8}\right) \left(\partial^2 \partial^\nu B_{\nu\mu}\right)
		- \frac{k^4}{4}\left(\frac{1}{2}g_1 \right)^2 (H^\dagger H) \left(\partial^\nu B_{\nu\mu}\right)
	\right] \nonumber \\
	& + \frac{1}{m_K^4}\left[
	\left(\frac{k^2}{2} - \frac{5k^4}{8}\right)\left(\partial^2 j_\mu\right)
	+ \frac{3k^4}{4}\left(\frac{1}{2}g_1 \right)^2(H^\dagger H) j_\mu
\right],
\end{align}
the Lagrangian becomes
\begin{align}\label{subleadingmatching}
	\Lagr_B =& -\frac{1}{4} B_{\mu\nu} B^{\mu\nu} + \sum_\psi \bar{\psi} \gamma^\mu i D_\mu \psi
	+ \left(D_\mu H\right)^\dagger \left(D^\mu H\right) \nonumber \\
	&- \frac{k^2}{2m_K^2} j_\mu j^\mu
	+ \frac{k^2-k^4}{2m_K^4}\left(\partial^2 j_\mu\right) j^\mu
	+ \frac{g_1^2 k^4}{4m_K^4} (H^\dagger H) j_\mu j^\mu,
\end{align}
where\footnote{We use a positive sign convention in the covariant derivative.}
\begin{align}
	j_\mu = \sum_\psi \left(-g_1 \hyp_\psi\right) \bar{\psi} \gamma_\mu \psi
	+ \left(-\frac{1}{2}g_1  \right) H^\dagger i \overset{\leftrightarrow}{D}_\mu H.
\end{align}
Up to $\mathcal{L}^{(8)}$ it is sufficient to use the marginal equations of motion to simplify the matching
to an operator basis consistent with the geoSMEFT formulation \cite{Helset:2020yio}
(and the Hilbert series~\cite{Lehman:2015via,Lehman:2015coa,Henning:2015daa,Henning:2015alf}).\\
\begin{table}
  \begin{center}
	  \caption{$\mathcal{L}^{(6)}$ matching coefficients; here $b_1 = k^2 - 2\lambda \, (k^2-k^4) \, \frac{\bar{v}_T^2}{m_K^2}$.
		Flavour indicies are suppressed and the heavy field does not violate $U(3)^5$ flavour symmetry. Fierz rearrangements of
		the four-fermion operators are allowed.}
    \label{tab:dim6Matching}
\begin{tabular}{l|c}
      \multicolumn{2}{c}{$H^2\psi^2 D$}   \\
      \toprule
      $C_{H\ell}^{1,(6)}$ & $-\frac{\hyp_\ell g_1^2}{2m_K^2}b_1$ \\
      \midrule
      $C_{He}^{(6)}$ & $-\frac{\hyp_e g_1^2}{2m_K^2}b_1$ \\
      \midrule
      $C_{Hq}^{1,(6)}$ & $-\frac{\hyp_q g_1^2}{2m_K^2}b_1$ \\
      \midrule
      $C_{Hu}^{(6)}$ & $-\frac{\hyp_u g_1^2}{2m_K^2}b_1$ \\
      \midrule
      $C_{Hd}^{(6)}$ & $-\frac{\hyp_d g_1^2}{2m_K^2}b_1$ \\
      \bottomrule
    \end{tabular}
    \quad
	  \begin{tabular}{l|c}
      \multicolumn{2}{c}{$H^4 D^2$}   \\
      \toprule
      $C_{H\Box}^{(6)}$ & $-\frac{g_1^2 k^2}{8m_K^2}$ \\
      \midrule
      $C_{HD}^{(6)}$ &  $-\frac{g_1^2 k^2}{2m_K^2}$ \\
      \bottomrule
    \multicolumn{2}{c}{}   \\
    \multicolumn{2}{c}{$\psi^4: (\bar{L}L)(\bar{L}L)$}   \\
      \toprule
      $C_{\ell\ell}^{(6)}$ & $-\frac{1}{8}\frac{g_1^2 k^2}{m_K^2}$ \\
      \midrule
      $C_{qq}^{1,(6)}$ & $-\frac{1}{72}\frac{g_1^2 k^2}{m_K^2}$ \\
      \midrule
      $C_{\ell q}^{1,(6)}$ &  $\frac{1}{12}\frac{g_1^2 k^2}{m_K^2}$ \\
      \bottomrule
    \end{tabular}
   	\quad
    \begin{tabular}{l|c}
      \multicolumn{2}{c}{$ \psi^4: (\bar{R}R)(\bar{R}R)$}   \\
      \toprule
      $C_{ee}^{(6)}$ & $-\frac{1}{2}\frac{g_1^2 k^2}{m_K^2}$ \\
      \midrule
      $C_{uu}^{(6)}$ &  $-\frac{2}{9}\frac{g_1^2 k^2}{m_K^2}$ \\
          \midrule
      $C_{dd}^{(6)}$ &  $-\frac{1}{18}\frac{g_1^2 k^2}{m_K^2}$ \\
      \midrule
      $C_{eu}^{(6)}$ &  $\frac{2}{3}\frac{g_1^2 k^2}{m_K^2}$ \\
      \midrule
      $C_{ed}^{(6)}$ &  $-\frac{1}{3}\frac{g_1^2 k^2}{m_K^2}$ \\
      \midrule
      $C_{ud}^{1,(6)}$ &  $\frac{2}{9}\frac{g_1^2 k^2}{m_K^2}$ \\
      \bottomrule
    \end{tabular}
\quad
    \begin{tabular}{l|c}
      \multicolumn{2}{c}{$ \psi^4: (\bar{L}L)(\bar{R}R)$}   \\
      \toprule
      $C_{\ell e}^{(6)}$ & $-\frac{1}{2}\frac{g_1^2 k^2}{m_K^2}$ \\
      \midrule
      $C_{\ell u}^{(6)}$ &  $\frac{1}{3}\frac{g_1^2 k^2}{m_K^2}$ \\
          \midrule
      $C_{\ell d}^{(6)}$ &  $-\frac{1}{6}\frac{g_1^2 k^2}{m_K^2}$ \\
      \midrule
      $C_{qe}^{(6)}$ &  $\frac{1}{6}\frac{g_1^2 k^2}{m_K^2}$ \\
      \midrule
      $C_{qu}^{1,(6)}$ &  $-\frac{1}{9}\frac{g_1^2 k^2}{m_K^2}$ \\
      \midrule
      $C_{qd}^{1,(6)}$ &  $\frac{1}{18}\frac{g_1^2 k^2}{m_K^2}$ \\
      \bottomrule
    \end{tabular}
  \end{center}
\end{table}
\\
The $\mathcal{L}^{(6)}$ matching is given in Table~\ref{tab:dim6Matching}.
The reduction of the derivative terms in the current at $\mathcal{L}^{(8)}$ requires non-trivial manipulations.
These terms can be reduced into the form
\begin{align}
\label{reducederivativecurrents}
j^\mu \partial^2 j_\mu &\simeq
g_1^2 \,\left[(D_\mu H^\dagger)(D_\nu H)(D^\mu H^\dagger)(D^\nu H)
- (D_\mu H^\dagger)(D_\nu H)(D^\nu H^\dagger)(D^\mu H)\right] \nn
&+ g_1^2 \left[
g_1  (H^\dagger H)\, B_{\mu \nu} (D^\mu H^\dagger)\, i \, (D^\nu H)
-  g_2 \, (H^\dagger  H) \, (D^\mu H^\dagger)\, i \, \sigma_a  \, (D^\nu H) \, W^a_{\mu \nu}\right]\nn
&+ \frac{g_1^2 \, g_2^2}{8} W^a_{\mu \nu} W_a^{\mu \nu} (H^\dagger H)^2
- \frac{g_1^4}{8} B_{\mu \nu} B^{\mu \nu} (H^\dagger H)^2
+ g_1^3 \hyp_\psi (\bar{\psi}\gamma^\mu \psi) \, B_{\mu \nu}\, D^\nu \left( H^\dagger H \right) \nn
&- \frac{g_1^2 \,g_2^2}{8} (\bar{\psi}_L \sigma_a \gamma^\mu \psi_L)
\left[(H^\dagger H) \left(H^\dagger  i \overset{\leftrightarrow}{D^a_\mu} H \right)
+ (H^\dagger \sigma_a H) \left(H^\dagger  i \overset{\leftrightarrow}{D}_\mu H \right)  \right] \nn
&-\frac{g_1^2 \,g_2^2}{4} \, \left[4 \, (H^\dagger H)^2 (D_\mu \, H^\dagger D^\mu H) +
\lambda \, (H^\dagger H)^3 \, (\bar{v}_T^2 - 2 (H^\dagger H))+
 H^\dagger \sigma_a H \, (D_\mu \, H^\dagger \sigma^a D^\mu H) \right] \nn
&+  2 i g_1^2 \hyp_\psi (\bar{\psi}\gamma^\mu \psi)
\left[\lambda(\bar{v}_T^2 - 2 H^\dagger H) \, \left(H^\dagger  \overset{\leftrightarrow}{D}_\mu H \right)+
(D_\nu H^\dagger) (D^\nu D_\mu H) -(D^\nu D_\mu H^\dagger) (D_\nu H) \right] \nn
&- g_1^2  \hyp_\psi (\bar{\psi}\gamma^\mu \psi) \, \left( H^\dagger H \right)
\left(\frac{1}{2}\left(g_1^2+ g_2^2\right)  \left(H^\dagger i \overset{\leftrightarrow}{D}_\mu H \right)
+ g_1^2 \,   \hyp_{\psi'} (\bar{\psi'}\gamma^\mu \psi') \right) \nn
&+ g_1^2  \hyp_\psi (\bar{\psi}\gamma^\mu \psi) \,
\left[- \frac{g_2^2}{2} \,  \bar{\psi}_L \gamma_\mu \sigma^a \psi_L \, (H^\dagger \sigma_a H)
+ g_2 \, W^a_{\mu \nu}\, D^\nu \left( H^\dagger \sigma_a H \right) \right],
\end{align}
where a sum is implied over all $\psi_L$, $\psi$, and $\psi'$ pairs, and terms proportional to Yukawa
couplings are neglected.  The conventions used for reducing to the operator basis in the $\mathcal{L}^{(8)}$
matching are those of the geoSMEFT formulation \cite{Helset:2020yio}, which allows all-orders results in
the $\bar{v}_T/\Lambda$ expansion to be defined.  In this convention derivatives have been moved
onto scalar fields and off of fermion fields.  A useful identity in deriving this result is
\bea
(H^\dagger \sigma_a H)(D^\mu H^\dagger) \, (D^\nu H) i W^a_{\mu \nu}
&=&- (H^\dagger H) (D^\mu H^\dagger) \,  \sigma_a (D^\nu H) i W^a_{\mu \nu}\\
&-&\frac{1}{4} D_\mu W^a_{\mu \nu} \left[ H^\dagger H \left(H^\dagger i \overset{\leftrightarrow}{D_\mu^a} H \right)
+ H^\dagger \sigma^a H  \left(H^\dagger i \overset{\leftrightarrow}{D}_\mu H \right)\right]\nn
&+& \frac{g_2}{8}\left[ (H^\dagger H)^2 W^a_{\mu \nu}  W^a_{\mu \nu}+
 (H^\dagger \sigma^a H)(H^\dagger \sigma^b H) W_a^{\mu \nu}  W_b^{\mu \nu}  \right] \nn
&+& \frac{g_1}{4} (H^\dagger \sigma^a H) (H^\dagger H) \, W_a^{\mu \nu} \, B_{\mu \nu}. \nonumber
\eea
The matching at $\mathcal{L}^{(8)}$ illustrates a number of interesting features:
\begin{itemize}
\item{At $\mathcal{L}^{(6)}$, the matching results are only dependent on the model parameters and the SM gauge coupling $g_1$.
This is consistent with naive expectations in a 
U(1) kinetic mixing model. At $\mathcal{L}^{(8)}$,
the result in Eqn.~\eqref{subleadingmatching} is expressed in terms of derivatives and 
U(1) currents.
Dependence on $g_2$ is introduced in the rearrangement of higher-derivative terms,
as required to be consistent with the geoSMEFT conventions. This coupling dependence comes about via
commutators of derivatives acting on the Higgs field.
Further dependence on $g_2$, and more $\mathcal{L}^{(8)}$ terms, are introduced through mapping the
SM gauge coupling $g_1$, present in the $\mathcal{L}^{(6)}$ matching, to input measurements, including SMEFT corrections.
As a result, the input-parameter scheme dependence is enhanced at ${\cal{O}}(v^4/\Lambda^4)$ in the SMEFT.}
\item{A naive interpretation of UV physics acting as a mediator leading to a $\mathcal{L}^{(6)}$ operator at tree level is frequently possible
	by inspection.
For example, the tree-level exchange of an $\rm SU(2)_L$ triplet field or singlet field leads to
\bea
Q_{H\psi}^{(3)}&=& \left(H^\dagger i \overset{\leftrightarrow}{D^a_\mu} H \right) \, \,  \bar{\psi} \, \gamma_\mu \,  \sigma^a \, \psi,\\
Q_{H\psi}&=& \left(H^\dagger i \overset{\leftrightarrow}{D}_\mu H \right) \, \, \bar{\psi} \gamma_\mu \psi,
\eea
respectively at $\mathcal{L}^{(6)}$. Such naive intuition fails at $\mathcal{L}^{(8)}$ and beyond. Specifically, at  $\mathcal{L}^{(8)}$ operators can be reduced due to the $\rm SU(2)_L$ completeness relations acting on the scalar
coordinates as
\bea
\bar{\psi} \gamma_\mu \psi \, H^\dagger \sigma^a H \, \left(H^\dagger i \overset{\leftrightarrow}{D^a}_\mu H \right)
\rightarrow \bar{\psi} \gamma_\mu \psi \, H^\dagger H \, \left(H^\dagger i \overset{\leftrightarrow}{D}_\mu H \right).
\eea
This rearrangement is present in the SMEFT at $\mathcal{L}^{(8)}$ when using a non-redundant operator basis.
Such simplifications lead to
Eq.~\eqref{reducederivativecurrents} in part. This reduces the
transparency of the underlying UV field content and the interactions leading to
tree-level matchings to higher-dimensional operators.
}
\begin{table}[ht!]
  \begin{center}
	  \caption{Matching coefficients onto operators in $\mathcal{L}^{(8)}$ relevant for
		$\Gamma(h \rightarrow \gamma \gamma)$ and $\Gamma(\mathcal{Z} \rightarrow \bar{\psi}\psi)$.
		In addition to these matching contributions, there are four-fermion operators and four-point contributions.
		See the results in Eqn.~\ref{reducederivativecurrents}, which include these terms and neglect only
                effects suppressed by Yukawa couplings. }
    \label{tab:dim8Matching}
\begin{tabular}{l|c}
      \multicolumn{2}{c}{$H^4\psi^2 D$}   \\
      \toprule
      $C_{H\ell}^{1,(8)}$ & $\frac{\hyp_\ell g_1^4}{4 \, m_K^4} \, k^4
			-\frac{g_1^2 \, \hyp_\ell}{m_K^4} (k^2-k^4) (2 \lambda + \frac{g_1^2+ g_2^2}{4}) $ \\
      \midrule
      $C_{He}^{1,(8)}$ & $\frac{\hyp_e g_1^4}{4 \, m_K^4} \, k^4
			-\frac{g_1^2 \, \hyp_e}{m_K^4}  (k^2-k^4) (2 \lambda + \frac{g_1^2+ g_2^2}{4}) $ \\
      \midrule
      $C_{Hq}^{1,(8)}$ & $\frac{\hyp_q g_1^4}{4 \, m_K^4} \, k^4
			-\frac{g_1^2 \, \hyp_q}{m_K^4} (k^2-k^4) (2 \lambda + \frac{g_1^2+ g_2^2}{4}) $ \\
      \midrule
      $C_{Hu}^{1,(8)}$ & 	$\frac{\hyp_u g_1^4}{4 \, m_K^4} \, k^4
			-\frac{g_1^2 \, \hyp_u}{m_K^4}  (k^2-k^4) (2 \lambda + \frac{g_1^2+ g_2^2}{4}) $ \\
      \midrule
      $C_{Hd}^{1,(8)}$ & $\frac{\hyp_d g_1^4}{4 \, m_K^4} \, k^4
			-\frac{g_1^2 \, \hyp_d}{m_K^4}  (k^2-k^4) (2 \lambda + \frac{g_1^2+ g_2^2}{4}) $ \\
			\midrule
      $C_{H\ell}^{2,(8)}$ & $-\frac{g_1^2 \, g_2^2}{16 \, m_K^4} (k^2-k^4) $ \\
			\midrule
      $C_{Hq}^{2,(8)}$ & $-\frac{g_1^2 \, g_2^2}{16 \, m_K^4} (k^2-k^4) $ \\
			\midrule
      $C_{H\ell}^{3,(8)}$ & $-\frac{g_1^2 \, g_2^2}{16 \, m_K^4} (k^2-k^4) $ \\
			\midrule
      $C_{Hq}^{3,(8)}$ & $-\frac{g_1^2 \, g_2^2}{16 \, m_K^4} (k^2-k^4) $ \\
      \bottomrule
    \end{tabular}
    \quad
	  \begin{tabular}{l|c}
      \multicolumn{2}{c}{$H^6 D^2$}   \\
      \toprule
      $C_{H,D2}^{(8)}$ & $\frac{g_1^4 \, k^4}{8 \, m_K^4} - \frac{g_1^2 \, g_2^2}{2 \, m_K^4} (k^2-k^4)$ \\
      \midrule
      $C_{HD}^{(8)}$ &  $\frac{3 \, g_1^4 \, k^4}{16 \, m_K^4} - \frac{g_1^2 \, g_2^2}{2 \, m_K^4} (k^2-k^4)$ \\
      \bottomrule
    \multicolumn{2}{c}{}   \\
    \multicolumn{2}{c}{$X^2 H^4$}   \\
      \toprule
        $C_{HB}^{(8)}$ & $- \frac{g_1^4}{16 \, m_K^4}(k^2-k^4)$ \\
				\midrule
				$C_{HW}^{(8)}$ & $ \frac{g_1^2 \,g_2^2}{16 \, m_K^4}(k^2-k^4)$ \\
      \bottomrule
    \end{tabular}
   	\end{center}
\end{table}
\item{At $\mathcal{L}^{(6)}$, there can be patterns that classify Wilson coefficients as tree-level or loop-level~\cite{Arzt:1994gp},
with the latter in particular applying to coefficients of operators with gauge field strengths.  This is an accidental pattern due to the
renormalizability of some UV physics models.  Such matching patterns are not present in non-renormalizable UV theories in
general \cite{Jenkins:2013fya}.  They also do not apply to operators with higher mass dimensions.  The result in
Eqn.~\eqref{reducederivativecurrents} shows that gauge field-strength operators can receive tree-level matching
contributions at $\mathcal{L}^{(8)}$ in a weakly-coupled renormalizable UV model.  This is consistent with the results in
Ref.~\cite{Jenkins:2013fya,Craig:2019wmo}.  At $\mathcal{L}^{(7)}$, the seesaw model also leads to operators with gauge field
strengths~\cite{Elgaard-Clausen:2017xkq} in tree-level matching.  These examples show that the operator normalization
pattern of Ref.~\cite{Arzt:1994gp} does not extend to operators of arbitrary mass dimension in the SMEFT.
}
\item{The rearrangement of derivative terms at $\mathcal{L}^{(8)}$ leads to matching coefficients
proportional to $\bar{v}_T^2/m_K^2$ for $\mathcal{L}^{(6)}$. Formally, an infinite series in $(\bar{v}_T^2/m_K^2)^n$
is present in matching coefficients for higher-dimensional operators.  This is due to rearranging matching terms
in the non-redundant operator basis.  However, as this dependence is an artifact of this particular basis we expect it to cancel in
the full result. This occurs as expected.
}
\end{itemize}
Restricting the results to the subset of operators that contribute to
$\Gamma(h \rightarrow \gamma \gamma)$ and $\Gamma(\mathcal{Z} \rightarrow \bar{\psi}\psi)$,
the matching results for $\mathcal{L}^{(8)}$ operators are given in Table~\ref{tab:dim8Matching}.

\subsection{Constraints to ${\cal{O}}(\bar{v}_T^4/m_K^4)$}

The kinetic mixing model allows a comparison of the constraints on an underlying UV physics model at different
orders in the SMEFT expansion, and for the partial-square calculation.  Consider an experimental bound on the
deviation of $\Gamma(h \rightarrow \gamma \gamma)$ from the SM prediction.  Substituting the results of
Table~\ref{tab:dim6Matching} into the partial-square $\Gamma(h \rightarrow \gamma \gamma)$ formula
Eqn.~\eqref{eq:hgamgamPSmW} yields no constraint on the model parameters, at least
when considering tree-level matching.
However, using Eqn.~\eqref{eq:hgamgamSMEFT} we find the partial width to be sensitive to this model at
${\cal{O}}(\bar{v}_T^4/m_K^4)$:
\bea
\frac{\Gamma_{\rm SMEFT}^{[\hat{m}_{W},\hat{\alpha}_{ew}]}(h \rightarrow \gamma \gamma)}{ \Gamma^{[\hat{m}_{W},\hat{\alpha}_{ew}]}_{\rm SM}(h \rightarrow \gamma \gamma)}
= 1 - [49.3, 46.8]  \, \frac{(k^4-k^2) \, g_1^2 (g_1^2-0.29 g_2^2)\, \bar{v}_T^4}{m_K^4}.
\eea

\begin{figure}[t!]
\includegraphics[height=0.29\textheight,width=0.55\textwidth]{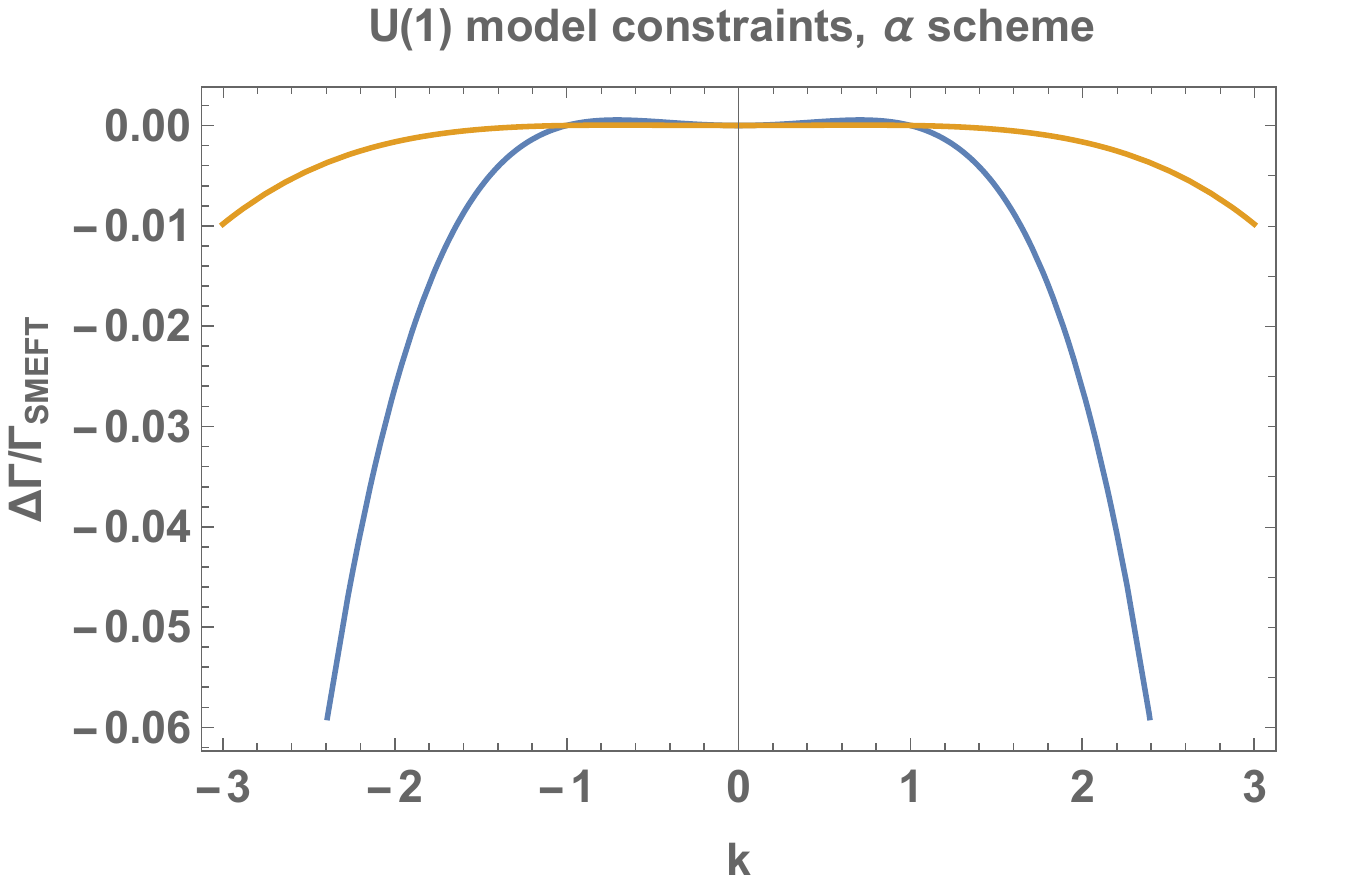}
\includegraphics[height=0.29\textheight,width=0.425\textwidth]{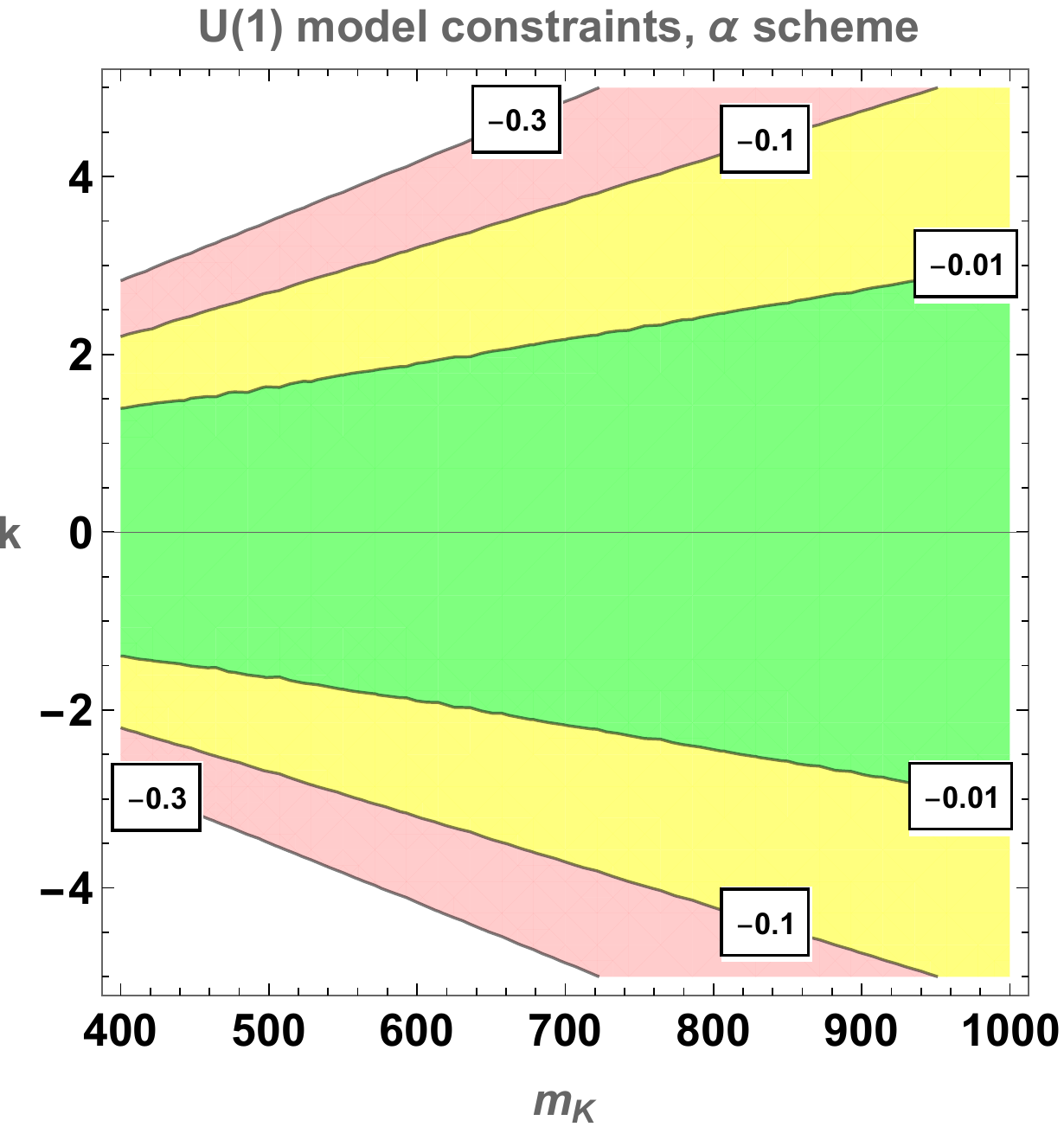}
\caption{The dependence of $\Gamma(h \rightarrow \gamma \gamma)$ on the parameters of a
U(1) mixing model using the SMEFT expansion to ${\cal{O}}(\bar{v}_T^4/m_K^4)$.
The left plot shows the dependence of the ratio
$\Delta \Gamma^{\hat{\alpha}_{ew}}_{\rm SMEFT}(h \rightarrow \gamma \gamma)/ \Gamma^{\hat{\alpha}_{ew}}_{\rm SM}(h \rightarrow \gamma \gamma)$
on the coupling parameter $k$ for $m_K=\{500,1000\} \, \, {\rm GeV}$, for the blue and orange curves respectively.  The right plot shows
the relative deviations $\{\pm 0.01, \pm 0.1, \pm 0.3\}$ of the partial width in the $\{m_K,k\}$ plane, with
intermediate deviations represented by coloured regions.  The results are shown in the $\hat\alpha_{ew}$
scheme, though results in the $\hat m_W$ scheme are qualitatively the same.  The partial width has no
sensitivity to the model at ${\cal{O}}(\bar{v}_T^2/m_K^2)$ with tree-level matching.  Direct experimental
bounds on $\Gamma(h \rightarrow \gamma \gamma)$ are not available since only ratios
of partial widths can be measured directly, e.g.
$\Gamma(h \rightarrow \gamma \gamma)/\Gamma(h \rightarrow 4 \ell)$~\cite{PhysRevD.101.012002}.\label{gamgaconstraintplotfig}} .
\end{figure}

\noindent
The width correction has $\approx 5\%$ scheme dependence, and we show its dependence on the model parameters
in Fig. \ref{gamgaconstraintplotfig} in the $\hat{\alpha}_{ew}$ scheme.  Direct bounds on
$\Gamma(h \rightarrow \gamma \gamma)$ are not available, since LHC cross sections
depend on the production interaction and the total Higgs width.  Ratios of partial widths are
available and could be applied as a constraint if a calculation of $h \rightarrow 4 \ell$ and/or $\Gamma_h$
were available in the SMEFT to this order.  A calculation has recently been performed to ${\cal{O}}(v^2/\Lambda^2)$
~\cite{Brivio:2019myy} and could be extended to ${\cal{O}}(v^4/\Lambda^4)$ with the geoSMEFT framework.
However, this is beyond the scope of this work.

Experimental constraints can be considered in the case of the total width of the $\mathcal{Z}$ boson, $\Gamma_Z = 2.4952 \pm 0.0023$~GeV.
Defining this quantity as the sum of the decay widths to each two-body final state, the partial-square calculations in the two input-parameter
schemes are
\bea
 \frac{\sum_{\psi} \bar{\Gamma}^{p.s., \hat{\alpha}_{ew}}_{\mathcal{Z} \rightarrow \bar{\psi}_p \psi_p}}{\sum_{\psi} \bar{\Gamma}^{{\rm SM}, \hat{\alpha}_{ew}}_{\mathcal{Z} \rightarrow \bar{\psi}_p \psi_p}}
&=& 1 + 4.5 \times 10^{-3} \, \frac{\bar{v}_T^2 \, k^2}{m_K^2} + 4.4 \times 10^{-3} \, k^4 \, \frac{\bar{v}_T^4}{m_K^4},
\eea
\bea
 \frac{\sum_{\psi} \bar{\Gamma}^{p.s., \hat{m}_{W}}_{\mathcal{Z} \rightarrow \bar{\psi}_p \psi_p}}{\sum_{\psi} \bar{\Gamma}^{{\rm SM},\hat{m}_{W}}_{\mathcal{Z} \rightarrow \bar{\psi}_p \psi_p}}
&=& 1 -3.1 \times 10^{-2} \, \frac{\bar{v}_T^2 \, k^2}{m_K^2} + 2.3 \times 10^{-4} \, k^4 \, \frac{\bar{v}_T^4}{m_K^4}.
\eea
The results show significant scheme dependence.  An interesting aspect of the scheme dependence
is the equivalence of the shifts in the partial widths in the $\hat{m}_W$ scheme, while the individual partial-width
corrections differ in the $\hat\alpha_{ew}$ scheme.

\begin{figure}[ht!]
\includegraphics[height=0.29\textheight,width=0.32\textwidth]{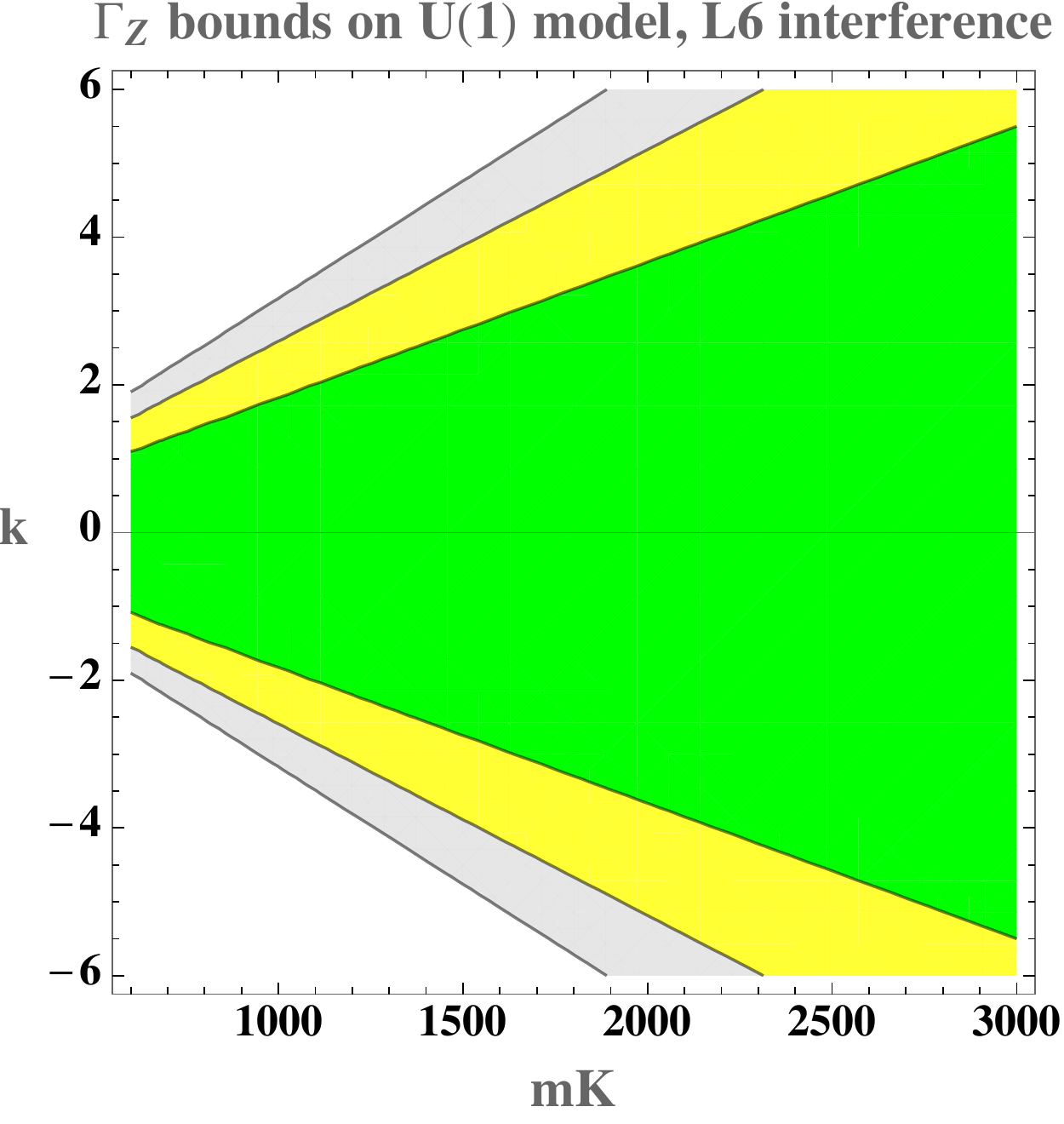}
\includegraphics[height=0.29\textheight,width=0.32\textwidth]{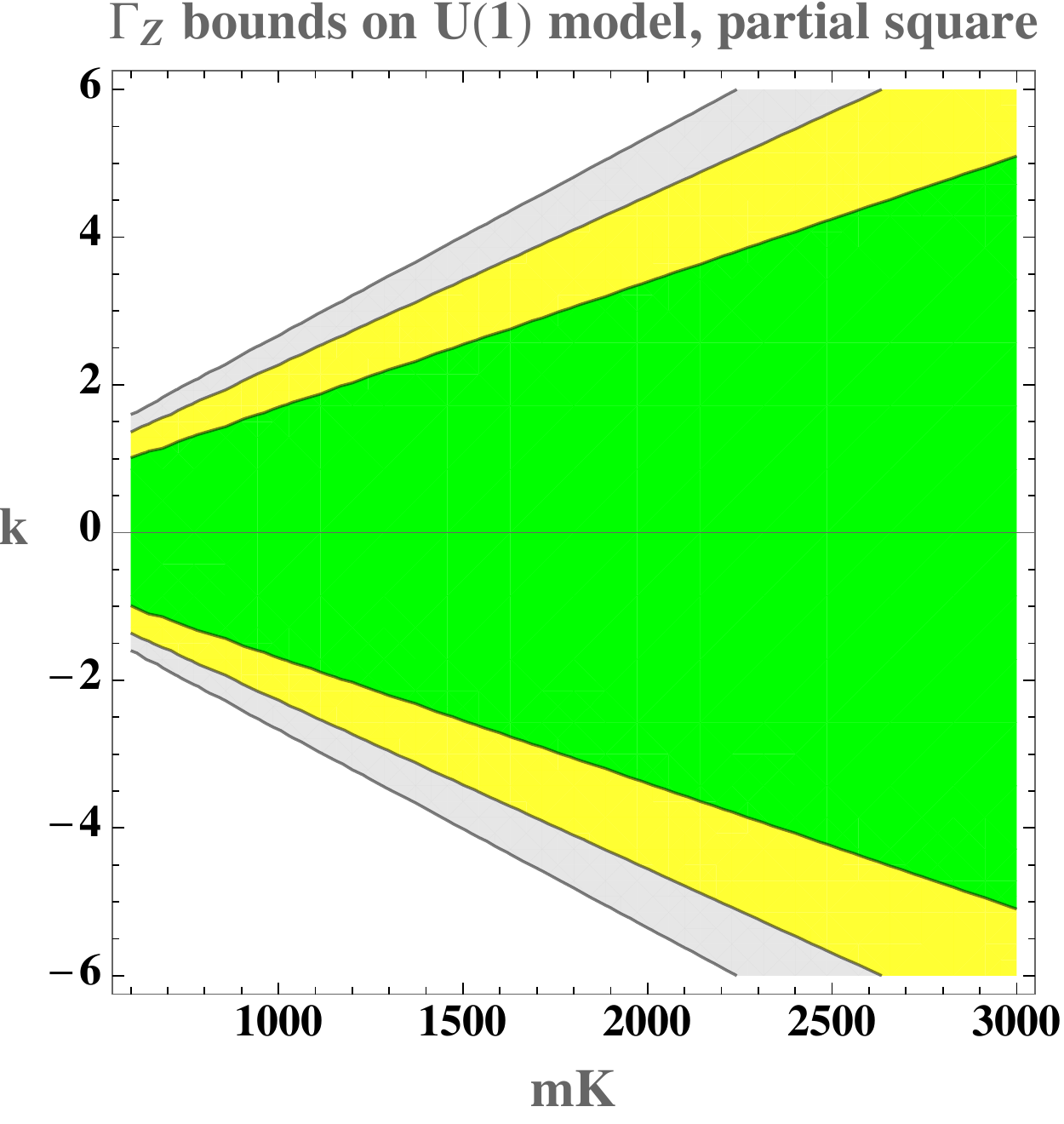}
\includegraphics[height=0.29\textheight,width=0.32\textwidth]{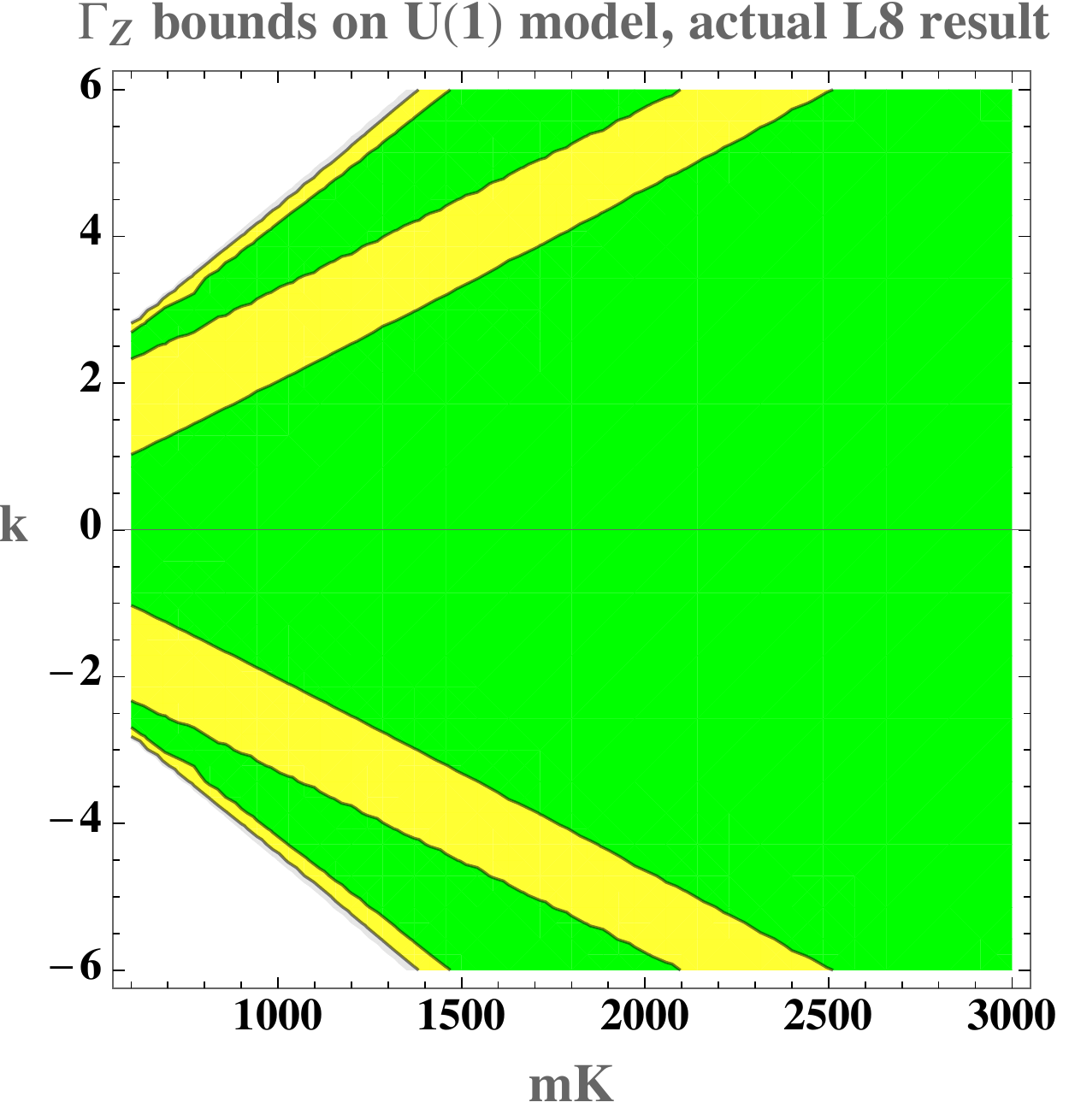}
\includegraphics[height=0.29\textheight,width=0.32\textwidth]{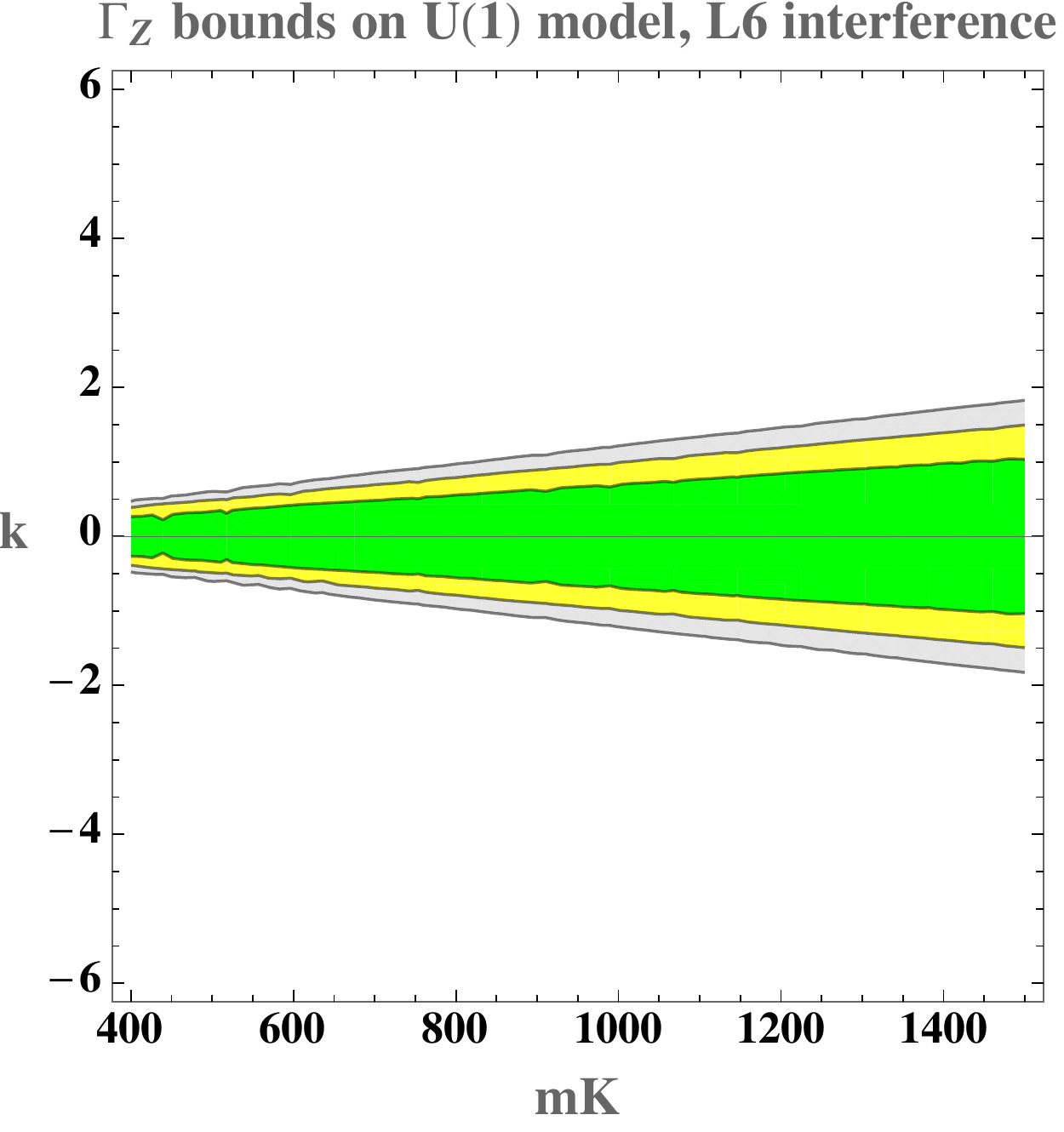}
\includegraphics[height=0.29\textheight,width=0.32\textwidth]{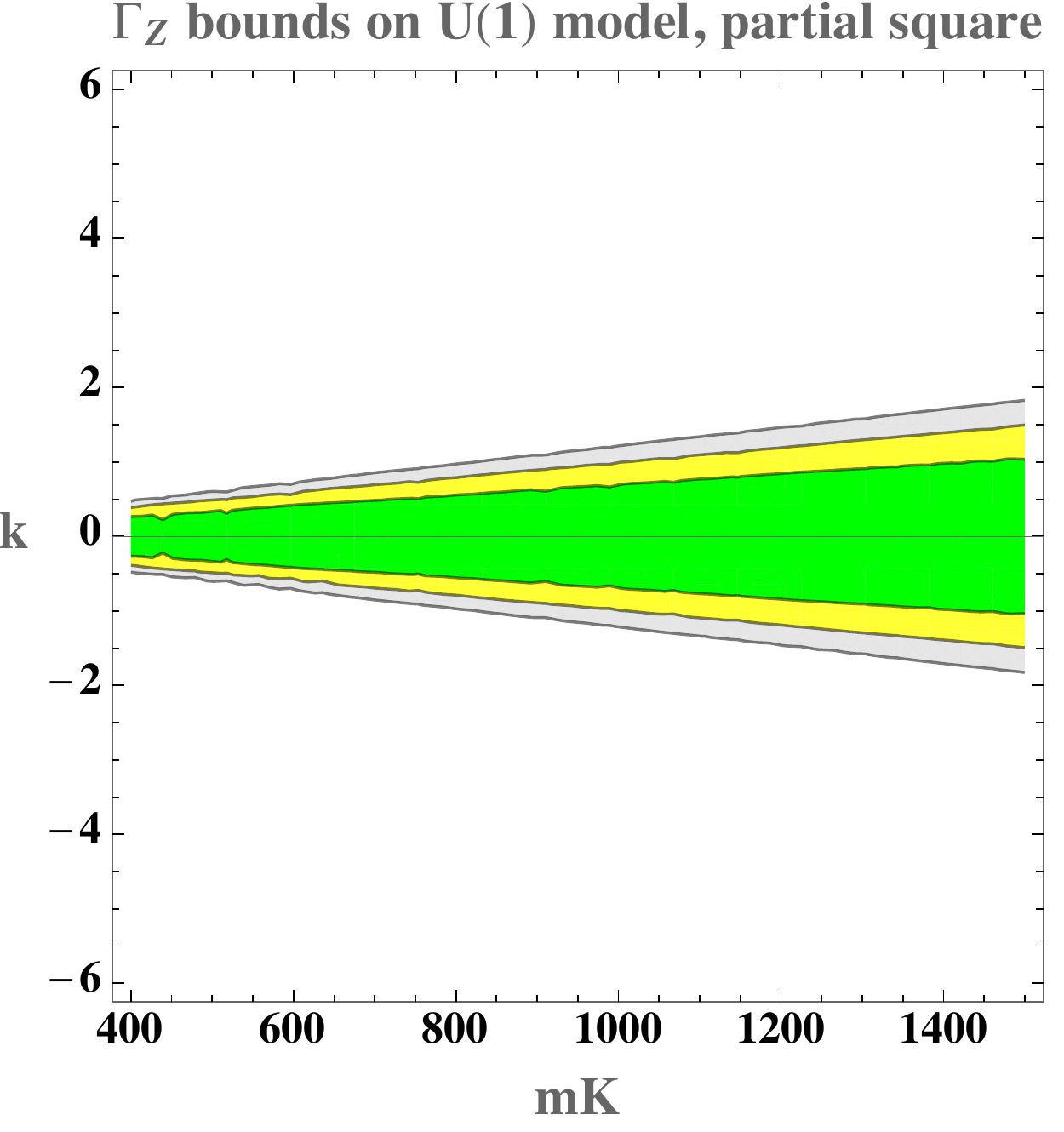}
\includegraphics[height=0.29\textheight,width=0.32\textwidth]{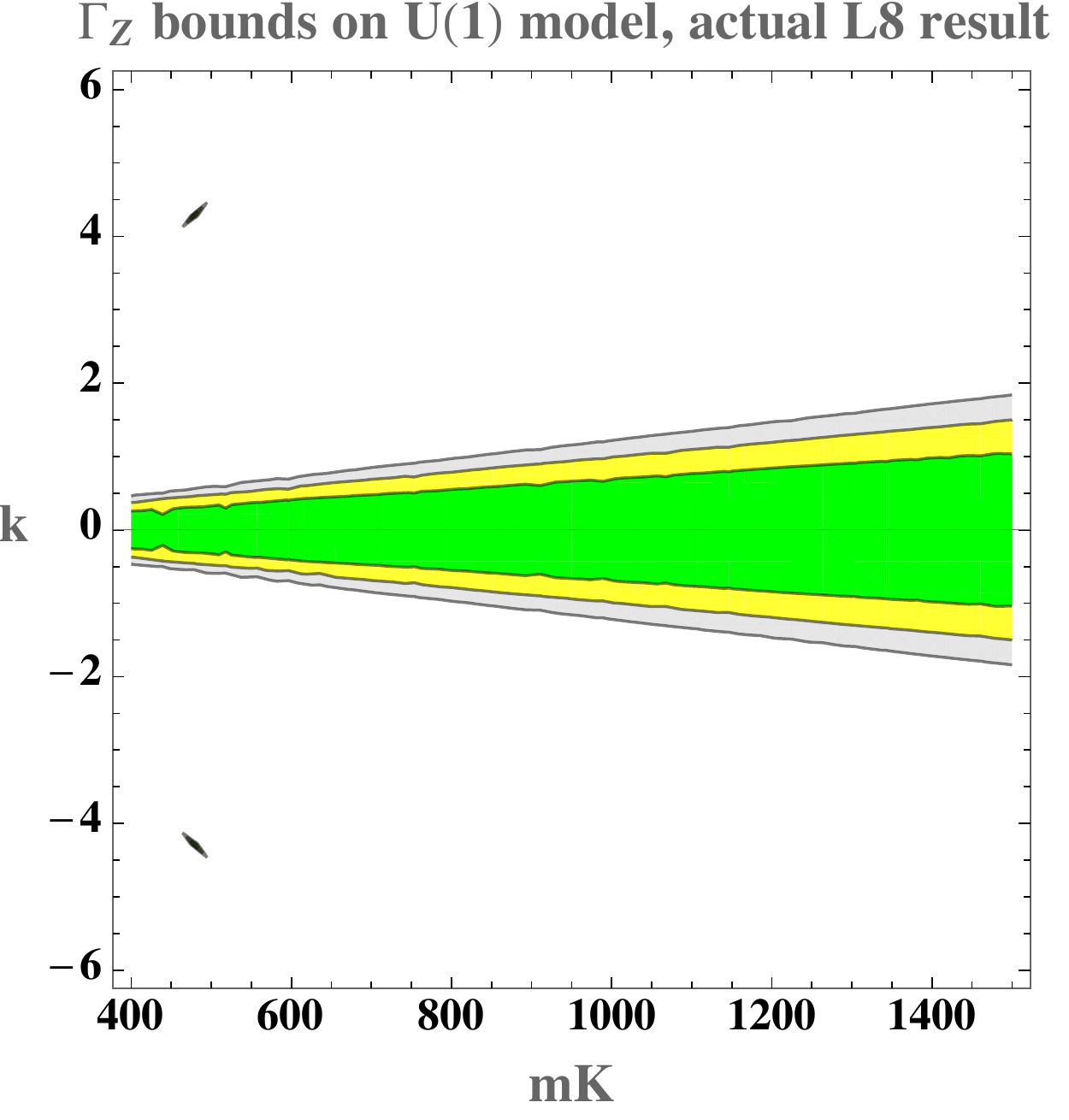}

\caption{Illustrative bounds on a 
U(1) mixing model parameters
due to bounds on $\Gamma_Z$. Shown is the $\{1,2,3\} \sigma$ allowed region in green, yellow, gray.
Here $1 \sigma$ for $\delta \Gamma_Z = 0.0023/2.4952$.
Results shown are for the $\hat\alpha_{ew}$ input-parameter scheme in the first row.
The results in the second row are in the $\hat m_W$ input-parameter scheme.\label{zdecayplotset1}}
\end{figure}

\begin{figure}[h!]
\includegraphics[height=0.25\textheight,width=0.32\textwidth]{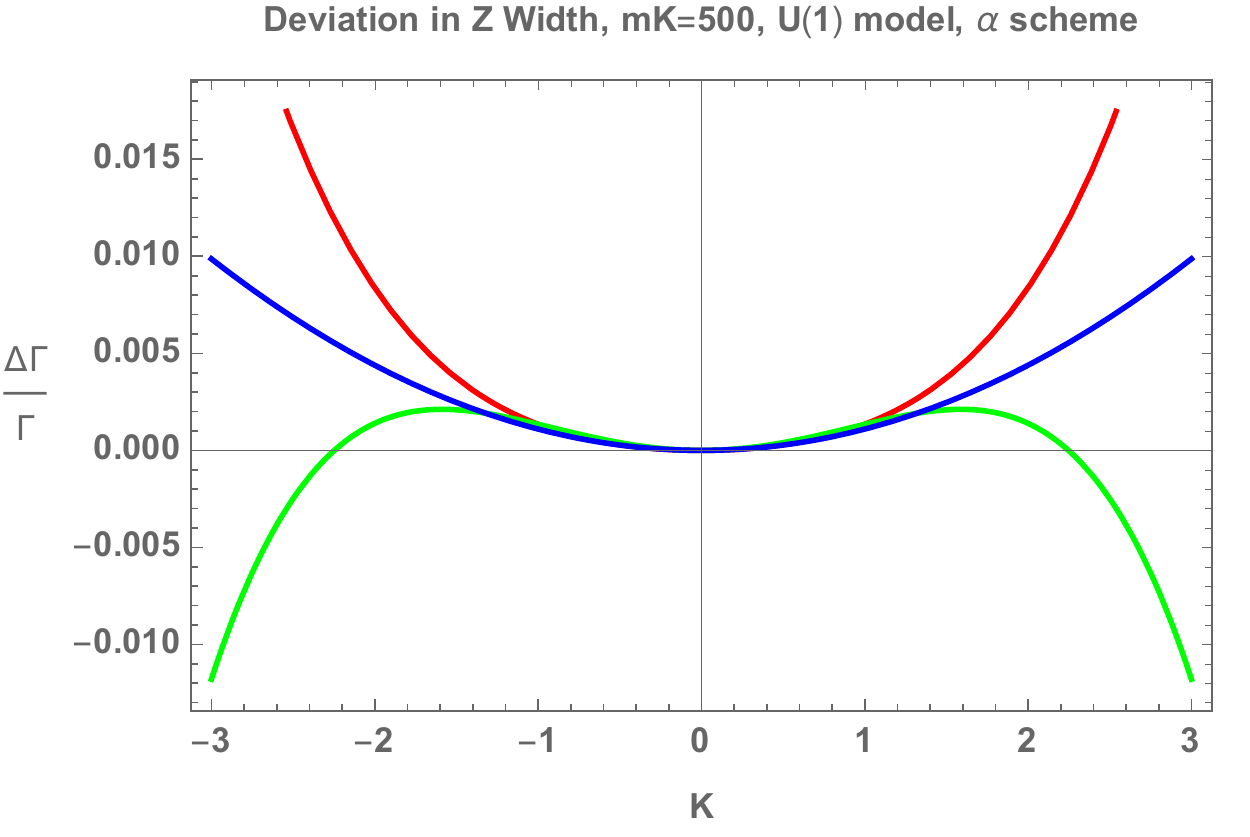}
\includegraphics[height=0.25\textheight,width=0.32\textwidth]{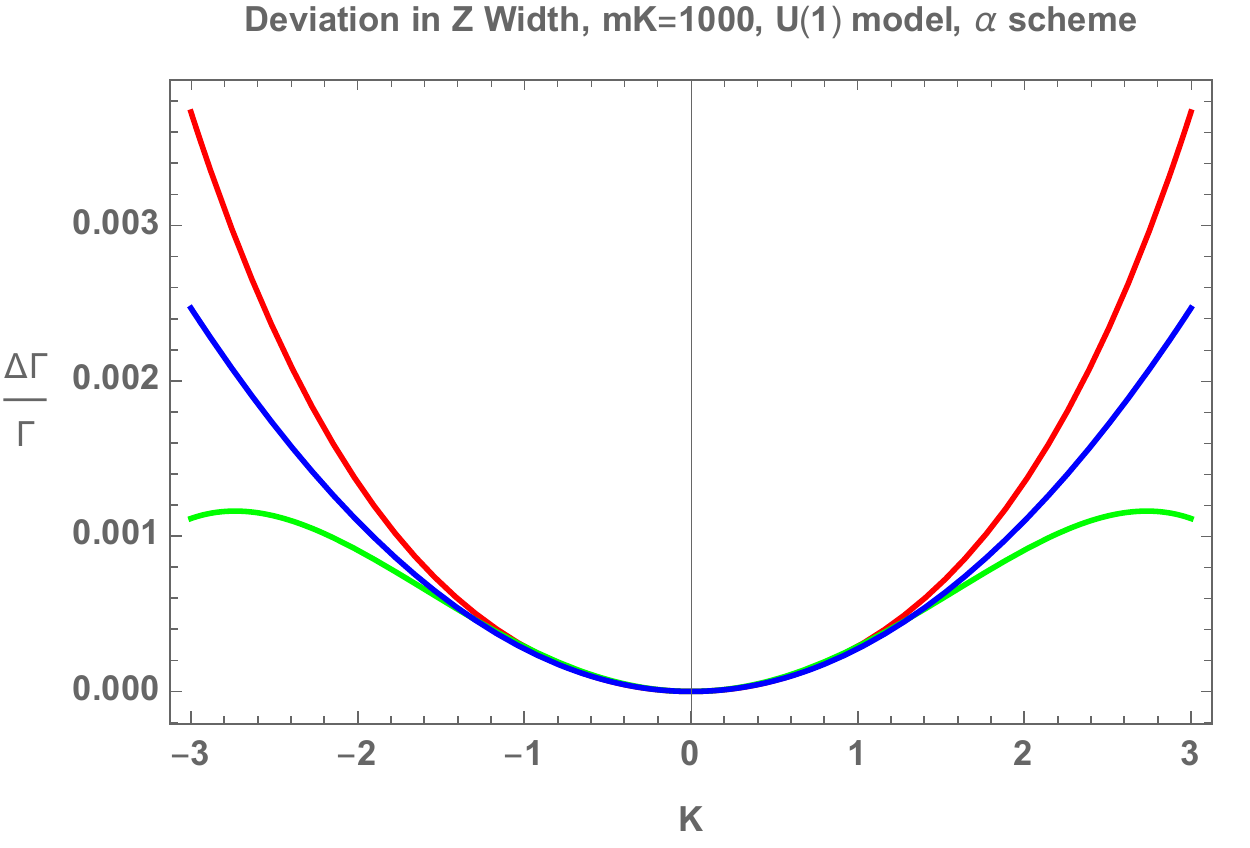}
\includegraphics[height=0.25\textheight,width=0.32\textwidth]{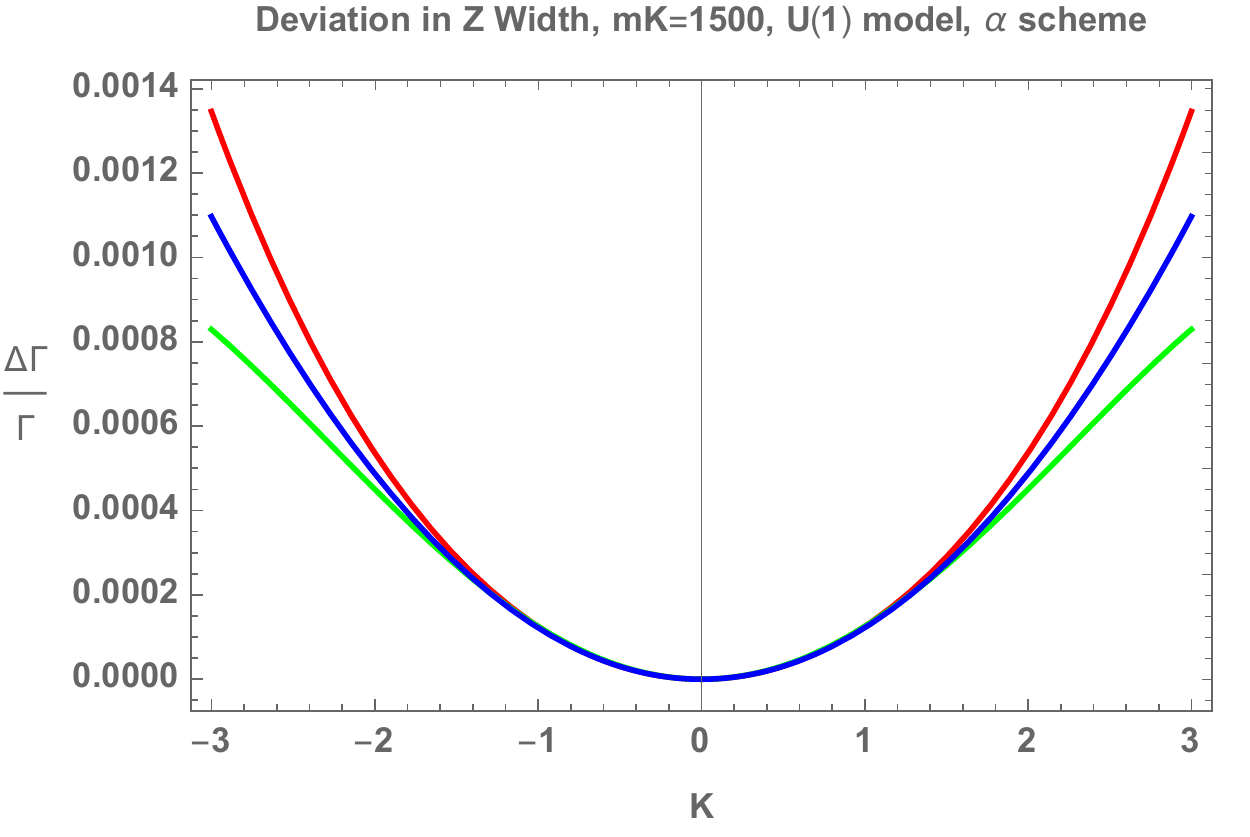}

\includegraphics[height=0.25\textheight,width=0.32\textwidth]{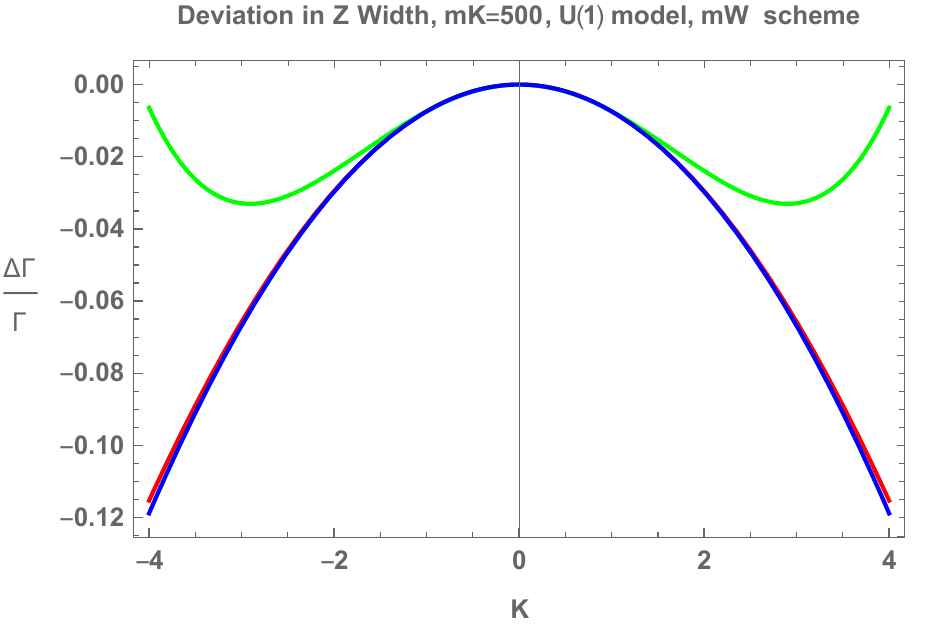}
\includegraphics[height=0.25\textheight,width=0.32\textwidth]{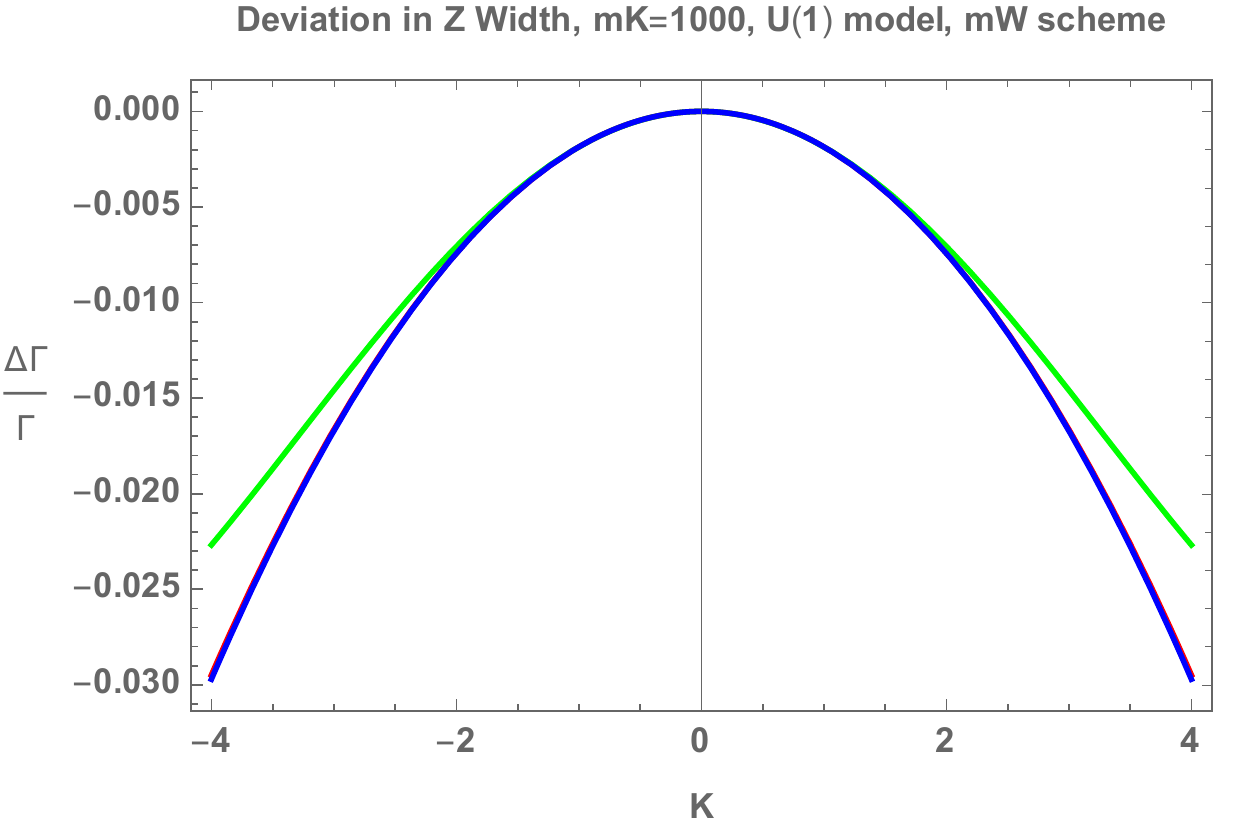}
\includegraphics[height=0.25\textheight,width=0.32\textwidth]{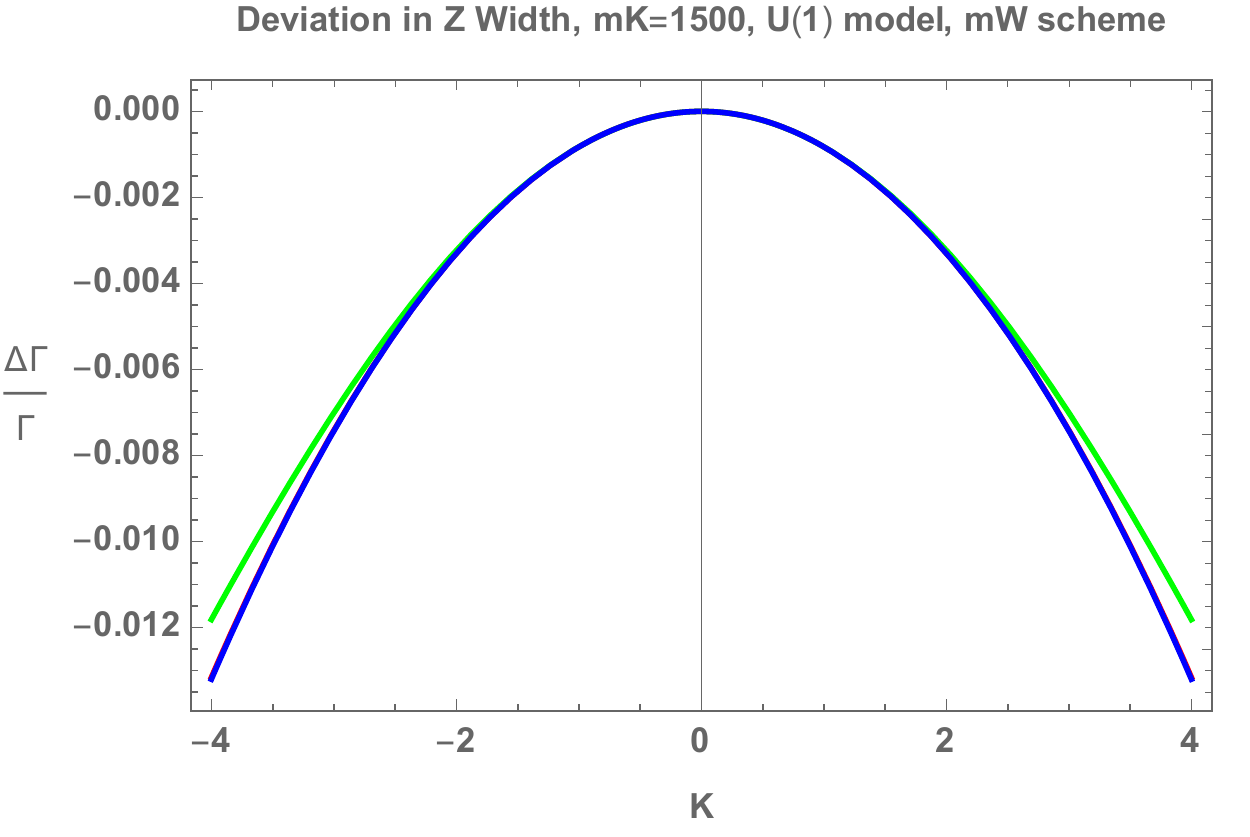}
\caption{Deviation in the $Z$ width in the 
U(1) mixing model for fixed $m_K$ comparing the partial-square result
in red and the full SMEFT result at $\mathcal{L}^{(8)}$ in green, and the full SMEFT result at $\mathcal{L}^{(6)}$ in blue.\label{zdecayplotset2}}
\end{figure}

The corresponding full SMEFT results matched onto the U(1) model at ${\cal{O}}(\bar{v}_T^4/m_K^4)$ are\footnote{For this result we
include the $\bar{v}_T^2/m_K^2$ correction in the parameter $b_1$ that is formally present as a contribution to matching onto
$\mathcal{L}^{(6)}$ operators. Doing so, the $\lambda$ dependence exactly cancels out, as expected.}
\bea
 \frac{\sum_{\psi} \bar{\Gamma}^{{\rm SMEFT},\hat{\alpha}_{ew}}_{\mathcal{Z} \rightarrow \bar{\psi}_p \psi_p}}{\sum_{\psi} \bar{\Gamma}^{{\rm SM}, \hat{\alpha}_{ew}}_{\mathcal{Z} \rightarrow \bar{\psi}_p \psi_p}}
&=& 1 + 4.5 \times 10^{-3}  \, \frac{\bar{v}_T^2 \, k^2}{m_K^2} - 5.7 \times 10^{-3}\, (k^4 - 1.74 k^2) \, \frac{\bar{v}_T^4}{m_K^4},
\eea
\bea
 \frac{\sum_{\psi} \bar{\Gamma}^{{\rm SMEFT}, \hat{m}_{W}}_{\mathcal{Z} \rightarrow \bar{\psi}_p \psi_p}}{\sum_{\psi} \bar{\Gamma}^{{\rm SM},\hat{m}_{W}}_{\mathcal{Z} \rightarrow \bar{\psi}_p \psi_p}}
&=& 1 -3.1 \times 10^{-2} \, \frac{\bar{v}_T^2 \, k^2}{m_K^2} + 7.9 \times 10^{-3} \, (k^4 - 0.88 k^2) \, \frac{\bar{v}_T^4}{m_K^4}.
\eea
All differences between partial-square and full SMEFT results are at order $\bar{v}_T^4/m_K^4$. Such differences are most
important when deviations from the SM are larger, e.g. for lower mass scales, where experimental analyses are more likely
to uncover deviations using the SMEFT formalism.  We show some of the implications of these results in
Figs.~\ref{zdecayplotset1} and \ref{zdecayplotset2}.  A number of conclusions are apparent:
\begin{itemize}
\item{The results show significant scheme dependence, which increases when a full SMEFT result is used. This is expected on
general grounds due to the decoupling theorem: low-energy measured parameters are absorbing the effects of high-scale physics.
Scheme dependence is expected to be reduced only through a global combination of constraining measurements.
}
\item{The model parameters extracted from the partial-square result for the $\hat\alpha_{ew}$ input-parameter scheme are constrained
more tightly than those at ${\cal{O}}(\bar{v}_T^4/m_K^4)$, given the $<0.1\%$ precision on the $\Gamma_Z$ measurement
(Fig.~\ref{zdecayplotset2}).  The ${\cal{O}}(\bar{v}_T^2/m_K^2)$ constraints are also overly tight, though less so.
}
\item{Results in the $\hat{m}_W$ scheme are more consistent across the different orders in the calculation.
The parameter constraints extracted from the partial-square and ${\cal{O}}(\bar{v}_T^2/m_K^2)$ calculations are essentially the same,
and slightly tighter than those from the full ${\cal{O}}(\bar{v}_T^4/m_K^4)$ calculation.
}
\item{At $\mathcal{L}^{(6)}$ there is no dependence on
$g_2$ in the matching. The dependence on $g_2$ at $\mathcal{L}^{(8)}$ comes
about due to the arrangement of operator forms in the middle term of Eqn.~\eqref{subleadingmatching}, and when inferring Lagrangian
parameter numerical values from input parameters.
The correction to the $\Gamma_Z$ width dependent on $g_2$ in the
U(1) model carries an overall $k^2-k^4$ dependance, suppressing the numerical dependence of the results on $g_2$.
We also find that the $\mathcal O(\lambda)$ contributions cancel in the ${\cal{O}}(\bar{v}_T^4/m_K^4)$ result, as
expected for a basis-independent result.}
\end{itemize}

These results are specific to the 
U(1) mixing model and should be considered as illustrative.  Nevertheless, they highlight the need to
combine multiple measurements to suppress scheme dependence, and they show that the inference of
model parameters from partial-square results can be less accurate than those from a ${\cal{O}}(v^2/\Lambda^2)$
calculation.  A consistent truncation order is preferred for measuring coefficients and matching to UV models.

\section{Conclusions}
Using the geoSMEFT formalism we have calculated the first complete results in the SMEFT to ${\cal{O}}(v^4/\Lambda^4)$.  We
have provided numerical expressions to this order for the operator dependence of the partial widths $\Gamma(h \rightarrow \gamma \gamma)$,
$\Gamma(h \rightarrow \mathcal{Z} \, \gamma)$, and $\Gamma(\mathcal{Z} \rightarrow \bar{\psi} \, \psi)$, for both the
$\hat m_W$ and $\hat{\alpha}_{ew}$ input-parameter schemes.  A necessary ingredient for these results is the theoretical formalism
of input-parameter schemes to all orders in the $\bar{v}_T/\Lambda$ expansion.

In addition to the full ${\cal{O}}(v^4/\Lambda^4)$ calculations, we have obtained numerical expressions for the expansion of each
partial width to ${\cal{O}}(v^2/\Lambda^2)$, and using a `partial-square' procedure whereby the amplitudes with dimension-6
operators are squared.  We have used these results to study the partial-width deviations from the SM for the different calculations
and a common set of parameters.  As expected the effect of the higher-order terms increase as the scale decreases, and are particularly
important for the (SM) loop-level widths $\Gamma(h \rightarrow \gamma \gamma)$ and $\Gamma(h \rightarrow \mathcal{Z} \, \gamma)$.  We have
investigated two procedures for estimating the effects of ${\cal{O}}(v^4/\Lambda^4)$ relative to ${\cal{O}}(v^2/\Lambda^2)$, and found
that the partial-square calculation provides a reasonable estimate of the truncation uncertainty for one of the operators affecting the
tree-level width of the $\mathcal{Z}$ boson, but both procedures underestimate the uncertainty for the loop-level partial widths.  Current
global fits find dimension-6 coefficients consistent with zero, so we recommend using an uncertainty based on the measurement precision
and expected $\Lambda$ dependence in order to minimize the effects of measurement noise.
A total uncertainty assignment, due to missing higher order effects, should also include an estimate for missing perturbative corrections.

We have performed a matching of operators up to $\mathcal{L}^{(8)}$ for a kinetic mixing model, and we have determined the differences
in inferred parameter values using the various calculations.  We have observed a significant dependence on the input-parameter scheme,
highlighting the importance of combining multiple measurements when fitting for Wilson coefficients.  The partial width
$\Gamma(h \rightarrow \gamma \gamma)$ is only affected at ${\cal{O}}(v^4/\Lambda^4)$, providing an example of the value of determining
coefficients to this order.  The total width $\Gamma_Z$ is affected at ${\cal{O}}(v^2/\Lambda^2)$, and the parameter constraints inferred
from a partial-square calculation are tighter than those inferred from either the ${\cal{O}}(v^2/\Lambda^2)$ or ${\cal{O}}(v^4/\Lambda^4)$
calculations in the $\hat{\alpha}_{ew}$ scheme.  A consistent expansion in the matching and the coefficient measurement is preferred on
general grounds, and this example demonstrates that the use of a partial-square calculation in a fit can lead to overly tight constraints.
While a partial-square procedure can provide an indication of the ${\cal{O}}(v^4/\Lambda^4)$ contributions, a full fixed-order calculation should be
used when measuring coefficients in data, or when matching to UV models.  We have demonstrated the power of the geoSMEFT formalism to expand the
canon of complete calculations at ${\cal{O}}(v^4/\Lambda^4)$ and open up a new avenue to
exploring the phenomenology of the SMEFT.

\acknowledgments
AH acknowledges support from the Carlsberg Foundation.
MT and AH acknowledge support from the Villum Fund, project number 00010102.
The work of AM is partially supported by the National Science Foundation under Grant No. Phy-1230860.
We thank Tyler Corbett and Jim Talbert for comments on the draft.

\appendix
\section{Gauge couplings and mixing angles}\label{appendixdefinitions}
The geometric Lagrangian parameters are
\begin{align}
\label{geometricelectric}
\bar{e} &= g_2\left(s_{\bar{\theta}} \sqrt{g}^{33} + c_{\bar{\theta}} \sqrt{g}^{34}  \right) = g_1\left(c_{\bar{\theta}} \sqrt{g}^{44} + s_{\bar{\theta}} \sqrt{g}^{34}  \right), \\
s_{\bar{\theta}}^2 &= \frac{(g_1 \sqrt{g}^{44} - g_2 \sqrt{g}^{34})^2}{g_1^2 [(\sqrt{g}^{34})^2+ (\sqrt{g}^{44})^2]+ g_2^2 [(\sqrt{g}^{33})^2+ (\sqrt{g}^{34})^2]
- 2 g_1 g_2 \sqrt{g}^{34} (\sqrt{g}^{33}+ \sqrt{g}^{44})},
\end{align}
and
\bea\label{geometricgz}
s_{\theta_Z}^2 &=& \frac{g_1 (\sqrt{g}^{44} s_{\bar{\theta}} - \sqrt{g}^{34} c_{\bar{\theta}})}{g_2 (\sqrt{g}^{33} c_{\bar{\theta}} - \sqrt{g}^{34} s_{\bar{\theta}})+ g_1 (\sqrt{g}^{44} s_{\bar{\theta}} - \sqrt{g}^{34} c_{\bar{\theta}})}, \\
\bar{g}_Z &=& \frac{g_2}{c_{\theta_Z}^2}\left(c_{\bar{\theta}} \sqrt{g}^{33} - s_{\bar{\theta}} \sqrt{g}^{34} \right) = \frac{g_1}{s_{\theta_Z}^2}\left(s_{\bar{\theta}} \sqrt{g}^{44} - c_{\bar{\theta}} \sqrt{g}^{34} \right).
\eea
A set of useful results for Lagrangian parameters expanded out to ${\cal{O}}(v^4/\Lambda^4)$
as
\begin{align}
\bar{P} = \bar{P}^{\rm SM} + \langle \bar{P} \rangle_{{\cal{O}}(v^2/\Lambda^2)}  + \langle \bar{P} \rangle_{{\cal{O}}(v^4/\Lambda^4)} + \dots
\end{align}
are

\begin{align}
	\overline{g}_2^{\rm SM} &= g_2,  \qquad\qquad \langle\overline{g}_2\rangle_{{\cal{O}}(v^2/\Lambda^2)}
	= g_2 \tilde C_{HW}^{(6)}, \quad \nn
	\langle\overline{g}_2\rangle_{{\cal{O}}(v^4/\Lambda^4)} &=  \frac{3}{2 \overline{g}_2^{\rm SM}}
\left(\langle\overline{g}_2\rangle_{{\cal{O}}(v^2/\Lambda^2)}\right)^2
+ \frac{1}{2}\langle\overline{g}_2\rangle_{{\cal{O}}(v^2/\Lambda^2)}\left|_{\tilde C_{HW}^{(6)} \rightarrow \tilde C_{HW}^{(8)}}\right. ,
\end{align}


\begin{align}
	\overline{e}^{\rm SM} &= \frac{g_1 g_2}{\sqrt{g_1^2 + g_2^2}}, & \quad
	\langle\overline{e}\rangle_{{\cal{O}}(v^2/\Lambda^2)}
	&= \frac{g_1 g_2\left( g_2^2 \tilde C_{HB}^{(6)} + g_1^2  \tilde C_{HW}^{(6)} - g_1 g_2 \tilde C_{HWB}^{(6)}\right)}{(g_1^2 + g_2^2)^{3/2}},
\end{align}
\begin{align}
	\langle\overline{e}\rangle_{{\cal{O}}(v^4/\Lambda^4)}
	&= \frac{3}{2 \overline{e}^{\rm SM}}
\left(\langle\overline{e}\rangle_{{\cal{O}}(v^2/\Lambda^2)}\right)^2
+
\langle\overline{e}\rangle_{{\cal{O}}(v^2/\Lambda^2)} \left|_{
\tilde C_{i}^{(6)} \rightarrow \tilde C_{i}^{(8)}}\right. ,
\end{align}


\begin{align}
	\overline{g}_Z^{\rm SM} &= \sqrt{g_1^2 + g_2^2}, & \quad
	\langle\overline{g}_Z\rangle_{{\cal{O}}(v^2/\Lambda^2)}
	&= \frac{g_1^2 \tilde C_{HB}^{(6)} + g_2^2  \tilde C_{HW}^{(6)} + g_1 g_2 \tilde C_{HWB}^{(6)}}{(g_1^2 + g_2^2)^{1/2}} ,
\end{align}

\begin{align}
	\langle\overline{g}_Z\rangle_{{\cal{O}}(v^4/\Lambda^4)}
	=& \frac{3}{2 \overline{g}_Z^{\rm SM}} \left(
\langle\overline{g}_Z\rangle_{{\cal{O}}(v^2/\Lambda^2)} \right)^2
+ \frac{\overline{g}_Z^{\rm SM}}{2 s_{\theta_Z}^{2,{\rm SM}} c_{\theta_Z}^{2,{\rm SM}}}
\left(\langle s_{\theta_Z}^2 \rangle_{{\cal{O}}(v^2/\Lambda^2)} \right)^2
+
\langle\overline{g}_Z\rangle_{{\cal{O}}(v^2/\Lambda^2)} \left|_{
\tilde C_{i}^{(6)} \rightarrow \tilde C_{i}^{(8)}}\right. ,
\end{align}


\begin{align}
	s_{\theta_Z}^{2,{\rm SM}} &= s_{\overline{\theta}}^{2,{\rm SM}} = \frac{g_1^2 }{g_1^2 + g_2^2},\\
   \langle s_{\theta_Z}^2 \rangle_{{\cal{O}}(v^2/\Lambda^2)} & = \langle s_{\overline{\theta}}^2 \rangle_{{\cal{O}}(v^2/\Lambda^2)}
	= \frac{g_1 g_2\left( 2g_1 g_2 \left(\tilde C_{HB}^{(6)} -  \tilde C_{HW}^{(6)}\right)
	- \left(g_1^2 - g_2^2\right) \tilde C_{HWB}^{(6)}\right)}{(g_1^2 + g_2^2)^{2}} ,
\end{align}

\begin{align}
	\langle s_{\theta_Z}^2 \rangle_{{\cal{O}}(v^4/\Lambda^4)}
	&= \frac{2}{ \overline{e}^{\rm SM}} \left(
\langle\overline{e}\rangle_{{\cal{O}}(v^2/\Lambda^2)} \right)
\left(\langle s_{\theta_Z}^2 \rangle_{{\cal{O}}(v^2/\Lambda^2)} \right)
+
\left(\langle s_{\theta_Z}^2\rangle_{{\cal{O}}(v^2/\Lambda^2)}\right)\left|_{
\tilde C_{i}^{(6)} \rightarrow \tilde C_{i}^{(8)}}\right. ,\\
\langle s_{\overline{\theta}}^2 \rangle_{{\cal{O}}(v^4/\Lambda^4)}
&= \langle s_{\theta_Z}^2 \rangle_{{\cal{O}}(v^4/\Lambda^4)} + \frac{\overline{g}_Z^{\rm SM} \tilde C_{HWB}^{(6)}}{4\overline{e}^{\rm SM}}
\left(\langle s_{\theta_Z}^2 \rangle_{{\cal{O}}(v^2/\Lambda^2)} \right).
\end{align}
It is also useful to note the relations
\bea
\langle s_{\theta_Z}^2 \rangle_{{\cal{O}}(v^2/\Lambda^2)} &=& - 2 \frac{g_1 \, g_2}{({g}^{\rm SM}_Z)^2} \, \langle h|\gamma Z\rangle_{\mathcal{L}^{(6)}}, \\
\langle s_{\theta_Z}^2 \rangle_{{\cal{O}}(v^4/\Lambda^4)} &=&
	2\left(\langle s_{\theta_Z}^2 \rangle_{{\cal{O}}(v^2/\Lambda^2)}\right)
 \langle h|\gamma \gamma \rangle_{\mathcal{L}^{(6)}}
	+
	\left.\left(
	\langle s_{\theta_Z}^2 \rangle_{{\cal{O}}(v^2/\Lambda^2)}\right)\right|_{C_i^{(6)}\rightarrow C_i^{(8)}},
\eea
where $\langle h|\gamma Z\rangle_{\mathcal{L}^{(6)}}$ is defined in Eqn.~\eqref{hzgamdim6}
and $\langle h|\gamma \gamma\rangle_{\mathcal{L}^{(6)}}$ is defined in Eqn.~\eqref{eq:hggDim6}.

\section{All-orders vev}\label{allordersvev}

An all-orders form of the vev can be constructed as an infinite series by
defining
\begin{align}
	\overline{v}_T^2 = v_{\rm SM}^2 \sum_{n=0}^{\infty} \sum_{m=0}^{\infty} \sum_{k=0}^{\infty}\dots
	\frac{ 4^n A_{m,k,\dots}(n+1)_p \left(\frac{q}{2}\right)_n}{(r)_n} \left(\frac{v_{\rm SM}^2}{\Lambda^2}\right)^j
	\left(\frac{6}{8\lambda}\tilde C_H^{(6)}\right)^n
	\left(\frac{8}{16\lambda}\tilde C_H^{(8)}\right)^m
	\left(\frac{10}{32\lambda}\tilde C_H^{(10)}\right)^k\dots
\end{align}
where $(x)_n$ is the Pochhammer symbol and
\begin{align}
	i &= n + m + k + \dots \\
	j&= n+2m+3m+\dots\\
	r &= 2-n + j \\
	p &= \frac{1}{4}\left[6m - 1 + (-1)^m\right] + 2k + \dots \\
	\frac{q}{2} &= \frac{1}{2} + \frac{1}{4}\left[6m + 1 - (-1)^m \right] + 2k \dots \\
	a &= \frac{1}{9} \left( 12 k + \sqrt{3} \sin\left(\frac{2\pi k}{3}\right) + 3 \cos \left(\frac{2\pi k}{3}\right) - 3 \right) + \dots \\
	b &= \frac{1}{9}\left(12 k - \sqrt{3}\sin\left(\frac{2\pi k}{3}\right) - \cos\left(\frac{2\pi k}{3}\right)+1\right) + \dots \\
	c &= \frac{1}{9}\left( 12k - 2 \cos\left(\frac{2\pi k}{3}\right) + 2\right) + \dots \\
	d &= \frac{1}{4}\left(6 k +1 - (-1)^k\right) + \dots \\
	e &= \frac{1}{4}\left( 6k + 1 + (-1)^k \right) + \dots \\
	A_{m,k,\dots} &= B_{k,\dots}\frac{ \left( \frac{27}{4}\right)^m \left(m+1\right)_a \left(\frac{1}{3}+b\right)_m
		\left(\frac{2}{3} + c\right)_m}{\left(1+d\right)_m \left(1+e \right)_m
	\left(1\right)_p}
	\\
	B_{k,\dots} &= C_{\dots}\frac{\left(\frac{256}{27}\right)^{k} \left(\frac{1}{4}\right)_{k}
		\left(\frac{1}{2}\right)_{k} \left(\frac{3}{4}\right)_{k}}{\left(\frac{2}{3}\right)_{k}
	\left(1\right)_{a}\left(1\right)_{k} \left(\frac{4}{3}\right)_{k}} \\
	C_{0,0,\dots} &= 1.
\end{align}
This does not violate the Abel impossibility theorem, as the solution is not a solution in radicals.

The same solution can be applied to solve for the Lagrangian parameter $\bar{v}_T$ in terms of the measured value of $\hat G_F$.

\section{$(\bar{v}^2_T/\Lambda^2)^n$ mappings of input parameters to Lagrangian parameters}\label{inputs}
We use the ``hat-bar" convention \cite{Brivio:2017vri,Alonso:2013hga,Berthier:2015oma} in our input parameter analysis.
Lagrangian parameters directly determined from the measured input parameters are defined as having hat superscripts.
Lagrangian parameters in the canonically normalized SMEFT are indicated with
bar superscripts. These parameters are the geoSMEFT mass eigenstate Lagrangian parameters.
A numerical value of an SM Lagrangian parameter ($P$) can be modified in the SMEFT, and the difference between these
Lagrangian parameters ($P$) is in general denoted as
$\delta P = \bar{P} - \hat{P}$. Note that defining the parameter shift in this manner introduces a sign convention
for $\delta P$. The Lagrangian parameters are defined to all orders in Appendix~\ref{appendixdefinitions}.

\subsection{Input parameter $\hat{G}_F$}\label{sec:GFredef}

The value of the vev of the Higgs field is obtained from the precise measurement of the decay $\Gamma(\mu^- \rightarrow e^- + \bar{\nu}_e + \nu_\mu)$.
It is sufficient
when considering corrections up to $\mathcal{L}^{(6)}$ to define the local effective interaction for muon decay as
\begin{align}\label{usualGF}
\mathcal{L}_{G_F} =  -\frac{4\mathcal{G}_F}{\sqrt{2}} \, \left(\bar{\nu}_\mu \, \gamma^\mu P_L \mu \right) \left(\bar{e} \, \gamma_\mu P_L \nu_e\right).
\end{align}
The modification of this input parameter in the SMEFT to $\mathcal{O}(v^2/\Lambda^2)$ in the Warsaw basis was given in Ref.~\cite{Alonso:2013hga}
\begin{align}
-\frac{4\mathcal{G}_F}{\sqrt{2}} &=  -\frac{2}{\bar{v}_T^2} +  \left(C^{(6)}_{\substack{ll \\ \mu ee \mu}} +  C^{(6)}_{\substack{ll \\ e \mu\mu e}}\right) - 2 \left(C^{3,(6)}_{\substack{Hl \\ ee }} +  C^{3,(6)}_{\substack{Hl \\ \mu\mu }}\right).
\label{gfermi}
\end{align}
At higher orders in the $\bar{v}_T/\Lambda$ expansion, the results in Ref.~\cite{Helset:2020yio} define the
contributions through a $W^\pm$ exchange
to Eqn.~\eqref{usualGF} via
\bea
\acontraction{\langle \bar{\psi} \psi|}{W}{\rangle \, \langle}{W}\langle \bar{\psi} \psi|W \rangle \, \langle W| \bar{\psi} \bar{\psi} \rangle
&\simeq& - \frac{2}{\sqrt{h_{11}}^2 \, \bar{v}_T^2} \, \left(\bar{\nu}_\mu \, \gamma^\mu P_L \mu \right) \left(\bar{e} \, \gamma_\mu P_L \nu_e\right)  \nn
&\,& \,
\times \left[1-\bar{v}_T \langle L^{\ell,22}_{1,1}\rangle   +  i \bar{v}_T
\langle L^{\ell,22}_{1,2} \rangle \right] \, \left[1-\bar{v}_T \langle L^{\ell,11}_{1,1}\rangle   -  i \bar{v}_T
\langle L^{\ell,11}_{1,2} \rangle \right].
\eea
Here we are neglecting corrections relatively suppressed by the light fermion masses and $\Gamma_W$.
The cross terms of the field-space connections $\langle L^{\ell}\rangle$ are examples of ``double insertions"
of higher-dimensional operators in the SMEFT, generically present when developing analyses to ${\cal{O}}(v^4/\Lambda^4)$
or higher.  To generalize to higher orders in the power counting expansion we define
\begin{align}
-\frac{4\mathcal{G}^W_F}{\sqrt{2}} &=  - \frac{2}{\sqrt{h_{11}}^2 \, \bar{v}_T^2} \,
\left[1-\bar{v}_T \langle L^{\ell,22}_{1,1}\rangle   +  i \bar{v}_T
\langle L^{\ell,22}_{1,2} \rangle \right] \, \left[1-\bar{v}_T \langle L^{\ell,11}_{1,1}\rangle   -  i \bar{v}_T
\langle L^{\ell,11}_{1,2} \rangle \right].
\label{gfermi2}
\end{align}
The tower of higher-dimensional operators that directly give a four-point function can interfere with the SM contribution.
Four-fermion operators that can contribute to muon decay have the chirality combinations $LLLL$ or $LLRR$.
At ${\cal{O}}(v^2/\Lambda^2)$ only $LLLL$ terms interfere with the SM amplitude, and such contributions are from $C_{ll}^{(6)}$ in Eqn.~\eqref{usualGF}.
When considering corrections to ${\cal{O}}(v^4/\Lambda^4)$ and higher, self-interference terms are also present in the Wick expansion that
need not interfere with \eqref{usualGF}. For example, in the Warsaw basis \cite{Grzadkowski:2010es} for $\mathcal{L}^{(6)}$, a contribution from
\bea
\mathcal{Q}_{\substack{le\\prst}}^{(6+ 2 n)} = (H^\dagger H)^n(\bar{\ell}_p \gamma_\mu \ell_r)(\bar{e}_s \gamma^\mu e_t)
\eea
is present when $n = 0$. Here $p,r,s,t$ are flavour indicies that run over $1,2,3$. The generalization for $LLRR$ operators to higher mass dimensions
introduces an additional operator of the form
\bea
\mathcal{Q}^{(8+ 2n)}_{\substack{\ell \cdot H e  \\ prst}} &=& (H^\dagger H)^{n} \, (H^\dagger \sigma^i H) \, (\bar{\ell}_p\,\gamma^{\mu}\, \sigma_i \,\ell_r) \,  (\bar{e}_s\,\gamma_{\mu}\, e_t).
\eea
We use a dot product in the operator label to indicate an $\rm SU(2)_L$ triplet contraction and
define a short-hand notation for these contributions
\bea
\mathcal{G}^{4\,pt}_{\substack{ee\\ pr}} &=& \frac{1}{\bar{v}_T^2}  \tilde{C}^{(6+ 2 n)}_{\substack{\ell e \\ pr12}}
-\frac{1}{\bar{v}_T^2}  \tilde{C}^{(8+ 2 n)}_{\substack{\ell \cdot H e  \\ pr12}}.
\eea

When considering four-fermion operators of the chirality $LLLL$ that can contribute to muon decay, it is sufficient to define
\begin{align}\label{LLLLgeneralization}
\mathcal{L}^{(d \geq 8)}_{\substack{4 \ell}} &=
C^{(8 + 2 n)}_{\substack{\ell \ell H \\ prst}} (H^{\dag}H)\, (H^{\dag}H)^{n}(\bar{\ell}_p\,\gamma^{\mu}\, \ell_r)(\bar{\ell}_s\, \gamma_{\mu}\, \ell_t) \nn
					     &+ C^{(8 + 2 n)}_{\substack{\ell \cdot H \ell H \\ prst}} (H^{\dag} \sigma^i \, H)\, (H^{\dag}H)^{n}(\bar{\ell}_p\,\gamma^{\mu}\, \sigma_i \, \ell_r)(\bar{\ell}_s\, \gamma_{\mu}\, \ell_t)  \nn
& +  C^{(10 + 2 n)}_{\substack{\ell \cdot \ell \cdot H \cdot H \\ prst}} \, (H^{\dag}H)^{n}  \, (H^{\dag} \sigma^i \, H)\, (H^{\dag} \sigma^j \, H)\, (\bar{\ell}_p\,\gamma^{\mu}\, \sigma_i \, \ell_r)(\bar{\ell}_s\, \gamma_{\mu}\, \sigma_j \, \ell_t).
\end{align}
where $\sigma^i$ are the Pauli matrices. Fermi statistics imposes a non-trivial counting in the allowed $prst$, as is also the case for $\mathcal{L}^{(6)}$, see
Refs.~\cite{Dashen:1993jt,Dashen:1994qi,Alonso:2013hga}.
An operator form with an explicit $\epsilon^{ijk}$ can be related to those above using the Pauli matrix commutation
and completeness relations. We also introduce the short-hand notation for the set of higher-dimensional-operator
contributions that interfere with the SM amplitude
\bea
\mathcal{G}^{4\,pt}_{\substack{\ell \ell \\rs}} &=&  \frac{1}{\bar{v}_T^2}  \left(\tilde{C}^{(6)}_{\substack{ll \\ rs e \mu}} +  \tilde{C}^{(6)}_{\substack{ll \\ e \mu rs}} +
\frac{1}{2^{1+n}} \left[\tilde{C}^{(8+ 2 n)}_{\substack{\ell \ell H \\ rs e \mu}}+ \tilde{C}^{(8+ 2 n)}_{\substack{\ell \ell H \\ e \mu rs}}\right]
+ \frac{1}{2^{1+n}} \left[\tilde{C}^{(8+ 2 n)}_{\substack{\ell \cdot H \ell H \\ e \mu rs }}- \tilde{C}^{(8+ 2 n)}_{\substack{\ell \cdot H \ell H \\ rs e \mu}}\right]\right) \nn
&-& \frac{1}{\bar{v}_T^2 \, 2^{2+n}}  \left[\tilde{C}^{(10+ 2 n)}_{\substack{\ell \cdot \ell \cdot H \cdot H \\ e \mu r s}}
+ \tilde{C}^{(10+ 2 n)}_{\substack{\ell \cdot \ell \cdot H \cdot H \\ rs e \mu}}\right],
\eea
which interferes with the SM when $r= 2, s=1$.

There are also self-interference terms present in the Wick expansion at ${\cal{O}}(v^4/\Lambda^4)$ and higher that need not interfere with Eqn.~\eqref{usualGF}.
For example, in the Warsaw basis \cite{Grzadkowski:2010es} contributions from $\mathcal{Q}^{(6)}_{\substack{ll \\ e \mu rs}}$
arise when $r \neq 2$,$s \neq 1$ \cite{Alonso:2013hga}.\footnote{
The muon decay width is measured without identification of the produced neutrino species, but the
SM weak interaction eigenstates are defined to be the flavour labels in Eqn.~\eqref{usualGF}, so only contributions from
the same weak neutrino eigenstates interfere with the SM contribution.}
These contributions are also given by $\mathcal{G}^{4\,pt}_{\substack{\ell \ell \\rs}}$,
when $r \neq2, s \neq 1$.

A measurement of the inclusive decay width of $\mu^-(p_1) \rightarrow e^-(p_4) + \bar{\nu}_e (p_3) + \nu_\mu (p_2)$ and an assumed value of the muon mass $\hat{m}_\mu$
yields
\bea
\hat{\Gamma}(\mu \rightarrow e \, \bar{\nu}_e \, \nu_\mu) = \frac{\hat{G}_F^2 \hat{m}^5_\mu}{192 \, \pi^3}.
\eea
The value of $G_F^2$ in a theoretical prediction of the right-hand side above
corresponds to the amplitude squared via
\bea
|\mathcal{A}|^2 =8 \, \left(\frac{4 \,\hat{G}_F}{\sqrt{2}}\right)^2 \, (p_1 \cdot p_3) (p_2 \cdot p_4)
\eea
The same sum over phase space and spin sum is present for all contributions
considered here. The mapping of these results to the Lagrangian parameters is then given by
\bea\label{allordersGF}
\left(\frac{4 \,\hat{G}_F}{\sqrt{2}}\right)^2 &=& \left(-\frac{4\mathcal{G}^W_F}{\sqrt{2}} + \mathcal{G}^{4\,pt}_{\substack{\ell \ell \\ 21}} \right)^2 +
\sum_{pr} \, \left(\mathcal{G}^{4\,pt}_{\substack{ee\\pr}}\right)^2 + \sum_{r,s, rs \neq 21} \left(\mathcal{G}^{4\,pt}_{\substack{\ell \ell \\rs}}\right)^2.
\eea
Inverting this equation and solving for $\bar{v}_T^2$ order by order in the $\bar{v}_T/\Lambda$ expansion
defines $\delta G_F$ order by order.
Consistent with past works \cite{Berthier:2015oma} we define this correction with the normalization
\bea
\bar{v}_T^2 = \frac{1}{\sqrt{2} \, \hat{G}_F} + \frac{\delta G_F}{\hat{G}_F},
\eea
which can be defined at any order in $(\bar{v}_T/\Lambda)^{(2n)}$ using \eqref{allordersGF}.
\subsection{Input parameter $\hat{\alpha}_{ew}$}\label{sec:alphadef}
For the 
U(1)$_{\rm{ew}}$
current we have
\bea
\mathcal{L}_{\mathcal{A}}  =
- \bar{e} \, \bar{\psi}_{\substack{p}} \, \slashed{\epsilon}_{\mathcal{A}} \, Q_\psi \, \delta_{pr}\, \psi_{\substack{r}}
\eea
The extraction of $\hat{\alpha}_{ew}$  occurs in the measurement of the Coulomb potential
of a charged particle in the low momentum limit ($q^2 \rightarrow 0$).
A low-scale measurement of this coupling must be run up through the hadronic resonance region
$q^2 \sim \Lambda_{\rm QCD}^2$ to be used at higher scales, and this introduces the dominant
error in the use of this input parameter.
See the discussion in Ref.~\cite{Brivio:2017btx} for more details.

\subsection{Input parameters $\hat{m}_Z,\hat{m}_W$}\label{sec:mzdef}
The remaining input parameters are more directly generalized to higher orders in the power counting.
The experimental measurements of these parameters are discussed in Ref.~\cite{Brivio:2017btx}.
For $m_{Z,W}$ we use the geometric definitions of the bar parameters,
which are valid to all orders in the $\bar{v}_T/\Lambda$ expansion:
\begin{align}
	\overline{m}_Z^2 = \frac{\overline{g}_Z^2}{4} \sqrt{h}_{33}^2 \overline{v}_T^2, \quad \quad \bar{m}_W^2 = \frac{\bar{g}_2^2}{4} \sqrt{h_{11}}^2 \bar{v}_T^2,
\end{align}
where in the Warsaw basis
\bea
\sqrt{h}_{33} &=& 1 + \frac{\tilde{C}_{HD}^{(6)}}{4} + \frac{\tilde{C}^{(8)}_{HD} +\tilde{C}^{(8)}_{HD,2}}{8} - \frac{(\tilde{C}_{HD}^{(6)})^2}{32} + \dots,\\
\sqrt{h}_{11} &=& 1 + \frac{\tilde{C}^{(8)}_{HD} - \tilde{C}^{(8)}_{HD,2}}{8} + \dots.
\eea

\section{\titlemath{$\{\hat{m}_W, \hat{m}_Z, \hat{G}_F\}$}{(aEM, MZ, GF, Mh)} input-parameter scheme at all orders in $(\bar{v}^2_T/\Lambda^2)^n$}
\label{zwidthmw}
In this scheme we can again use Eqn.~\eqref{gzshift} to define a shift to $\bar{g}_Z$.
We also use
\begin{align}
	\overline{g}_2 = g_2 \sqrt{g}^{11} = \frac{2 \hat m_W}{\sqrt{h}_{11}\overline{v}_T}.
\end{align}
and \begin{align}
	g_1 = g_2 \frac{\left( s_{\overline{\theta}} \sqrt{g}^{33} + c_{\overline{\theta}}\sqrt{g}^{34}\right)}
	{\left(c_{\overline{\theta}} \sqrt{g}^{44} + s_{\overline{\theta}} \sqrt{g}^{34}\right)}
\end{align}
to solve for $s_{\overline{\theta}}^2$ via
\begin{align}
s_{\overline{\theta}}^2 &=\frac{1}{\left[ \left(\sqrt{g}^{44}\right)^2 + \left(\sqrt{g}^{34}\right)^2\right]^2}
	\left\{ -\left( \frac{g_2 \sqrt{g}_-}{\overline{g}_Z}\right)^2\left[\left(\sqrt{g}^{44}\right)^2
		- \left(\sqrt{g}^{34}\right)^2\right]
		+ \left(\sqrt{g}^{44}\right)^2\left[\left(\sqrt{g}^{44}\right)^2 + \left(\sqrt{g}^{34}\right)^2\right]  \right. \nonumber\\
   &-
\left.		 2 \left(\frac{g_2 \sqrt{g}_-}{\overline{g}_Z}\right)
		\sqrt{ (\sqrt{g}^{44})^2 (\sqrt{g}^{34})^2 \left[\left(\sqrt{g}^{44}\right)^2 + \left(\sqrt{g}^{34}\right)^2
	- \left( \frac{g_2\sqrt{g}_-}{\overline{g}_Z}\right)^2\right]} \right\}.
\end{align}
The remaining Lagrangian parameters can then be defined via
\bea
\bar{e}=  \frac{\bar{g}_2}{\sqrt{g}^{11}} \left(s_{\bar{\theta}} \sqrt{g}^{33} + c_{\bar{\theta}} \sqrt{g}^{34} \right),
\eea
and
\bea
s_{\theta_Z}^2 = \frac{\overline{e}}{\overline{g}_Z} \frac{\left(s_{\overline{\theta}}\sqrt{g}^{44}
  - c_{\overline{\theta}}\sqrt{g}^{34}\right)}{\left(c_{\overline{\theta}}\sqrt{g}^{44}
+ s_{\overline{\theta}}\sqrt{g}^{34}\right)}.
\eea
In both schemes, $\bar{g}_Z$ and $s_{\theta_Z}^2$ have the same definition in terms or other ``barred" Lagrangian parameters.

\subsection{$\Gamma(\mathcal{Z} \rightarrow \bar{b} b)$}
The effective coupling results to ${\cal{O}}(v^4/\Lambda^4)$ in the case of down quarks are given by
\bea
\langle g_{\rm SM,pp}^{\mathcal{Z},d_{L}}\rangle &=& 0.32, \\
\langle g_{\rm eff,pp}^{\mathcal{Z},d_{L}}\rangle_{{\cal{O}}(v^2/\Lambda^2)}
&=&0.02 \tilde{C}_{HD}^{(6)} +0.10 \,  \tilde{C}_{HWB}^{(6)} -0.22 \, \delta G_{F}^{(6)}
+ 0.37 \, (\tilde{C}_{\substack{Hq\\pp}}^{(6)}+ \tilde{C}_{\substack{Hq\\pp}}^{3,(6)}), \\
\langle g_{\rm eff,pp}^{\mathcal{Z},d_{L}}\rangle_{{\cal{O}}(v^4/\Lambda^4)}
&=& -\left(\frac{\tilde{C}_{HD}^{(6)}}{4}+ \frac{\delta G_{F}^{(6)}}{\sqrt{2}}\right) \langle g_{\rm eff,pp}^{\mathcal{Z},d_{L}}\rangle_{{\cal{O}}(v^2/\Lambda^2)}
+  \tilde{C}_{HWB}^{(6)} \left(-0.06 \tilde{C}_{HD}^{(6)} +0.10 (\tilde{C}_{HB}^{(6)} +\tilde{C}_{HW}^{(6)})  \right) \nn
&+&0.01 (\tilde{C}_{HD}^{(6)})^2 - 0.06 \tilde{C}_{HD}^{(6)} \, \delta G_{F}^{(6)} -0.04 \tilde{C}_{HD}^{(8)}
+0.06 \tilde{C}_{H,D2}^{(8)} +0.05 \tilde{C}_{HWB}^{(8)}\nn
&+& 0.19 \tilde{C}_{HW,2}^{(8)} + \frac{0.37}{2} (\tilde{C}_{\substack{Hq\\pp}}^{2,(8)} + \tilde{C}_{\substack{Hq\\pp}}^{3, (8)}+ \tilde{C}_{\substack{Hq\\pp}}^{(8)})
+0.08 (\delta G_{F}^{(6)})^2
-0.22 \delta G_{F}^{(8)}, \\
\langle g_{\rm SM,pp}^{\mathcal{Z},d_{R}}\rangle &=& -0.06,\\
\langle g_{\rm eff,pp}^{\mathcal{Z},d_{R}}\rangle_{{\cal{O}}(v^2/\Lambda^2)}&=&
0.11 \, \tilde{C}_{HD}^{(6)} +0.10 \,  \tilde{C}_{HWB}^{(6)} +0.04 \, \delta G_{F}^{(6)}
+ 0.37 \, \tilde{C}_{\substack{Hd\\pp}}^{(6)}, \\
\langle g_{\rm eff,pp}^{\mathcal{Z},d_{R}}\rangle_{{\cal{O}}(v^4/\Lambda^4)}
&=& -\left(\frac{\tilde{C}_{HD}^{(6)}}{4}+ \frac{\delta G_{F}^{(6)}}{\sqrt{2}}\right)\langle g_{\rm eff,pp}^{\mathcal{Z},d_{R}}\rangle_{{\cal{O}}(v^2/\Lambda^2)}
+  \tilde{C}_{HWB}^{(6)} \left(-0.06\tilde{C}_{HD}^{(6)} +0.10(\tilde{C}_{HB}^{(6)} +\tilde{C}_{HW}^{(6)})  \right) \nn
&-&0.002 (\tilde{C}_{HD}^{(6)})^2 +0.01 \tilde{C}_{HD}^{(6)} \, \delta G_{F}^{(6)} +0.01 \tilde{C}_{HD}^{(8)}
+0.10 \tilde{C}_{H,D2}^{(8)} +0.05 \tilde{C}_{HWB}^{(8)} \nn
&+& 0.19 \tilde{C}_{HW,2}^{(8)} + \frac{0.37}{2} \, \tilde{C}_{\substack{Hd\\pp}}^{(8)}  -0.01 (\delta G_{F}^{(6)})^2 +0.04 \delta G_{F}^{(8)}.
\eea

\subsection{$\Gamma(\mathcal{Z} \rightarrow \bar{\ell} \ell)$}
The effective coupling results to ${\cal{O}}(v^4/\Lambda^4)$ in the case of charged leptons are given by
\bea
\langle g_{\rm SM,pp}^{\mathcal{Z},\ell_{L}}\rangle &=& 0.20, \\
\langle g_{\rm eff,pp}^{\mathcal{Z},\ell_{L}}\rangle_{{\cal{O}}(v^2/\Lambda^2)}
&=&0.24 \tilde{C}_{HD}^{(6)} +0.31 \,  \tilde{C}_{HWB}^{(6)} -0.15 \, \delta G_{F}^{(6)}
+ 0.37 \, (\tilde{C}_{\substack{H\ell\\pp}}^{(6)}+ \tilde{C}_{\substack{H\ell\\pp}}^{3,(6)}), \\
\langle g_{\rm eff,pp}^{\mathcal{Z},\ell_{L}}\rangle_{{\cal{O}}(v^4/\Lambda^4)}
&=& -\left(\frac{\tilde{C}_{HD}^{(6)}}{4}+ \frac{\delta G_{F}^{(6)}}{\sqrt{2}}\right) \langle g_{\rm eff,pp}^{\mathcal{Z},\ell_{L}}\rangle_{{\cal{O}}(v^2/\Lambda^2)}
+  \tilde{C}_{HWB}^{(6)} \left(-0.19 \tilde{C}_{HD}^{(6)} +0.31 (\tilde{C}_{HB}^{(6)} +\tilde{C}_{HW}^{(6)})  \right) \nn
&+&0.01(\tilde{C}_{HD}^{(6)})^2 - 0.04 \tilde{C}_{HD}^{(6)} \, \delta G_{F}^{(6)} -0.03 \tilde{C}_{HD}^{(8)}
+0.26 \tilde{C}_{H,D2}^{(8)} +0.15\tilde{C}_{HWB}^{(8)} \nn
&+& 0.58 \tilde{C}_{HW,2}^{(8)} + \frac{0.37}{2} (\tilde{C}_{\substack{H\ell\\pp}}^{2,(8)} + \tilde{C}_{\substack{H\ell\\pp}}^{3, (8)}+ \tilde{C}_{\substack{H\ell\\pp}}^{(8)})  +0.05 (\delta G_{F}^{(6)})^2 -0.15 \delta G_{F}^{(8)},
\eea
\bea
\langle g_{\rm SM,pp}^{\mathcal{Z},\ell_{R}}\rangle &=& -0.17,\\
\langle g_{\rm eff,pp}^{\mathcal{Z},\ell_{R}}\rangle_{{\cal{O}}(v^2/\Lambda^2)}&=&
0.33 \, \tilde{C}_{HD}^{(6)} +0.31 \,  \tilde{C}_{HWB}^{(6)} +0.12 \, \delta G_{F}^{(6)}
+ 0.37 \, \tilde{C}_{\substack{He\\pp}}^{(6)}, \\
\langle g_{\rm eff,pp}^{\mathcal{Z},\ell_{R}}\rangle_{{\cal{O}}(v^4/\Lambda^4)}
&=& -\left(\frac{\tilde{C}_{HD}^{(6)}}{4}+ \frac{\delta G_{F}^{(6)}}{\sqrt{2}}\right)\langle g_{\rm eff,pp}^{\mathcal{Z},\ell_{R}}\rangle_{{\cal{O}}(v^2/\Lambda^2)}
+  \tilde{C}_{HWB}^{(6)} \left(-0.19\tilde{C}_{HD}^{(6)} +0.31(\tilde{C}_{HB}^{(6)} +\tilde{C}_{HW}^{(6)})  \right) \nn
&-&0.01 (\tilde{C}_{HD}^{(6)})^2 +0.03 \tilde{C}_{HD}^{(6)} \, \delta G_{F}^{(6)} +0.02 \tilde{C}_{HD}^{(8)}
+0.31 \tilde{C}_{H,D2}^{(8)} +0.15 \tilde{C}_{HWB}^{(8)} \nn
&+& 0.58 \tilde{C}_{HW,2}^{(8)} + 0.19 \, \tilde{C}_{\substack{He\\pp}}^{(8)}  -0.04 (\delta G_{F}^{(6)})^2 +0.12 \delta G_{F}^{(8)}.
\eea
\subsection{$\Gamma(\mathcal{Z} \rightarrow \bar{\nu} \nu)$}
The effective coupling results to ${\cal{O}}(v^4/\Lambda^4)$ are given by
\bea
\langle g_{\rm SM,pp}^{\mathcal{Z},\nu_{L}}\rangle &=& -0.37, \\
\langle g_{\rm eff,pp}^{\mathcal{Z},\nu_{L}}\rangle_{{\cal{O}}(v^2/\Lambda^2)}
&=&0.09 \tilde{C}_{HD}^{(6)}  +0.26 \, \delta G_{F}^{(6)}
+ 0.37 \, (\tilde{C}_{\substack{H\ell\\pp}}^{(6)}- \tilde{C}_{\substack{H\ell\\pp}}^{3,(6)}), \\
\langle g_{\rm eff,pp}^{\mathcal{Z},\nu_{L}}\rangle_{{\cal{O}}(v^4/\Lambda^4)}
&=& -\left(\frac{\tilde{C}_{HD}^{(6)}}{4}+ \frac{\delta G_{F}^{(6)}}{\sqrt{2}}\right) \langle g_{\rm eff,pp}^{\mathcal{Z},\ell_{L}}\rangle_{{\cal{O}}(v^2/\Lambda^2)}
 -0.09 (\delta G_{F}^{(6)})^2 +0.26 \delta G_{F}^{(8)} \nn
&-&0.01 (\tilde{C}_{HD}^{(6)})^2 +0.07 \tilde{C}_{HD}^{(6)} \, \delta G_{F}^{(6)} +0.05 \tilde{C}_{HD}^{(8)}
+0.05 \tilde{C}_{H,D2}^{(8)} \nn
&-& \frac{0.37}{2} (\tilde{C}_{\substack{H\ell\\pp}}^{2,(8)} + \tilde{C}_{\substack{H\ell\\pp}}^{3, (8)}- \tilde{C}_{\substack{H\ell\\pp}}^{(8)}).
\eea
\section{\titlemath{$\{\hat{\alpha}_{ew}, \hat{m}_Z, \hat{G}_F \}$}{(aEM, MZ, GF, Mh)} input-parameter scheme at all orders in $(\bar{v}^2_T/\Lambda^2)^n$}
\label{zwidthalphaem}
For the $\{\hat{\alpha}_{ew}, \hat{m}_Z, \hat{G}_F \}$ input-parameter scheme we use the inputs \cite{Berthier:2015oma,Brivio:2017btx}
\begin{equation}
 \begin{aligned}
  \hat e &=\sqrt{4\pi\hat{\a}_{ew}}, \quad
 & \hat v_T &= \frac{1}{2^{1/4}\sqrt{\hat G_F}}, \quad \hat m_Z^2,
 \end{aligned}
 \end{equation}
 to numerically fix values for the 
U(1)$_{\rm{ew}}$ coupling, the vev, and the $\mathcal{Z}$ and Higgs pole masses.
From these inferred numerical values for Lagrangian parameters, one derives numerical values for
the remaining Lagrangian parameters. When considering the geoSMEFT formalism an efficient
way to derive the shifts in the remaining Lagrangian parameters is to first determine
\begin{align}\label{gzshift}
	\overline{g}_Z &= \frac{2 \bar M_Z}{\sqrt{h}_{33}\overline{v}_T}  \\
   &= \hat{g}_Z + \delta g_Z
\end{align}
to a desired order in $(\bar{v}_T/\Lambda)^{(2n)}$.\footnote{When comparing to past work
\cite{Brivio:2017vri,Alonso:2013hga,Berthier:2015oma} the sign of a defined $\hat{g}_Z$ has conventionally
absorbed a factor of $i^2 = -1$ when only considering $\mathcal{L}^{(6)}$ corrections.}
Then we express $g_1$ and $g_2$ in terms of $\overline{e}$, $s_{\overline{\theta}}$ and the $g^{AB}$ metric
using Eqn.~\eqref{geometricelectric},
\begin{align}
	\label{eq:g1g2Input}
	g_1 = \frac{\overline{e}}{c_{\overline{\theta}}\sqrt{g}^{44} + s_{\overline{\theta}}\sqrt{g}^{34}},
	\qquad\qquad
	g_2 = \frac{\overline{e}}{s_{\overline{\theta}}\sqrt{g}^{33} + c_{\overline{\theta}}\sqrt{g}^{34}}.
\end{align}
Further, Eqn.~\eqref{geometricgz} combined with Eqn.~\eqref{eq:g1g2Input} allows $s_{\overline{\theta}}^2$
to be defined in terms of quantities related to input parameters via
\begin{align}
	s_{\overline{\theta}}^2 &= \frac{\overline{e}}{2\overline{g}_Z
		\left[\left(\sqrt{g}^{33}\right)^2+\left(\sqrt{g}^{34}\right)^2\right]
		\left[\left(\sqrt{g}^{44}\right)^2+\left(\sqrt{g}^{34}\right)^2\right]}\left\{
		2\sqrt{g}^{34}\left(\sqrt{g}^{33}-\sqrt{g}^{44}\right)\sqrt{g}_- \right.\nonumber \\&
		+ \frac{\overline{g}_Z}{\overline{e}}\left[
			\left(\sqrt{g}^{33}\right)^2 \left(\sqrt{g}^{44}\right)^2
			+2\left(\sqrt{g}^{44}\right)^2 \left(\sqrt{g}^{34}\right)^2
		+\left(\sqrt{g}^{34}\right)^4\right] \nonumber \\ &
		\left.
		- \sqrt{g}_+ \sqrt{g}_- \sqrt{ \left(\frac{\overline{g}_Z}{\overline{e}}\right)^2
			+ 4 \frac{\overline{g}_Z}{\overline{e}\sqrt{g}_-}
	\sqrt{g}^{34} \left(\sqrt{g}^{33}+\sqrt{g}^{44}\right)-4}\right\}
\end{align}
where $\sqrt{g}_\pm = \sqrt{g}^{33}\sqrt{g}^{44} \pm \left(\sqrt{g}^{34}\right)^2$.

The remaining electroweak Lagrangian parameters are then determined in terms of the inputs by using
\begin{align}
	s_{\theta_Z}^2 &= \frac{\overline{e}}{\overline{g}_Z} \frac{\left(s_{\overline{\theta}}\sqrt{g}^{44}
		- c_{\overline{\theta}}\sqrt{g}^{34}\right)}{\left(c_{\overline{\theta}}\sqrt{g}^{44}
	+ s_{\overline{\theta}}\sqrt{g}^{34}\right)}, \\
	\bar{g}_2 &= \frac{\overline{e} \sqrt{g}^{11}}{\left(s_{\overline{\theta}}\sqrt{g}^{33}
	+ c_{\overline{\theta}}\sqrt{g}^{34}\right)}, \\
	\overline{m}_W^2 &= \frac{\overline{g}_2^2}{4}\left(\sqrt{h}_{11}\right)^2 \overline{v}_T^2
\end{align}
To utilize these relations, one expands out to a fixed order in
$(\bar{v}_T/\Lambda)^{(2n)}$, thereby relating each ``barred" Lagrangian parameter
to parameters defined by input measurements via $\bar{P} = \hat{P} + \delta P$.
This determines $\delta P$ up to the fixed order in $(\bar{v}_T/\Lambda)^{(2n)}$ one is examining.

The following subsections provide numerical results for the partial widths in the SMEFT in this scheme.  For
reference the corresponding SM predictions are given by
\begin{align}
\Gamma^{\rm \hat{\alpha}_{ew}}_{\rm SM}(h \rightarrow \gamma \gamma) &= 1.08 \times 10^{-5} \, {\rm GeV}, & \quad
 \Gamma^{\rm \hat{\alpha}_{ew}}_{\rm SM}(h \rightarrow \mathcal{Z} \gamma)& = 6.61 \times 10^{-6}\, {\rm GeV}, \\
\bar{\Gamma}^{\rm \hat{\alpha}_{ew}}_{\rm SM} (\mathcal{Z} \rightarrow \bar{u} u) &= 0.28 \, {\rm GeV}, & \quad
\bar{\Gamma}^{\rm \hat{\alpha}_{ew}}_{\rm SM} (\mathcal{Z} \rightarrow \bar{d} d) &= 0.36 \, {\rm GeV}, \\
\bar{\Gamma}^{\rm \hat{\alpha}_{ew}}_{\rm SM} (\mathcal{Z} \rightarrow \bar{\ell} \ell) &= 0.08 \, {\rm GeV}, & \quad
\bar{\Gamma}^{\rm \hat{\alpha}_{ew}}_{\rm SM} (\mathcal{Z} \rightarrow \bar{\nu} \nu) &= 0.17 \, {\rm GeV}.
\end{align}

\subsection{$\Gamma(h \rightarrow \gamma \gamma)$}
The partial-square calculation of the partial decay width $\Gamma(h \rightarrow \gamma \gamma)$ is
\begin{align}
&\frac{\Gamma_{p.s.}^{\hat{\alpha}_{ew}}(h \rightarrow \gamma \gamma)}{ \Gamma^{\hat{\alpha}_{ew}}_{\rm SM}(h \rightarrow \gamma \gamma)} \simeq
1 -  748 f^{\hat{\alpha}_{ew}}_1
+374^2 \, (f^{\hat{\alpha}_{ew}}_1)^2
- 349 \, (\tilde{C}_{HW}^{(6)} - \tilde{C}_{HB}^{(6)})^2  \\
&- 53.3\,\tilde{C}_{HWB}^{(6)} \, \left(\tilde{C}_{HB}^{(6)} +8.5 \tilde{C}_{HW}^{(6)} -10.6 \tilde{C}_{HWB}^{(6)} \right)
- 13.9\,\tilde{C}_{HD}^{(6)} \, \left(\tilde{C}_{HB}^{(6)} +18 \tilde{C}_{HW}^{(6)} -17 \tilde{C}_{HWB}^{(6)} \right) \nn
&+ 1370\, \delta G_F^{(6)} \, \left(\tilde{C}_{HB}^{(6)}-0.14 \tilde{C}_{HW}^{(6)} -0.15 \tilde{C}_{HWB}^{(6)} \right),\nonumber
\label{eq:hgamgamPSalpha}
\end{align}
while the full ${\cal{O}}(v^4/\Lambda^4)$ SMEFT result is
\bea
\frac{\Gamma_{\rm SMEFT}^{\hat{\alpha}_{ew}}(h \rightarrow \gamma \gamma)}{ \Gamma^{\hat{\alpha}_{ew}}_{\rm SM}(h \rightarrow \gamma \gamma)}
= \frac{\Gamma_{p.s.}^{\hat{\alpha}_{ew}}(h \rightarrow \gamma \gamma)}{ \Gamma^{\hat{\alpha}_{ew}}_{\rm SM}(h \rightarrow \gamma \gamma)}
-  748 \left[\left(\tilde C_{H\Box}^{(6)} -\frac{\tilde C_{HD}^{(6)}}{4}\right) f^{\hat{\alpha}_{ew}}_1+ f^{\hat{\alpha}_{ew}}_2\right] - 1147 \, (f^{\hat{\alpha}_{ew}}_1)^2. \nn
\eea

\subsection{$\Gamma(h \rightarrow \mathcal{Z} \gamma)$}
The partial-square calculation of the partial decay width $\Gamma(h \rightarrow \mathcal{Z}\gamma)$ is
\bea
\frac{\Gamma_{p.s.}^{\hat{\alpha}_{W}}(h \rightarrow \mathcal{Z} \gamma)}{ \Gamma^{\hat{m}_{W}}_{\rm SM}(h \rightarrow \mathcal{Z} \gamma)} &\simeq&
1 -239 f^{\hat{\alpha}_{W}}_3
+120^2 \, (f^{\hat{\alpha}_{W}}_3)^2
-128 \, (\tilde{C}_{HB}^{(6)} - \,\tilde{C}_{HW}^{(6)})^2 \nn
&+& 27.8\,\tilde{C}_{HD}^{(6)} \, \left(\tilde{C}_{HB}^{(6)}- \tilde{C}_{HW}^{(6)} -4.1 \tilde{C}_{HWB}^{(6)} \right)
-17.1\,\tilde{C}_{HWB}^{(6)} \, \left(\tilde{C}_{HB}^{(6)} - \tilde{C}_{HW}^{(6)} -33 \tilde{C}_{HWB}^{(6)} \right)  \nn
&-& 444\, \delta G_F^{(6)} \, \left(\tilde{C}_{HB}^{(6)}- \tilde{C}_{HW}^{(6)} +1.5 \tilde{C}_{HWB}^{(6)} \right),\nonumber
\eea

while the full ${\cal{O}}(v^4/\Lambda^4)$ SMEFT result is
\bea
\frac{\Gamma_{\rm SMEFT}^{\hat{\alpha}_{ew}}(h \rightarrow \mathcal{Z} \gamma)}{ \Gamma^{\hat{\alpha}_{ew}}_{\rm SM}(h \rightarrow \mathcal{Z} \gamma)}
&=&\frac{\Gamma_{p.s.}^{\hat{\alpha}_{ew}}(h \rightarrow  \mathcal{Z} \gamma)}{ \Gamma^{\hat{\alpha}_{ew}}_{\rm SM}(h \rightarrow \mathcal{Z} \gamma)}
-239 \left[\left(\tilde C_{H\Box}^{(6)} -\frac{\tilde C_{HD}^{(6)}}{4}\right) \, f^{\hat{\alpha}_{ew}}_3+ f^{\hat{\alpha}_{ew}}_4\right]  \nn
&-& 321 \, (f^{\hat{\alpha}_{ew}}_3)^2-239 f^{\hat{\alpha}_{ew}}_3 \left(3.9 \tilde{C}_{HB}^{(6)}+ 0.12 \tilde{C}_{HW}^{(6)} \right).
\eea
\subsection{$\Gamma(\mathcal{Z} \rightarrow\bar{u} u)$}
The effective-coupling results in the $\hat{\alpha}_{ew}$ scheme dictating $\mathcal{Z}$ boson decay to up-type quarks are
\begin{align}
\langle g_{\rm SM,pp}^{\mathcal{Z},u_{L}}\rangle &= -0.26, \\
\langle g_{\rm eff,pp}^{\mathcal{Z},u_{L}}\rangle_{{\cal{O}}(v^2/\Lambda^2)}
&= 0.15 \tilde{C}_{HD}^{(6)} +0.39 \,  \tilde{C}_{HWB}^{(6)} + 0.41 \, \delta G_{F}^{(6)}
+ 0.37 \, (\tilde{C}_{\substack{Hq\\pp}}^{(6)}- \tilde{C}_{\substack{Hq\\pp}}^{3,(6)}), \\
\langle g_{\rm eff,pp}^{\mathcal{Z},u_{L}}\rangle_{{\cal{O}}(v^4/\Lambda^4)}
&= -\left(\frac{\tilde{C}_{HD}^{(6)}}{4}+ \frac{\delta G_{F}^{(6)}}{\sqrt{2}}\right) \langle g_{\rm eff,pp}^{\mathcal{Z},u_{L}}\rangle_{{\cal{O}}(v^2/\Lambda^2)}
\\ &+  \tilde{C}_{HWB}^{(6)} \left(0.35 \tilde{C}_{HD}^{(6)} +0.39 (\tilde{C}_{HB}^{(6)} +\tilde{C}_{HW}^{(6)})+ 0.98 \delta G_{F}^{(6)}  \right) \nn
&+0.02 (\tilde{C}_{HD}^{(6)})^2 + 0.31 \tilde{C}_{HD}^{(6)} \, \delta G_{F}^{(6)} +0.07 \tilde{C}_{HD}^{(8)}
+0.07 \tilde{C}_{H,D2}^{(8)} +0.58 (\tilde{C}_{HWB}^{(6)})^2 \nn
&+ 0.20 \tilde{C}_{HWB}^{(8)}
-  \frac{0.37}{2} (\tilde{C}_{\substack{Hq\\pp}}^{2,(8)} + \tilde{C}_{\substack{Hq\\pp}}^{3, (8)}- \tilde{C}_{\substack{Hq\\pp}}^{(8)})
 +0.15 (\delta G_{F}^{(6)})^2 + 0.41 \delta G_{F}^{(8)}, \nonumber
\end{align}
\begin{align}
\langle g_{\rm SM,pp}^{\mathcal{Z},u_{R}}\rangle &=  0.12,\\
\langle g_{\rm eff,pp}^{\mathcal{Z},u_{R}}\rangle_{{\cal{O}}(v^2/\Lambda^2)}&=
0.05\, \tilde{C}_{HD}^{(6)} +0.39 \,  \tilde{C}_{HWB}^{(6)} + 0.15 \, \delta G_{F}^{(6)}
+ 0.37 \, \tilde{C}_{\substack{Hu\\pp}}^{(6)}, \\
\langle g_{\rm eff,pp}^{\mathcal{Z},u_{R}}\rangle_{{\cal{O}}(v^4/\Lambda^4)}
&= -\left(\frac{\tilde{C}_{HD}^{(6)}}{4}+ \frac{\delta G_{F}^{(6)}}{\sqrt{2}}\right) \langle g_{\rm eff,pp}^{\mathcal{Z},u_{R}}\rangle_{{\cal{O}}(v^2/\Lambda^2)}
\\ &+  \tilde{C}_{HWB}^{(6)} \left(0.35 \tilde{C}_{HD}^{(6)} +0.39 (\tilde{C}_{HB}^{(6)} +\tilde{C}_{HW}^{(6)}) + 0.98 \delta G_{F}^{(6)} \right) \nn
&+0.03 (\tilde{C}_{HD}^{(6)})^2 +0.24 \tilde{C}_{HD}^{(6)} \, \delta G_{F}^{(6)} +0.03 \tilde{C}_{HD}^{(8)}
+0.03 \tilde{C}_{H,D2}^{(8)} + 0.58 (\tilde{C}_{HWB}^{(6)})^2 \nn
&+ 0.20 \tilde{C}_{HWB}^{(8)}
+  \frac{0.37}{2} \, \tilde{C}_{\substack{Hu\\pp}}^{(8)}  +0.24 (\delta G_{F}^{(6)})^2 + 0.15 \delta G_{F}^{(8)}. \nonumber
\end{align}

\subsection{$\Gamma(\mathcal{Z} \rightarrow \bar{b} b)$}
The effective coupling results defining $\mathcal{Z}$ decay to ${\cal{O}}(v^4/\Lambda^4)$
in the case of down quarks are given by
\begin{align}
\langle g_{\rm SM,pp}^{\mathcal{Z},d_{L}}\rangle &= 0.31, \\
\langle g_{\rm eff,pp}^{\mathcal{Z},d_{L}}\rangle_{{\cal{O}}(v^2/\Lambda^2)}
&=-0.12 \tilde{C}_{HD}^{(6)} -0.20 \,  \tilde{C}_{HWB}^{(6)} -0.34 \, \delta G_{F}^{(6)}
+ 0.37 \, (\tilde{C}_{\substack{Hq\\pp}}^{(6)}+ \tilde{C}_{\substack{Hq\\pp}}^{3,(6)}), \\
\langle g_{\rm eff,pp}^{\mathcal{Z},d_{L}}\rangle_{{\cal{O}}(v^4/\Lambda^4)}
&= -\left(\frac{\tilde{C}_{HD}^{(6)}}{4}+ \frac{\delta G_{F}^{(6)}}{\sqrt{2}}\right) \langle g_{\rm eff,pp}^{\mathcal{Z},d_{L}}\rangle_{{\cal{O}}(v^2/\Lambda^2)}
\\ &+  \tilde{C}_{HWB}^{(6)} \left(-0.17 \tilde{C}_{HD}^{(6)} -0.20 (\tilde{C}_{HB}^{(6)} +\tilde{C}_{HW}^{(6)}) -0.49 \delta G_{F}^{(6)} \right) \nn
&-0.003 (\tilde{C}_{HD}^{(6)})^2 - 0.19 \tilde{C}_{HD}^{(6)} \, \delta G_{F}^{(6)} -0.06 \tilde{C}_{HD}^{(8)}
-0.06 \tilde{C}_{H,D2}^{(8)} -0.29 (\tilde{C}_{HWB}^{(6)})^2  \nn
&-0.10 \tilde{C}_{HWB}^{(8)}
+ \frac{0.37}{2} (\tilde{C}_{\substack{Hq\\pp}}^{2,(8)} + \tilde{C}_{\substack{Hq\\pp}}^{3, (8)}+ \tilde{C}_{\substack{Hq\\pp}}^{(8)})  -0.03 (\delta G_{F}^{(6)})^2 -0.34 \delta G_{F}^{(8)}, \nn
\langle g_{\rm SM,pp}^{\mathcal{Z},d_{R}}\rangle &= -0.06,\\
\langle g_{\rm eff,pp}^{\mathcal{Z},d_{R}}\rangle_{{\cal{O}}(v^2/\Lambda^2)}&=
-0.03 \, \tilde{C}_{HD}^{(6)}-0.20 \,  \tilde{C}_{HWB}^{(6)}-0.08\, \delta G_{F}^{(6)}
+ 0.37 \, \tilde{C}_{\substack{Hd\\pp}}^{(6)}, \\
\langle g_{\rm eff,pp}^{\mathcal{Z},d_{R}}\rangle_{{\cal{O}}(v^4/\Lambda^4)}
&= -\left(\frac{\tilde{C}_{HD}^{(6)}}{4}+ \frac{\delta G_{F}^{(6)}}{\sqrt{2}}\right)\langle g_{\rm eff,pp}^{\mathcal{Z},d_{R}}\rangle_{{\cal{O}}(v^2/\Lambda^2)}
\\ &+  \tilde{C}_{HWB}^{(6)} \left(-0.17\tilde{C}_{HD}^{(6)} -0.20(\tilde{C}_{HB}^{(6)} +\tilde{C}_{HW}^{(6)}) -0.49 \delta G_{F}^{(6)} \right) \nn
&-0.015 (\tilde{C}_{HD}^{(6)})^2 -0.12 \tilde{C}_{HD}^{(6)} \, \delta G_{F}^{(6)} -0.01 \tilde{C}_{HD}^{(8)}
-0.01 \tilde{C}_{H,D2}^{(8)} - 0.29(\tilde{C}_{HWB}^{(6)})^2  \nn
&-0.10 \tilde{C}_{HWB}^{(8)}
+  \frac{0.37}{2} \, \tilde{C}_{\substack{Hd\\pp}}^{(8)}  -0.12 (\delta G_{F}^{(6)})^2 -0.08 \delta G_{F}^{(8)}. \nonumber
\end{align}

\subsection{$\Gamma(\mathcal{Z} \rightarrow \bar{\ell} \ell)$}
The effective coupling results to $1/\Lambda^4$ in the case of charged leptons are given by
\begin{align}
\langle g_{\rm SM,pp}^{\mathcal{Z},\ell_{L}}\rangle &= 0.21, \\
\langle g_{\rm eff,pp}^{\mathcal{Z},\ell_{L}}\rangle_{{\cal{O}}(v^2/\Lambda^2)}
&=-0.17\tilde{C}_{HD}^{(6)} -0.59 \,  \tilde{C}_{HWB}^{(6)} -0.49 \, \delta G_{F}^{(6)}
+ 0.37 \, (\tilde{C}_{\substack{H\ell\\pp}}^{(6)}+ \tilde{C}_{\substack{H\ell\\pp}}^{3,(6)}), \\
\langle g_{\rm eff,pp}^{\mathcal{Z},\ell_{L}}\rangle_{{\cal{O}}(v^4/\Lambda^4)}
&= -\left(\frac{\tilde{C}_{HD}^{(6)}}{4}+ \frac{\delta G_{F}^{(6)}}{\sqrt{2}}\right) \langle g_{\rm eff,pp}^{\mathcal{Z},\ell_{L}}\rangle_{{\cal{O}}(v^2/\Lambda^2)}
\\ &+  \tilde{C}_{HWB}^{(6)} \left(-0.52 \tilde{C}_{HD}^{(6)} -0.59 (\tilde{C}_{HB}^{(6)} +\tilde{C}_{HW}^{(6)}) -1.46 \delta G_{F}^{(6)} \right) \nn
&-0.03 (\tilde{C}_{HD}^{(6)})^2 - 0.43 \tilde{C}_{HD}^{(6)} \, \delta G_{F}^{(6)} -0.09 \tilde{C}_{HD}^{(8)}
-0.09 \tilde{C}_{H,D2}^{(8)} - 0.88 (\tilde{C}_{HWB}^{(6)})^2 \nn
&-0.29 \tilde{C}_{HWB}^{(8)}
+ \frac{0.37}{2} (\tilde{C}_{\substack{H\ell\\pp}}^{2,(8)} + \tilde{C}_{\substack{H\ell\\pp}}^{3, (8)}+ \tilde{C}_{\substack{H\ell\\pp}}^{(8)})  -0.26 (\delta G_{F}^{(6)})^2 -0.49 \delta G_{F}^{(8)},\nonumber
\end{align}
\begin{align}
\langle g_{\rm SM,pp}^{\mathcal{Z},\ell_{R}}\rangle &= -0.17,\\
\langle g_{\rm eff,pp}^{\mathcal{Z},\ell_{R}}\rangle_{{\cal{O}}(v^2/\Lambda^2)}&=
-0.08 \, \tilde{C}_{HD}^{(6)} -0.59 \,  \tilde{C}_{HWB}^{(6)} -0.23 \, \delta G_{F}^{(6)}
+ 0.37 \, \tilde{C}_{\substack{He\\pp}}^{(6)}, \\
\langle g_{\rm eff,pp}^{\mathcal{Z},\ell_{R}}\rangle_{{\cal{O}}(v^4/\Lambda^4)}
&= -\left(\frac{\tilde{C}_{HD}^{(6)}}{4}+ \frac{\delta G_{F}^{(6)}}{\sqrt{2}}\right)\langle g_{\rm eff,pp}^{\mathcal{Z},\ell_{R}}\rangle_{{\cal{O}}(v^2/\Lambda^2)}
\\ &+  \tilde{C}_{HWB}^{(6)} \left(-0.52\tilde{C}_{HD}^{(6)} -0.59(\tilde{C}_{HB}^{(6)} +\tilde{C}_{HW}^{(6)})  -1.46 \delta G_{F}^{(6)}\right) \nn
&-0.04 (\tilde{C}_{HD}^{(6)})^2 -0.37 \tilde{C}_{HD}^{(6)} \, \delta G_{F}^{(6)} -0.04\tilde{C}_{HD}^{(8)}
-0.04 \tilde{C}_{H,D2}^{(8)}- 0.88(\tilde{C}_{HWB}^{(6)})^2  \nn
&-0.29 \tilde{C}_{HWB}^{(8)}
+ 0.19 \, \tilde{C}_{\substack{He\\pp}}^{(8)}  -0.36 (\delta G_{F}^{(6)})^2 -0.23 \delta G_{F}^{(8)}.
\nonumber
\end{align}
\subsection{$\Gamma(\mathcal{Z} \rightarrow \bar{\nu} \nu)$}
The effective coupling results to ${\cal{O}}(v^4/\Lambda^4)$ are given by
\bea
\langle g_{\rm SM,pp}^{\mathcal{Z},\nu_{L}}\rangle &=& -0.37, \\
\langle g_{\rm eff,pp}^{\mathcal{Z},\nu_{L}}\rangle_{{\cal{O}}(v^2/\Lambda^2)}
&=&0.09 \tilde{C}_{HD}^{(6)}  +0.26 \, \delta G_{F}^{(6)}
+ 0.37 \, (\tilde{C}_{\substack{H\ell\\pp}}^{(6)}- \tilde{C}_{\substack{H\ell\\pp}}^{3,(6)}), \\
\langle g_{\rm eff,pp}^{\mathcal{Z},\nu_{L}}\rangle_{{\cal{O}}(v^4/\Lambda^4)}
&=& -\left(\frac{\tilde{C}_{HD}^{(6)}}{4}+ \frac{\delta G_{F}^{(6)}}{\sqrt{2}}\right) \langle g_{\rm eff,pp}^{\mathcal{Z},\ell_{L}}\rangle_{{\cal{O}}(v^2/\Lambda^2)}
 -0.09 (\delta G_{F}^{(6)})^2 +0.26 \delta G_{F}^{(8)} \nn
&-&0.01 (\tilde{C}_{HD}^{(6)})^2 +0.07 \tilde{C}_{HD}^{(6)} \, \delta G_{F}^{(6)} +0.05 \tilde{C}_{HD}^{(8)}
+0.05 \tilde{C}_{H,D2}^{(8)} \nn
&-& \frac{0.37}{2} (\tilde{C}_{\substack{H\ell\\pp}}^{2,(8)} + \tilde{C}_{\substack{H\ell\\pp}}^{3, (8)}- \tilde{C}_{\substack{H\ell\\pp}}^{(8)}).
\eea

\bibliographystyle{JHEP}
\bibliography{bibliography.bib}

\end{document}